\documentclass[12pt]{article}

\usepackage{CJKutf8}
\usepackage{cite}
\usepackage{graphicx}
\usepackage{amsmath}
\usepackage{amsthm}

\usepackage{geometry}
\usepackage{framed}
\usepackage{multirow}
\usepackage{rotating}
\usepackage{amsfonts}
\usepackage{longtable}
\usepackage{array} 

\usepackage{fancyhdr}
\begin{document}

\title{State-Space Modelling and Analysis}
\date{}

\author{Hao Li 
\thanks{Namely \begin{CJK}{UTF8}{gbsn}李颢\end{CJK}, the same author of the works \cite{Li2026ACTPA_SJTU_2, Li2026ACTPA_SJTU_1}.} }

\maketitle

\begin{abstract}
\textbf{Control science} is a core representative of the third industrial revolution and is so important to modern civilization. \textbf{Control systems} are the main subject of control science and may involve many aspects of consideration, such as hardware consideration, software consideration, operation consideration, maintenance consideration, economy consideration, society consideration. However, besides all such aspects of consideration, one aspect that is most essential to the control system is methodology consideration in mathematical sense, knowledge on which is what we refer to as \textbf{control theory}. Besides its importance from the mathematical perspective, control theory is even more charming as it is deeply rooted in practical applications. Charms of control theory consist in both \textit{know-why} and \textit{know-how} and it is the fusion of control theory and practical applications that highlights such charms. Control theory for practical applications, especially when somewhat with so-called ``advanced'' flavour, involves several fundamental aspects. This article introduces the \textit{State-Space Modelling and Analysis} aspect of \textit{Advanced Control Theory for Practical Applications} \cite{Li2026ACTPA_SJTU_2, Li2026ACTPA_SJTU_1}.
\end{abstract}

\section{State-space modelling}  \label{sec:state_space_modelling}

\textit{State-space modelling and analysis} is closely related to \textit{modern control} in contrast with \textit{classical control}. As explained in the previous book \textit{Control Theory For Practical Applications} \cite{Li2024CTPA_Springer, Li2024CTPA_SJTU_1}, 
there is no strict and distinct categorization of classical control and modern control. Roughly speaking, from the perspective of history, classical control appeared before and flourished during the second world war, whereas modern control made its debut after the second world war especially during 1960s. From the perspective of mathematics, classical control normally involves $s$-domain analysis based on the Laplace transform, whereas modern control normally involves state-space analysis and a much wider range of mathematical techniques. From the perspective of problem complexity, classical control normally applies to \textit{single-input-single-output} control problems which are comparatively simple, whereas modern control can apply to more complicated control problems such as \textit{single-input-multiple-output} and \textit{multiple-input-multiple-output} control problems.

\subsection{State differential equation}

For a control system with certain target process
\footnote{If actuator dynamics is not negligible with respect to process dynamics, then the actuator can be incorporated into the target process.},
the set of properties that characterizes the target process is called its \textbf{process state} (from the perspective of the target process itself), or its \textbf{system state} (from the perspective of the control system holistically), or simply its \textbf{state}.

Generally, given a control system with its state denoted as $\mathbf{x}$ and its control input to the target process denoted as $\mathbf{u}$, dynamics of the state $\mathbf{x}$ can be modelled generically by a \textbf{state differential equation} as
\begin{equation}  \label{eq:state_differential_equation}
\frac{\mathrm{d}}{\mathrm{d} t} \mathbf{x} = f(\mathbf{x}, \mathbf{u}),
\end{equation}
which reflects the spirit of \textbf{state-space modelling} for the control system.

\subsubsection*{Application: single inverted pendulum}

\begin{figure}[h!]
\begin{center}
\includegraphics[width=0.25\columnwidth]{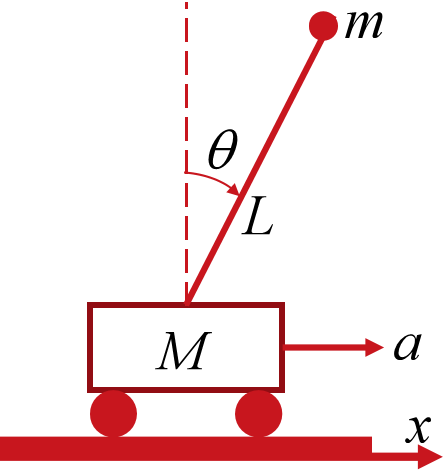}
\end{center}
\caption{Inverted pendulum control (cart acceleration $a$, cart position $x$, and inverted pendulum angle $\theta$)}
\label{fig:inverted_pendulum_control}
\end{figure}

Single inverted pendulum control aims at moving a cart to a target position while balancing an inverted pendulum on the cart, as illustrated in Figure \ref{fig:inverted_pendulum_control}. The process is movement of the cart as well as the inverted pendulum. The concerned process output is the single inverted pendulum state that consists of the cart position, the cart speed, the inverted pendulum angle, and the inverted pendulum angular speed. The expected inverted pendulum state consists of the target cart position, zero cart speed, zero inverted pendulum angle, and zero inverted pendulum angular speed.

Dynamics of the single inverted pendulum control system's state $\mathbf{x}$ is modelled by a state differential equation as
\begin{equation}  \label{eq:SIP_state_DE}
\frac{\mathrm{d}}{\mathrm{d} t} \mathbf{x} \equiv \frac{\mathrm{d}}{\mathrm{d} t} \begin{bmatrix} \theta \\ \frac{\mathrm{d} \theta}{\mathrm{d} t} \\ x \\ \frac{\mathrm{d} x}{\mathrm{d} t} \end{bmatrix} = \begin{bmatrix} \frac{\mathrm{d} \theta}{\mathrm{d} t} \\ \frac{\sin \theta}{L} g - \frac{\cos \theta}{L} a \\ \frac{\mathrm{d} x}{\mathrm{d} t} \\ a \end{bmatrix} \equiv f(\mathbf{x}, a),
\end{equation}
where the state 
\begin{align*}
\mathbf{x} \equiv \begin{bmatrix} \theta & \frac{\mathrm{d} \theta}{\mathrm{d} t} & x & \frac{\mathrm{d} x}{\mathrm{d} t} \end{bmatrix}^\mathrm{T}
\end{align*}
consists of the inverted pendulum angle and angular velocity, and the cart position and velocity. The control input is the single-input of cart acceleration $a$. Refer to Section \ref{sec:SIP_dynamics} in Appendix \ref{app:system_dynamics} for derivation details.

\subsubsection*{Application: double inverted pendulum}

\begin{figure}[h!]
\begin{center}
\includegraphics[width=0.25\columnwidth]{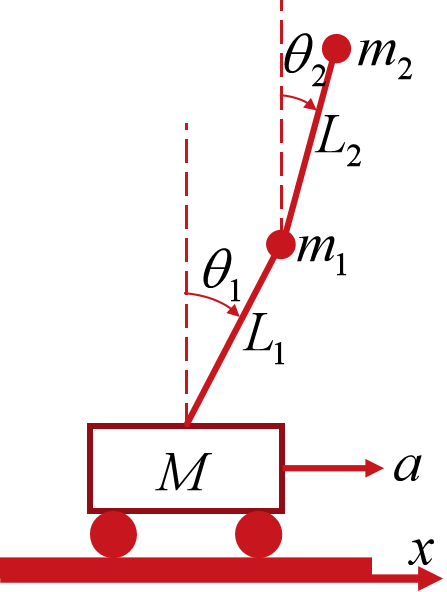}
\end{center}
\caption{Double inverted pendulum control}
\label{fig:double_inverted_pendulum_control}
\end{figure}

Elegance and difficulty of single inverted pendulum control have already been explained in the previous book \textit{Control Theory For Practical Applications} \cite{Li2024CTPA_Springer, Li2024CTPA_SJTU_1}. If the single inverted pendulum control problem was not challenging enough to readers, one might consider an even more challenging problem, namely the double inverted pendulum control problem, as illustrated in Figure \ref{fig:double_inverted_pendulum_control}. The double inverted pendulum control system aims at controlling the cart acceleration $a$ so that the cart is kept at a specific position $x$ and both inverted pendulum angles $\theta_1$ and $\theta_2$ are kept zero.

Dynamics of the double inverted pendulum control system's state $\mathbf{x}$ is modelled by a state differential equation as
\begin{align}  \label{eq:DIP_state_DE}
\frac{\mathrm{d}}{\mathrm{d} t} \mathbf{x} &\equiv \frac{\mathrm{d}}{\mathrm{d} t} \begin{bmatrix} \theta_1 \\ \frac{\mathrm{d} \theta_1}{\mathrm{d} t} \\ \theta_2 \\ \frac{\mathrm{d} \theta_2}{\mathrm{d} t} \\ x \\ \frac{\mathrm{d} x}{\mathrm{d} t} \end{bmatrix} = \begin{bmatrix} \frac{\mathrm{d} \theta_1}{\mathrm{d} t} \\
\frac{(\sin \theta_1 + \frac{m_2}{m_1} \sin \Delta \theta \cos \theta_2)}{1 + \frac{m_2}{m_1} (\sin \Delta \theta)^2} \frac{g}{L_1} - \frac{(\cos \theta_1 - \frac{m_2}{m_1} \sin \Delta \theta \sin \theta_2)}{1 + \frac{m_2}{m_1} (\sin \Delta \theta)^2} \frac{a}{L_1} \\
\frac{\mathrm{d} \theta_2}{\mathrm{d} t} \\
- \frac{\sin \Delta \theta \cos \theta_1}{1 + \frac{m_2}{m_1} (\sin \Delta \theta)^2} (1 + \frac{m_2}{m_1}) \frac{g}{L_2} - \frac{\sin \Delta \theta \sin \theta_1}{1 + \frac{m_2}{m_1} (\sin \Delta \theta)^2} (1 + \frac{m_2}{m_1}) \frac{a}{L_2} \\
\frac{\mathrm{d} x}{\mathrm{d} t} \\ a \end{bmatrix}  \nonumber  \\
  &= \begin{bmatrix} \frac{\mathrm{d} \theta_1}{\mathrm{d} t} \\ 
\frac{(\sin \theta_1 + \frac{m_2}{m_1} \sin \Delta \theta \cos \theta_2)}{1 + \frac{m_2}{m_1} (\sin \Delta \theta)^2} \frac{g}{L_1} \\ 
\frac{\mathrm{d} \theta_2}{\mathrm{d} t} \\ 
- \frac{\sin \Delta \theta \cos \theta_1}{1 + \frac{m_2}{m_1} (\sin \Delta \theta)^2} (1 + \frac{m_2}{m_1}) \frac{g}{L_2}  \\ 
\frac{\mathrm{d} x}{\mathrm{d} t} \\ 0 \end{bmatrix} 
  + \begin{bmatrix} 0 \\ 
- \frac{(\cos \theta_1 - \frac{m_2}{m_1} \sin \Delta \theta \sin \theta_2)}{1 + \frac{m_2}{m_1} (\sin \Delta \theta)^2} \frac{1}{L_1} \\ 
0 \\ 
- \frac{\sin \Delta \theta \sin \theta_1}{1 + \frac{m_2}{m_1} (\sin \Delta \theta)^2} (1 + \frac{m_2}{m_1}) \frac{1}{L_2}  \\ 
0 \\ 1 \end{bmatrix} a \equiv f(\mathbf{x}, a),
\end{align}
where 
\begin{align*}
\Delta \theta \equiv \theta_1 - \theta_2
\end{align*}
and the state 
\begin{align*}
\mathbf{x} \equiv \begin{bmatrix} \theta_1 & \frac{\mathrm{d} \theta_1}{\mathrm{d} t} & \theta_2 & \frac{\mathrm{d} \theta_2}{\mathrm{d} t} & x & \frac{\mathrm{d} x}{\mathrm{d} t} \end{bmatrix}^\mathrm{T}
\end{align*}
consists of the first inverted pendulum angle and angular velocity, the second inverted pendulum angle and angular velocity, and the cart position and velocity. The control input is the single-input of cart acceleration $a$ as well.

\begin{figure}[h!]
\begin{center}
\includegraphics[width=0.25\columnwidth]{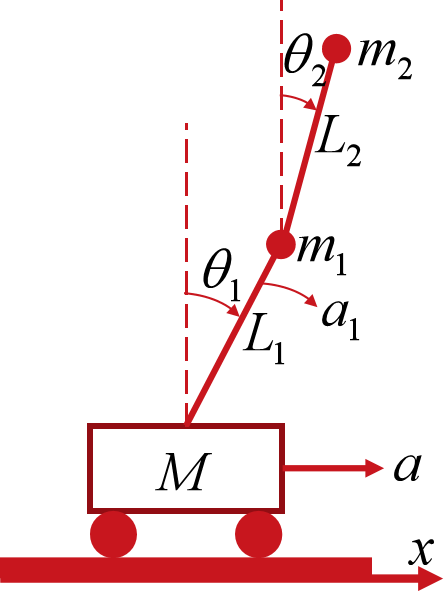}
\end{center}
\caption{Variant of double inverted pendulum control}
\label{fig:double_inverted_pendulum_variant_control}
\end{figure}

Consider a variant of the double inverted pendulum control problem, as illustrated in Figure \ref{fig:double_inverted_pendulum_variant_control}. Compared with the original double inverted pendulum control problem, the variant of the double inverted pendulum control problem shares all configurations except that its control input to the target process is no longer the \textit{single-input} of cart acceleration $a$ but the \textit{multiple-input} of both cart acceleration $a$ and first inverted pendulum angular acceleration $a_1$. The variant of the double inverted pendulum control system aims at controlling the cart acceleration $a$ and the first inverted pendulum angular acceleration $a_1$ simultaneously so that the cart is kept at a specific position $x$ and both inverted pendulum angles $\theta_1$ and $\theta_2$ are kept zero.

For the variant of the double inverted pendulum control system, dynamics of the state $\mathbf{x}$ (the same to that defined above) is modelled by a state differential equation as
\begin{align}  \label{eq:DIP_state_DE_MIMO}
\frac{\mathrm{d}}{\mathrm{d} t} \mathbf{x} &\equiv \frac{\mathrm{d}}{\mathrm{d} t} \begin{bmatrix} \theta_1 \\ \frac{\mathrm{d} \theta_1}{\mathrm{d} t} \\ \theta_2 \\ \frac{\mathrm{d} \theta_2}{\mathrm{d} t} \\ x \\ \frac{\mathrm{d} x}{\mathrm{d} t} \end{bmatrix} = \begin{bmatrix} \frac{\mathrm{d} \theta_1}{\mathrm{d} t} \\ 
\frac{(\sin \theta_1 + \frac{m_2}{m_1} \sin \Delta \theta \cos \theta_2)}{1 + \frac{m_2}{m_1} (\sin \Delta \theta)^2} \frac{g}{L_1} - \frac{(\cos \theta_1 - \frac{m_2}{m_1} \sin \Delta \theta \sin \theta_2)}{1 + \frac{m_2}{m_1} (\sin \Delta \theta)^2} \frac{a}{L_1} + \frac{a_1}{1 + \frac{m_2}{m_1} (\sin \Delta \theta)^2} \\ 
\frac{\mathrm{d} \theta_2}{\mathrm{d} t} \\ 
- \frac{\sin \Delta \theta \cos \theta_1}{1 + \frac{m_2}{m_1} (\sin \Delta \theta)^2} (1 + \frac{m_2}{m_1}) \frac{g}{L_2} - \frac{\sin \Delta \theta \sin \theta_1}{1 + \frac{m_2}{m_1} (\sin \Delta \theta)^2} (1 + \frac{m_2}{m_1}) \frac{a}{L_2} - \frac{\cos \Delta \theta \frac{L_1}{L_2} a_1}{1 + \frac{m_2}{m_1} (\sin \Delta \theta)^2}  \\ 
\frac{\mathrm{d} x}{\mathrm{d} t} \\ a \end{bmatrix}  \nonumber  \\
  &= \begin{bmatrix} \frac{\mathrm{d} \theta_1}{\mathrm{d} t} \\ 
\frac{(\sin \theta_1 + \frac{m_2}{m_1} \sin \Delta \theta \cos \theta_2)}{1 + \frac{m_2}{m_1} (\sin \Delta \theta)^2} \frac{g}{L_1} \\ 
\frac{\mathrm{d} \theta_2}{\mathrm{d} t} \\ 
- \frac{\sin \Delta \theta \cos \theta_1}{1 + \frac{m_2}{m_1} (\sin \Delta \theta)^2} (1 + \frac{m_2}{m_1}) \frac{g}{L_2}  \\ 
\frac{\mathrm{d} x}{\mathrm{d} t} \\ 0 \end{bmatrix} 
  - \begin{bmatrix} 0 \\ 
\frac{(\cos \theta_1 - \frac{m_2}{m_1} \sin \Delta \theta \sin \theta_2)}{1 + \frac{m_2}{m_1} (\sin \Delta \theta)^2} \frac{1}{L_1} \\ 
0 \\ 
\frac{\sin \Delta \theta \sin \theta_1}{1 + \frac{m_2}{m_1} (\sin \Delta \theta)^2} (1 + \frac{m_2}{m_1}) \frac{1}{L_2}  \\ 
0 \\ - 1 \end{bmatrix} a 
  + \begin{bmatrix} 0 \\ \frac{1}{1 + \frac{m_2}{m_1} (\sin \Delta \theta)^2} \\ 0 \\ - \frac{\cos \Delta \theta \frac{L_1}{L_2}}{1 + \frac{m_2}{m_1} (\sin \Delta \theta)^2} \\ 0 \\ 0 \end{bmatrix} a_1   \nonumber  \\
  &\equiv f(\mathbf{x}) - g(\mathbf{x}) a + g_1(\mathbf{x}) a_1.
\end{align}
Refer to Section \ref{sec:DIP_dynamics} in Appendix \ref{app:system_dynamics} for derivation details.

\subsubsection*{Application: low-speed autonomous vehicle}

\begin{figure}[h!]
\begin{center}
\includegraphics[width=0.6\columnwidth]{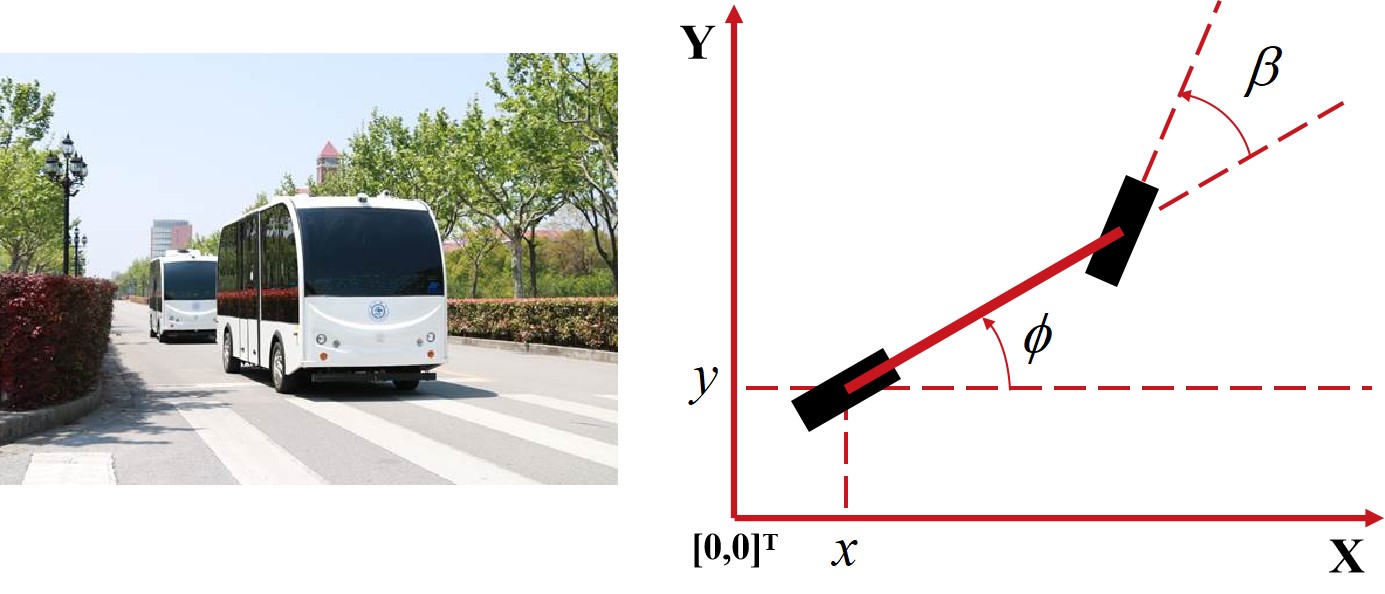}
\end{center}
\caption{Low-speed autonomous vehicle navigation}
\label{fig:low_speed_autonomous_vehicle}
\end{figure}

As illustrated in Figure \ref{fig:low_speed_autonomous_vehicle}, dynamics of the autonomous vehicle control system's state $\mathbf{x}$ (at constant low-speed $v$) is modelled by a state differential equation as
\begin{equation}  \label{eq:bicycle_kinematics_model}
\frac{\mathrm{d}}{\mathrm{d} t} \mathbf{x} \equiv \frac{\mathrm{d}}{\mathrm{d} t} \begin{bmatrix} x \\ y \\ \phi \\ \beta \end{bmatrix} = \begin{bmatrix} v \cos \phi \\ v \sin \phi \\ \frac{v}{L} \tan \beta \\ \frac{1}{\tau_{\beta}} (\beta_I - \beta) \end{bmatrix} \equiv f(\mathbf{x}, \beta_I),
\end{equation}
where the state 
\begin{align*}
\mathbf{x} \equiv \begin{bmatrix} x & y & \phi & \beta \end{bmatrix}^\mathrm{T}
\end{align*}
consists of the vehicle longitudinal position, the vehicle lateral position, the vehicle orientation or heading angle, and the vehicle steering angle. $L$ denotes the vehicle wheel-base, $\tau_{\beta}$ denotes the time-constant of the steer controller, and $\beta_I$ denotes the vehicle steering angle command which serves as control input. The model described in (\ref{eq:bicycle_kinematics_model}) is called the \textit{bicycle kinematics model}
\footnote{Sometimes, the model may be reduced to
\begin{align*}
\left\{
\begin{array}{l l}
\frac{\mathrm{d}}{\mathrm{d} t} x &= v \cos \phi \\
\frac{\mathrm{d}}{\mathrm{d} t} y &= v \sin \phi \\
\frac{\mathrm{d}}{\mathrm{d} t} \phi &= \frac{v}{L} \tan \beta 
\end{array}
\right.
\end{align*}
where vehicle steering dynamics is neglected. In the reduced model, the steering angle $\beta$ serves directly as control input to the vehicle.}. 

Vehicle lateral control i.e. steering control is the core control part of an autonomous vehicle. The vehicle lateral dynamics is extracted as
\begin{align*}
\left\{
\begin{array}{l l}
\frac{\mathrm{d}}{\mathrm{d} t} y &= v \sin \phi \\
\frac{\mathrm{d}}{\mathrm{d} t} \phi &= \frac{v}{L} \tan \beta \\
\frac{\mathrm{d}}{\mathrm{d} t} \beta &= \frac{1}{\tau_{\beta}} (\beta_I - \beta) 
\end{array}
\right.
\end{align*}
and formalized by a state differential equation as
\begin{equation}  \label{eq:vehicle_lateral_control}
\frac{\mathrm{d}}{\mathrm{d} t} \mathbf{x} \equiv \frac{\mathrm{d}}{\mathrm{d} t} \begin{bmatrix} y \\ \phi \\ \beta \end{bmatrix} = \begin{bmatrix} v \sin \phi \\ \frac{v}{L} \tan \beta \\ \frac{1}{\tau_{\beta}} (\beta_I - \beta) \end{bmatrix} \equiv f(\mathbf{x}, \beta_I),
\end{equation}
where 
\begin{align*}
\mathbf{x} \equiv \begin{bmatrix} y & \phi & \beta \end{bmatrix}^\mathrm{T}
\end{align*}
denotes the vehicle lateral state
\footnote{For analysis of vehicle lateral dynamics, the vehicle lateral position $y$ and the vehicle orientation angle $\phi$ actually refer to the lateral position and orientation angle of the vehicle with respect to certain local road reference. In other words, $y$ and $\phi$ here refer to the relative lateral position and orientation angle in certain local road reference, instead of absolute ones in the global world reference.}
and the vehicle steering angle command $\beta_I$ serves as control input to the vehicle. The model described in (\ref{eq:vehicle_lateral_control}) is called the \textit{bicycle lateral kinematics model}. Refer to Section \ref{sec:vehicle_dynamics_no_slip} in Appendix \ref{app:system_dynamics} for derivation details.

\subsubsection*{Application: high-speed autonomous vehicle}

As just mentioned above, vehicle lateral control is the core control part of an autonomous vehicle. Dynamics of the autonomous vehicle control system's lateral state $\mathbf{x}$ (at constant high-speed $v$) is modelled by a state differential equation as
\begin{equation}  \label{eq:vehicle_lateral_dynamics_with_slip}
\frac{\mathrm{d}}{\mathrm{d} t} \mathbf{x} \equiv \frac{\mathrm{d}}{\mathrm{d} t} \begin{bmatrix} y \\ \frac{\mathrm{d} y}{\mathrm{d} t} \\ \phi \\ \frac{\mathrm{d} \phi}{\mathrm{d} t} \\ \beta \end{bmatrix} 
= \begin{bmatrix} \frac{\mathrm{d} y}{\mathrm{d} t} \\ 
\frac{2 C_f}{m} \frac{v (\beta - \arctan \frac{\dot y - v \phi + L_f \dot \phi}{v})}{\sqrt{(\dot y - v \phi + L_f \dot \phi)^2 + v^2}} + \frac{2 C_r}{m} \frac{v (- \arctan \frac{\dot y - v \phi - L_r \dot \phi}{v})}{\sqrt{(\dot y - v \phi - L_r \dot \phi)^2 + v^2}} \\ 
\frac{\mathrm{d} \phi}{\mathrm{d} t} \\ 
\frac{2 C_f L_f}{J} \frac{v (\beta - \arctan \frac{\dot y - v \phi + L_f \dot \phi}{v})}{\sqrt{(\dot y - v \phi + L_f \dot \phi)^2 + v^2}} -  \frac{2 C_r L_r}{J} \frac{v (- \arctan \frac{\dot y - v \phi - L_r \dot \phi}{v})}{\sqrt{(\dot y - v \phi - L_r \dot \phi)^2 + v^2}} \\
\frac{1}{\tau_{\beta}} (\beta_I - \beta) \end{bmatrix} \equiv f(\mathbf{x}, \beta_I),
\end{equation}
where the vehicle lateral state 
\begin{align*}
\mathbf{x} \equiv \begin{bmatrix} y & \frac{\mathrm{d} y}{\mathrm{d} t} & \phi & \frac{\mathrm{d} \phi}{\mathrm{d} t} & \beta \end{bmatrix}^\mathrm{T}
\end{align*}
namely
\begin{align*}
\mathbf{x} \equiv \begin{bmatrix} y & \dot y & \phi & \dot \phi & \beta \end{bmatrix}^\mathrm{T}
\end{align*}
consists of the vehicle lateral position with respect to certain local road reference, the vehicle lateral velocity in the local road reference, the vehicle orientation or heading angle (namely yaw angle) with respect to the local road reference, the vehicle yaw rate, and the vehicle steering angle. In (\ref{eq:vehicle_lateral_dynamics_with_slip}), $m$ denotes the vehicle mass, $J$ denotes the rotating inertia of the vehicle, $L_f$ denotes the length between the vehicle mass center or gravity center and the front wheel, $L_r$ denotes the length between the vehicle gravity center and the rear wheel
\footnote{The sum $L = L_f + L_r$ denotes the vehicle wheel-base.}, 
$C_f$ denotes the front tyre cornering stiffness, $C_r$ denotes the rear tyre cornering stiffness, $\tau_{\beta}$ denotes the time-constant of the steer controller, and $\beta_I$ denotes the vehicle steering angle command which serves as control input. Refer to Section \ref{sec:vehicle_dynamics_with_slip} in Appendix \ref{app:system_dynamics} for derivation details.

\subsubsection*{Application: autonomous motorcycle (or bicycle)}

\begin{figure}[h!]
\begin{center}
\includegraphics[width=0.32\columnwidth]{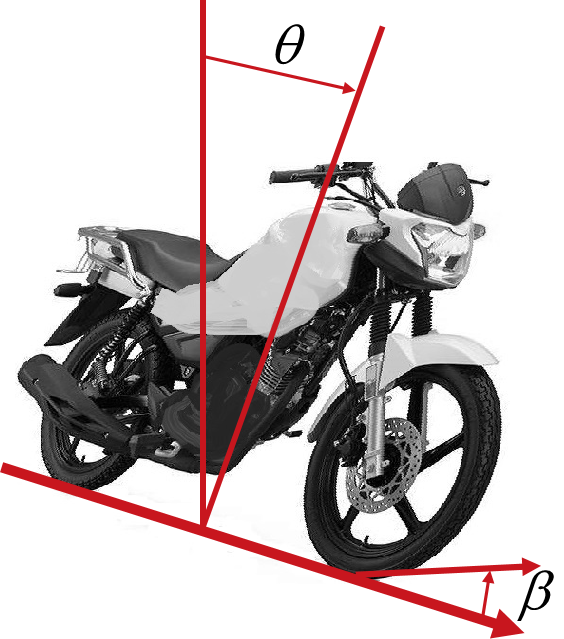}
\end{center}
\caption{Motorcycle control (steering angle $\beta$ and roll angle $\theta$)}
\label{fig:motorcycle_control}
\end{figure}

Motorcycle (or bicycle) control consists of motorcycle longitudinal control and motorcycle lateral control. The former is essentially the same to that for an autonomous vehicle and comparatively simple, whereas the latter is more complicated and plays a core role for the autonomous motorcycle. Motorcycle lateral control aims at steering the front wheel such that the motorcycle can maintain a specific lateral position (say the lane center) as well as its vertical balance, as illustrated in Figure \ref{fig:motorcycle_control}.

Dynamics of the motorcycle control system's state $\mathbf{x}$ is modelled by a state differential equation as
\begin{equation}  \label{eq:motorcycle_state_DE}
\frac{\mathrm{d}}{\mathrm{d} t} \mathbf{x} \equiv \frac{\mathrm{d}}{\mathrm{d} t} \begin{bmatrix} x \\ y \\ \phi \\ \beta \\ \theta \\ \frac{\mathrm{d} \theta}{\mathrm{d} t} \end{bmatrix} 
= \begin{bmatrix} v \cos \phi \\ v \sin \phi \\ \frac{v}{L} \tan \beta \\ \frac{1}{\tau_{\beta}} (\beta_I - \beta) \\ \frac{\mathrm{d} \theta}{\mathrm{d} t} \\
\frac{\sin \theta}{H} g - \frac{\cos \theta}{H} \frac{v^2}{L} \tan \beta \end{bmatrix} \equiv f(\mathbf{x}, \beta_I),
\end{equation}
where the state 
\begin{align*}
\mathbf{x} \equiv \begin{bmatrix} x & y & \phi & \beta & \theta & \frac{\mathrm{d} \theta}{\mathrm{d} t} \end{bmatrix}^\mathrm{T}
\end{align*}
consists of the motorcycle longitudinal position, the motorcycle lateral position, the motorcycle orientation or heading angle (namely yaw angle), the motorcycle steering angle, the motorcycle vertical angle (namely roll angle), and the motorcycle vertical angular velocity. The control input is the motorcycle steering angle command $\beta_I$. Besides, for motorcycle parameters, $L$ denotes the motorcycle wheel-base, $H$ denotes the height of the motorcycle gravity center, and $\tau_{\beta}$ denotes the time-constant of the steer controller. 

Motorcycle lateral control is the core control part of an autonomous motorcycle. The motorcycle lateral state 
\begin{align*}
\mathbf{x} \equiv \begin{bmatrix} y & \phi & \beta & \theta & \frac{\mathrm{d} \theta}{\mathrm{d} t} \end{bmatrix}^\mathrm{T}
\end{align*}
is extracted from the original one by removing the motorcycle longitudinal position $x$. Dynamics of the motorcycle lateral state is extracted as well and formalized by a state differential equation as
\begin{equation}  \label{eq:motorcycle_lateral_control}
\frac{\mathrm{d}}{\mathrm{d} t} \mathbf{x} \equiv \frac{\mathrm{d}}{\mathrm{d} t} \begin{bmatrix} y \\ \phi \\ \beta \\ \theta \\ \frac{\mathrm{d} \theta}{\mathrm{d} t} \end{bmatrix}
= \begin{bmatrix} v \sin \phi \\ \frac{v}{L} \tan \beta \\ - \frac{1}{\tau_{\beta}} \beta \\ \frac{\mathrm{d} \theta}{\mathrm{d} t} \\
\frac{\sin \theta}{H} g - \frac{\cos \theta}{H} \frac{v^2}{L} \tan \beta \end{bmatrix} + \begin{bmatrix} 0 \\ 0 \\ \frac{1}{\tau_{\beta}} \\ 0 \\ 0 \end{bmatrix} \beta_I \equiv f(\begin{bmatrix} y \\ \phi \\ \beta \\ \theta \\ \frac{\mathrm{d} \theta}{\mathrm{d} t} \end{bmatrix}) + \begin{bmatrix} 0 \\ 0 \\ \frac{1}{\tau_{\beta}} \\ 0 \\ 0 \end{bmatrix} \beta_I.
\end{equation}
Refer to Section \ref{sec:motorcycle_dynamics} in Appendix \ref{app:system_dynamics} for derivation details.

\subsection{Linear state differential equation}  \label{sec:linear_state_differential_equation}

Generally, given a control system with its state denoted as $\mathbf{x}$ and its control input to the target process denoted as $\mathbf{u}$. Suppose its state $\mathbf{x}$ is within a range about certain operation point $\mathbf{x}_0$ (usually an equilibrium state) in which its state differential equation is linear or can be fairly linearized, then dynamics of the state $\mathbf{x}$ can be approximately modelled by a \textbf{linear state differential equation} as
\begin{align}  \label{eq:state_differential_equation_linear0}
\frac{\mathrm{d}}{\mathrm{d} t} \bar{\mathbf{x}} = f(\mathbf{x}, \mathbf{u}) - f(\mathbf{x}_0, \mathbf{0}) \approx \mathbf{A} \bar{\mathbf{x}} + \mathbf{B} \mathbf{u},
\end{align}
where
\begin{align*}
\bar{\mathbf{x}} \equiv \mathbf{x} - \mathbf{x}_0, \quad \mathbf{A} = \frac{\partial f(\mathbf{x}, \mathbf{u})}{\partial \mathbf{x}}|_{\mathbf{x} = \mathbf{x}_0, \mathbf{u}  = \mathbf{0}}, \quad \mathbf{B} = \frac{\partial f(\mathbf{x}, \mathbf{u})}{\partial \mathbf{u}}|_{\mathbf{x} = \mathbf{x}_0, \mathbf{u}  = \mathbf{0}}.
\end{align*}
For formalism simplicity, we still abuse $\mathbf{x}$ to denote 
\begin{align*}
\bar{\mathbf{x}} \equiv \mathbf{x} - \mathbf{x}_0
\end{align*}
in (\ref{eq:state_differential_equation_linear0}) and obtain
\begin{align}  \label{eq:state_differential_equation_linear}
\frac{\mathrm{d}}{\mathrm{d} t} \mathbf{x} = \mathbf{A} \mathbf{x} + \mathbf{B} \mathbf{u},
\end{align}
which is the generic formalism of \textbf{linear state-space modelling}. For the linear control system, the square matrix $\mathbf{A}$ is called the \textbf{state transition matrix} and the matrix $\mathbf{B}$ is called the \textbf{control input matrix}.

\subsubsection*{Application: single inverted pendulum}

The single inverted pendulum control system is illustrated in Figure \ref{fig:inverted_pendulum_control}, with dynamics of its state modelled by the state differential equation (\ref{eq:SIP_state_DE}). If the inverted pendulum angle $\theta$ is close to zero, then the state differential equation described in (\ref{eq:SIP_state_DE}) can be fairly linearized about the equilibrium state and simplified into a linear state differential equation as
\begin{equation}  \label{eq:SIP_state_DE_linear}
\frac{\mathrm{d}}{\mathrm{d} t} \mathbf{x} = \begin{bmatrix} 0 & 1 & 0 & 0 \\ \frac{g}{L} & 0 & 0 & 0 \\ 0 & 0 & 0 & 1  \\ 0 & 0 & 0 & 0 \end{bmatrix} \mathbf{x} + \begin{bmatrix} 0 \\ -\frac{1}{L} \\ 0 \\ 1 \end{bmatrix} a \equiv \mathbf{A} \mathbf{x} + \mathbf{B} a,
\end{equation}
where the state 
\begin{align*}
\mathbf{x} \equiv \begin{bmatrix} \theta & \frac{\mathrm{d} \theta}{\mathrm{d} t} & x & \frac{\mathrm{d} x}{\mathrm{d} t} \end{bmatrix}^\mathrm{T}
\end{align*}
is the same to that specified in (\ref{eq:SIP_state_DE}).

\subsubsection*{Application: double inverted pendulum}

The double inverted pendulum control system is illustrated in Figure \ref{fig:double_inverted_pendulum_control}, with dynamics of its state modelled by the state differential equation (\ref{eq:DIP_state_DE}). If both inverted pendulum angles $\theta_1$ and $\theta_2$ are close to zero, then the state differential equation described in (\ref{eq:DIP_state_DE}) can be fairly linearized about the equilibrium state and simplified into a linear state differential equation as 
\begin{equation}  \label{eq:DIP_state_DE_linear}
\frac{\mathrm{d}}{\mathrm{d} t} \mathbf{x} = \begin{bmatrix} 0 & 1 & 0 & 0 & 0 & 0 \\
(1 + \frac{m_2}{m_1}) \frac{g}{L_1} & 0 & -\frac{m_2}{m_1} \frac{g}{L_1} & 0 & 0 & 0 \\
0 & 0 & 0 & 1 & 0 & 0 \\
-(1 + \frac{m_2}{m_1}) \frac{g}{L_2} & 0 & (1 + \frac{m_2}{m_1}) \frac{g}{L_2} & 0 & 0 & 0 \\
0 & 0 & 0 & 0 & 0 & 1  \\ 0 & 0 & 0 & 0 & 0 & 0 \end{bmatrix} \mathbf{x} + \begin{bmatrix} 0 \\ -\frac{1}{L_1} \\ 0 \\ 0 \\ 0 \\ 1 \end{bmatrix} a \equiv \mathbf{A} \mathbf{x} + \mathbf{B} a.
\end{equation}
where the state 
\begin{align*}
\mathbf{x} \equiv \begin{bmatrix} \theta_1 & \frac{\mathrm{d} \theta_1}{\mathrm{d} t} & \theta_2 & \frac{\mathrm{d} \theta_2}{\mathrm{d} t} & x & \frac{\mathrm{d} x}{\mathrm{d} t} \end{bmatrix}^\mathrm{T}
\end{align*}
is the same to that specified in (\ref{eq:DIP_state_DE}).

The variant of the double inverted pendulum control system is illustrated in Figure \ref{fig:double_inverted_pendulum_variant_control}, with dynamics of its state modelled by the state differential equation (\ref{eq:DIP_state_DE_MIMO}). The linear counterpart of (\ref{eq:DIP_state_DE_MIMO}) is
\begin{align}  \label{eq:DIP_state_DE_linear_MIMO}
\frac{\mathrm{d}}{\mathrm{d} t} \mathbf{x} &= \begin{bmatrix} 0 & 1 & 0 & 0 & 0 & 0 \\
(1 + \frac{m_2}{m_1}) \frac{g}{L_1} & 0 & -\frac{m_2}{m_1} \frac{g}{L_1} & 0 & 0 & 0 \\
0 & 0 & 0 & 1 & 0 & 0 \\
-(1 + \frac{m_2}{m_1}) \frac{g}{L_2} & 0 & (1 + \frac{m_2}{m_1}) \frac{g}{L_2} & 0 & 0 & 0 \\
0 & 0 & 0 & 0 & 0 & 1  \\ 0 & 0 & 0 & 0 & 0 & 0 \end{bmatrix} \mathbf{x} + \begin{bmatrix} 0 & 0 \\ -\frac{1}{L_1} & 1 \\ 0 & 0 \\ 0 & -\frac{L_1}{L_2} \\ 0 & 0 \\ 1 & 0 \end{bmatrix} \begin{bmatrix} a \\ a_1 \end{bmatrix} \\
  &\equiv \mathbf{A} \mathbf{x} + \mathbf{B} \mathbf{u}.  \nonumber
\end{align}

\subsubsection*{Application: low-speed autonomous vehicle}

Low-speed autonomous vehicle navigation is illustrated in Figure \ref{fig:low_speed_autonomous_vehicle}, with dynamics of the vehicle lateral state modelled by the state differential equation (\ref{eq:vehicle_lateral_control}). In practical applications, both the vehicle orientation angle $\theta$ and the vehicle steering angle $\beta$ are usually close to zero, then the state differential equation described in (\ref{eq:vehicle_lateral_control}) can be fairly linearized about the equilibrium state and simplified into a linear state differential equation as
\begin{align}  \label{eq:vehicle_lateral_control_approximation}
\frac{\mathrm{d}}{\mathrm{d} t} \mathbf{x} = \begin{bmatrix} 0 & v & 0 \\ 0 & 0 & \frac{v}{L} \\ 0 & 0 & -\frac{1}{\tau_{\beta}} \end{bmatrix} \mathbf{x} + \begin{bmatrix} 0 \\ 0 \\ \frac{1}{\tau_{\beta}} \end{bmatrix} \beta_I \equiv \mathbf{A} \mathbf{x} + \mathbf{B} \beta_I,
\end{align}
where the vehicle lateral state 
\begin{align*}
\mathbf{x} \equiv \begin{bmatrix} y & \theta & \beta \end{bmatrix}^\mathrm{T}
\end{align*}
is the same to that specified in (\ref{eq:vehicle_lateral_control}).

\subsubsection*{Application: high-speed autonomous vehicle}

In practical applications, relevant angles involved in (\ref{eq:vehicle_lateral_dynamics_with_slip}) are usually close to zero, then the state differential equation described in (\ref{eq:vehicle_lateral_dynamics_with_slip}) can be fairly linearized about the equilibrium state and simplified into a linear state differential equation as
\begin{align}  \label{eq:vehicle_lateral_dynamics_with_slip_linear}
\frac{\mathrm{d}}{\mathrm{d} t} \mathbf{x} &= \begin{bmatrix} 0 & 1 & 0 & 0 & 0 \\
0 & - \frac{2 C_f + 2 C_r}{m v} & \frac{2 C_f + 2 C_r}{m} & - \frac{2 C_f L_f - 2 C_r L_r}{m v} & \frac{2 C_f}{m} \\ 
0 & 0 & 0 & 1 & 0 \\
0 & - \frac{2 C_f L_f - 2 C_r L_r}{J v} & \frac{2 C_f L_f - 2 C_r L_r}{J} & - \frac{2 C_f L_f^2 + 2 C_r L_r^2}{J v} & \frac{2 C_f L_f}{J} \\
0 & 0 & 0 & 0 & - \frac{1}{\tau_{\beta}} \end{bmatrix} \mathbf{x} + \begin{bmatrix} 0 \\ 0 \\ 0 \\ 0 \\ \frac{1}{\tau_{\beta}} \end{bmatrix} \beta_I  \\ 
  &\equiv \mathbf{A} \mathbf{x} + \mathbf{B} \beta_I,  \nonumber
\end{align}
where the vehicle lateral state 
\begin{align*}
\mathbf{x} \equiv \begin{bmatrix} y & \frac{\mathrm{d} y}{\mathrm{d} t} & \phi & \frac{\mathrm{d} \phi}{\mathrm{d} t} & \beta \end{bmatrix}^\mathrm{T}
\end{align*}
is the same to that specified in (\ref{eq:vehicle_lateral_dynamics_with_slip}).

\subsubsection*{Application: autonomous motorcycle (or bicycle)}

Dynamics of the motorcycle lateral state is modelled by the state differential equation (\ref{eq:motorcycle_lateral_control}). In practical applications, the motorcycle orientation angle $\phi$ (namely yaw angle), the motorcycle vertical angle $\theta$ (namely roll angle), and the motorcycle steering angle $\beta$ are usually close to zero, then the state differential equation described in (\ref{eq:motorcycle_lateral_control}) can be fairly linearized about the equilibrium state and simplified into a linear state differential equation as
\begin{align}  \label{eq:motorcycle_lateral_control_approximation}
\frac{\mathrm{d}}{\mathrm{d} t} \mathbf{x} = \begin{bmatrix} 0 & v & 0 & 0 & 0 \\ 0 & 0 & \frac{v}{L} & 0 & 0 \\ 0 & 0 & -\frac{1}{\tau_{\beta}} & 0 & 0 \\ 0 & 0 & 0 & 0 & 1 \\ 0 & 0 & -\frac{v^2}{H L} & \frac{g}{H} & 0 \end{bmatrix} \mathbf{x} + \begin{bmatrix} 0 \\ 0 \\ \frac{1}{\tau_{\beta}} \\ 0 \\ 0 \end{bmatrix} \beta_I \equiv \mathbf{A} \mathbf{x} + \mathbf{B} \beta_I,
\end{align}
where the motorcycle lateral state 
\begin{align*}
\mathbf{x} \equiv \begin{bmatrix} y & \phi & \beta & \theta & \frac{\mathrm{d} \theta}{\mathrm{d} t} \end{bmatrix}^\mathrm{T}
\end{align*}
is the same to that specified in (\ref{eq:motorcycle_lateral_control}).

\section{Self-evolutionary systems}  \label{sec:self_evo_systems}

Feedback is the soul of control science. Most control systems encountered in practical applications are closed-loop feedback control systems. For a closed-loop feedback control system with the state $\mathbf{x}$ characterizing the target process, the control input $\mathbf{u}$ is normally generated according to feedback of the state $\mathbf{x}$ which is compared with certain expected state $\mathbf{x}_\mathrm{E}$. 

If the expected state $\mathbf{x}_\mathrm{E}$ is time-invariant, then let the feedback control law be generically denoted as
\begin{align*}
\mathbf{u} = g(\mathbf{x} - \mathbf{x}_\mathrm{E}), 
\end{align*}
substitute it into (\ref{eq:state_differential_equation}) and obtain
\begin{equation}  \label{eq:closed-loop_feedback_SDE1}
\frac{\mathrm{d}}{\mathrm{d} t} \mathbf{x} = f(\mathbf{x}, g(\mathbf{x} - \mathbf{x}_\mathrm{E})) \equiv f_c(\mathbf{x} - \mathbf{x}_\mathrm{E}).
\end{equation}
If 
\begin{align*}
\mathbf{x}_\mathrm{E} \not = \mathbf{0}, 
\end{align*}
we can shift the state $\mathbf{x}$ by an offset of $\mathbf{x}_\mathrm{E}$ and transform (\ref{eq:closed-loop_feedback_SDE1}) into
\begin{equation}  \label{eq:closed-loop_feedback_SDE2}
\frac{\mathrm{d}}{\mathrm{d} t} (\mathbf{x} - \mathbf{x}_\mathrm{E}) = \frac{\mathrm{d}}{\mathrm{d} t} \mathbf{x} = f_c(\mathbf{x} - \mathbf{x}_\mathrm{E}),
\end{equation}
where 
\begin{align*}
\bar{\mathbf{x}} \equiv \mathbf{x} - \mathbf{x}_\mathrm{E}
\end{align*}
can be regarded as a new representation of the state and the expected new state is 
\begin{align*}
\bar{\mathbf{x}}_\mathrm{E} \equiv \mathbf{x}_\mathrm{E} - \mathbf{x}_\mathrm{E} = \mathbf{0}. 
\end{align*}
Therefore, for analysis simplicity yet without influencing analysis essence, assume 
\begin{align*}
\mathbf{x}_\mathrm{E} = \mathbf{0}
\end{align*}
by default in (\ref{eq:closed-loop_feedback_SDE1}) and (\ref{eq:closed-loop_feedback_SDE2}) and hence obtain
\begin{equation}  \label{eq:closed-loop_feedback_SDE}
\frac{\mathrm{d}}{\mathrm{d} t} \mathbf{x} = f(\mathbf{x}, g(\mathbf{x})) \equiv f_c(\mathbf{x}).
\end{equation}

If the expected state $\mathbf{x}_\mathrm{E}$ is not constant but is a time-variant function specified explicitly as
\begin{align*}
\mathbf{x}_\mathrm{E} = h(t),
\end{align*}
then the feedback control law turns to be
\begin{align*}
\mathbf{u} = g(\mathbf{x} - h(t)) \equiv g_h(\mathbf{x}). 
\end{align*}
Although we cannot shift the state $\mathbf{x}$ by certain constant state offset, we can treat the expected state $\mathbf{x}_\mathrm{E}$ as a set of known time-variant parameters in the feedback control law and further treat
\begin{align*}
\mathbf{u} = g_h(\mathbf{x})
\end{align*}
as a control law depending on the state $\mathbf{x}$ only. Substitute it into (\ref{eq:state_differential_equation}) and obtain
\begin{align*}
\frac{\mathrm{d}}{\mathrm{d} t} \mathbf{x} = f(\mathbf{x}, g_h(\mathbf{x})) \equiv f_c(\mathbf{x}),
\end{align*}
where we abuse the generic functional notation $f_c(\cdot)$ as in (\ref{eq:closed-loop_feedback_SDE}) for the closed-loop feedback control system.

Even if the expected state $\mathbf{x}_\mathrm{E}$ is a time-variant function that may partially be specified explicitly and partially be determined implicitly according to the state $\mathbf{x}$ which is unknown \textit{a priori}
\footnote{This can take place in practical applications, especially when \textit{dynamical motion planning} \cite{LaValle2006} is involved.},
formalized as
\begin{align*}
\mathbf{x}_\mathrm{E} = h(\mathbf{x}, t),
\end{align*}
then the feedback control law
\begin{align*}
\mathbf{u} = g(\mathbf{x} - h(\mathbf{x}, t)) \equiv g_h(\mathbf{x}) 
\end{align*}
can still be treated as a control law depending on the state $\mathbf{x}$ only. Substitute it into (\ref{eq:state_differential_equation}) and obtain again a generic formalism
\begin{align*}
\frac{\mathrm{d}}{\mathrm{d} t} \mathbf{x} = f(\mathbf{x}, g_h(\mathbf{x})) \equiv f_c(\mathbf{x})
\end{align*}
as in (\ref{eq:closed-loop_feedback_SDE}) for the closed-loop feedback control system.

Therefore, once the feedback control law is determined, dynamics of the closed-loop feedback control system is equivalent to dynamics of a \textbf{self-evolutionary system} or \textbf{autonomous system} namely a system whose state evolution depends on its own state completely, as modelled by (\ref{eq:closed-loop_feedback_SDE}). The state differential equation described in (\ref{eq:closed-loop_feedback_SDE}) is called the \textbf{closed-loop feedback state differential equation} of the control system.

Similarly, for the generic formalism of linear state-space modelling described in (\ref{eq:state_differential_equation_linear}), its corresponding \textbf{linear closed-loop feedback state differential equation} is a generic \textit{homogeneous linear state differential equation} as
\begin{equation}  \label{eq:closed-loop_feedback_SDE_linear}
\frac{\mathrm{d}}{\mathrm{d} t} \mathbf{x} = \mathbf{A}_c \mathbf{x},
\end{equation}
which is equivalent to the state differential equation of a \textbf{linear self-evolutionary system}. The square matrix $\mathbf{A}_c$ is the \textbf{state transition matrix} of the linear self-evolutionary system.

\section{Stability analysis}

\subsection{Stability criterion for linear self-evolutionary systems}  \label{sec:stability_criterion_linear}

We first focus on the stability criterion for linear self-evolutionary systems which can be generically modelled by the homogeneous linear state differential equation described in (\ref{eq:closed-loop_feedback_SDE_linear}). 

If (\ref{eq:closed-loop_feedback_SDE_linear}) gives a scalar differential equation
\begin{align*}
\frac{\mathrm{d}}{\mathrm{d} t} x = a x,
\end{align*}
then it can be solved as
\begin{align*}
\frac{\mathrm{d} x}{x} = a \mathrm{d} t \iff \ln x = a t + C \implies x = \mathrm{e}^{a t} \mathrm{e}^C = \mathrm{e}^{a t} x_0, 
\end{align*}
where the initial condition of $x$ is assumed to be $x_0$. However, for the generic state differential equation
\begin{align*}
\frac{\mathrm{d}}{\mathrm{d} t} \mathbf{x} = \mathbf{A} \mathbf{x},
\end{align*}
it cannot be solved in above way, because there is neither natural definition of \textit{exponential} nor that of \textit{logarithmic} for matrices. To solve the state differential equation, express the state $\mathbf{x}$ as an infinite series
\begin{align*}
\mathbf{x} = \mathbf{x}_0 + \mathbf{x}_1 t + \mathbf{x}_2 t^2 + \mathbf{x}_3 t^3 + \cdots = \sum_{k=0}^{\infty} \mathbf{x}_k t^k, 
\end{align*}
substitute it into the original equation and compare corresponding terms on both sides
\begin{align*}
&\mathbf{x}_1 + 2 \mathbf{x}_2 t + 3 \mathbf{x}_3 t^2 + \cdots = \mathbf{A} (\mathbf{x}_0 + \mathbf{x}_1 t + \mathbf{x}_2 t^2 + \mathbf{x}_3 t^3 + \cdots)  \\
\iff& \mathbf{x}_1 = \mathbf{A} \mathbf{x}_0, \quad \mathbf{x}_2 = \frac{1}{2 !} \mathbf{A}^2 \mathbf{x}_0, \quad \mathbf{x}_3 = \frac{1}{3 !} \mathbf{A}^3 \mathbf{x}_0, \quad \cdots
\end{align*}
So
\begin{align*}
\mathbf{x} = (\sum_{k=0}^{\infty} \frac{1}{k !} \mathbf{A}^k t^k) \mathbf{x}_0 = \mathrm{e}^{\mathbf{A} t} \mathbf{x}_0.
\end{align*}

Replace $\mathbf{A}$ by $\mathbf{A}_c$ in above result and obtain the solution of the state differential equation described in (\ref{eq:closed-loop_feedback_SDE_linear}) as
\begin{equation}  \label{eq:closed-loop_feedback_SDE_linear_solution}
\mathbf{x} = \mathrm{e}^{\mathbf{A}_c t} \mathbf{x}_0,
\end{equation}
where 
\begin{align*}
\mathbf{x}_0 \equiv \mathbf{x}(0)
\end{align*}
and the \textit{matrix exponential function} is defined as
\begin{align*}
\mathrm{e}^{\mathbf{M}} \equiv \mathbf{I} + \mathbf{M} + \frac{\mathbf{M}^2}{2 !} + \frac{\mathbf{M}^3}{3 !} + \cdots = \sum_{k=0}^{\infty} \frac{\mathbf{M}^k}{k !}.
\end{align*}

\subsubsection*{Application: rotating disk position open-loop initial response analysis}

\begin{figure}[h!]
\begin{center}
\includegraphics[width=0.8\columnwidth]{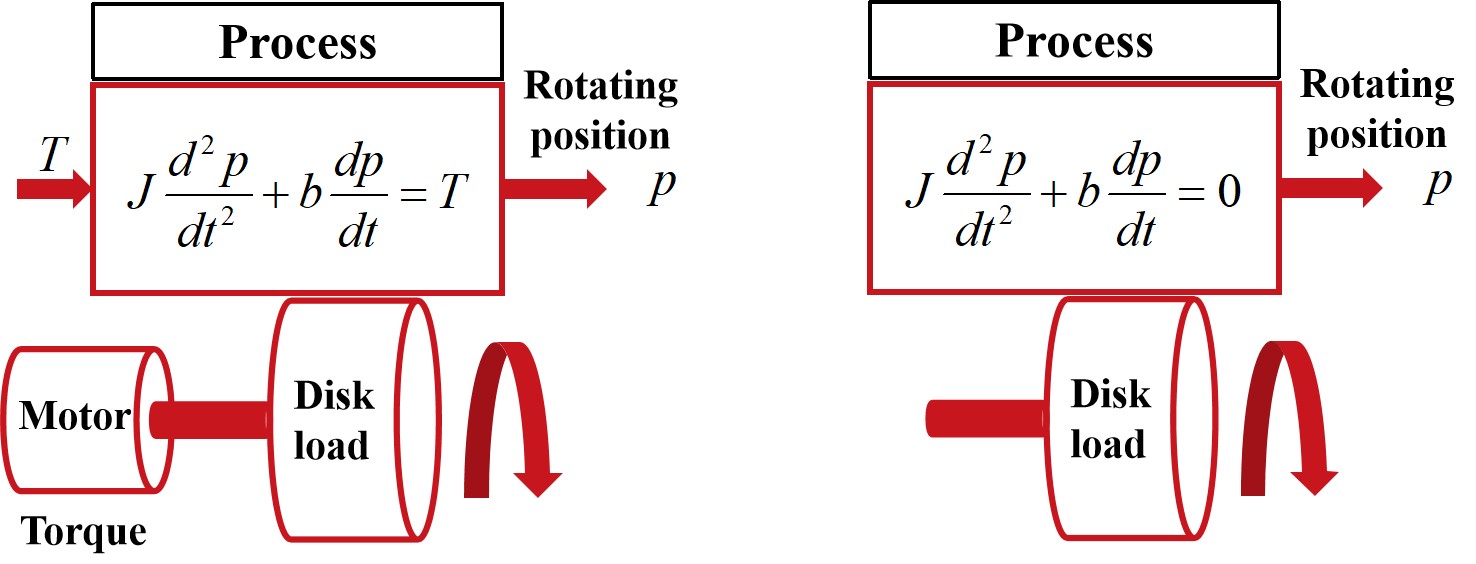}
\end{center}
\caption{Rotating disk dynamics: (left) rotating disk position open-loop control; (right) rotating disk position open-loop initial response (when $T = 0$)}
\label{fig:RDP_open_loop_initial_response}
\end{figure}

Consider the rotating disk position open-loop control system illustrated in the left sub-figure of Figure \ref{fig:RDP_open_loop_initial_response}. Consider the rotating disk state 
\begin{align*}
\mathbf{x} \equiv \begin{bmatrix} p & v \end{bmatrix}^\mathrm{T}
\end{align*}
which consists of the rotating disk position $p$ and the rotating disk speed 
\begin{align*}
v \equiv \frac{\mathrm{d} p}{\mathrm{d} t}. 
\end{align*}
Dynamics of the rotating disk state $\mathbf{x}$ is modelled by a linear state differential equation as
\begin{equation}  \label{eq:RDP_state_DE_linear}
\frac{\mathrm{d}}{\mathrm{d} t} \mathbf{x} \equiv \frac{\mathrm{d}}{\mathrm{d} t} \begin{bmatrix} p \\ v \end{bmatrix} = \begin{bmatrix} 0 & 1 \\ 0 & -\frac{b}{J} \end{bmatrix} \begin{bmatrix} p \\ v \end{bmatrix} + \begin{bmatrix} 0 \\ \frac{1}{J} \end{bmatrix} T \equiv \mathbf{A} \mathbf{x} + \mathbf{B} T,
\end{equation}
where $T$ denotes the control input torque, $J$ denotes the inertia of the rotating disk load, and $b$ denotes the friction coefficient.

The rotating disk position open-loop initial response namely the system response under zero control input torque, as illustrated in the right sub-figure of Figure \ref{fig:RDP_open_loop_initial_response}, is determined by an equivalent self-evolutionary system described by the linear state differential equation
\begin{equation}  \label{eq:RDP_state_DE_initial_response}
\frac{\mathrm{d}}{\mathrm{d} t} \mathbf{x} \equiv \frac{\mathrm{d}}{\mathrm{d} t} \begin{bmatrix} p \\ v \end{bmatrix} = \begin{bmatrix} 0 & 1 \\ 0 & -\frac{b}{J} \end{bmatrix} \begin{bmatrix} p \\ v \end{bmatrix} \equiv \mathbf{A}_c \mathbf{x}.
\end{equation}
For concrete configuration of rotating disk parameters
\footnote{Like in the previous book \textit{Control Theory For Practical Applications} \cite{Li2024CTPA_Springer, Li2024CTPA_SJTU_1}, the author assumes that readers are familiar with fundamental physics. So for expression conciseness throughout this book, the author omits variable physical units which can be easily completed according to fundamental physics knowledge \cite{Feynman2004}.},
set 
\begin{align*}  
J = 5, \quad b = 5. 
\end{align*}
Compute the rotating disk position open-loop initial response via (\ref{eq:closed-loop_feedback_SDE_linear_solution}) as
\begin{align*}
\begin{bmatrix} p \\ v \end{bmatrix} &= \mathrm{e}^{\mathbf{A}_c t} \begin{bmatrix} p_0 \\ v_0 \end{bmatrix} = \mathrm{e}^{\begin{bmatrix} 0 & 1 \\ 0 & -1 \end{bmatrix} t} \begin{bmatrix} p_0 \\ v_0 \end{bmatrix} = (\sum_{k=0}^{\infty} \frac{\begin{bmatrix} 0 & 1 \\ 0 & -1 \end{bmatrix}^k t^k}{k !}) \begin{bmatrix} p_0 \\ v_0 \end{bmatrix}  \\
  &= (\begin{bmatrix} 1 & 0 \\ 0 & 1 \end{bmatrix} + \begin{bmatrix} 0 & 1 \\ 0 & -1 \end{bmatrix} \sum_{k=1}^{\infty} \frac{(-1)^{k-1} t^k}{k !}) \begin{bmatrix} p_0 \\ v_0 \end{bmatrix} \qquad (\mbox{note } \mathbf{A}_c^2 = - \mathbf{A}_c)  \\
  &= (\begin{bmatrix} 1 & 0 \\ 0 & 1 \end{bmatrix} + \begin{bmatrix} 0 & 1 \\ 0 & -1 \end{bmatrix} (1 - \mathrm{e}^{-t})) \begin{bmatrix} p_0 \\ v_0 \end{bmatrix} = \begin{bmatrix} 1 & 1 - \mathrm{e}^{-t} \\ 0 & \mathrm{e}^{-t} \end{bmatrix} \begin{bmatrix} p_0 \\ v_0 \end{bmatrix}.
\end{align*}
So
\begin{align*}
p &= p_0 + (1 - \mathrm{e}^{-t}) v_0,  \\
v &= \mathrm{e}^{-t} v_0,
\end{align*}
which satisfy the linear state differential equation (\ref{eq:RDP_state_DE_initial_response}) with the initial condition $\{p_0, v_0\}$ and hence are indeed the solution of (\ref{eq:RDP_state_DE_initial_response}).

\subsubsection*{Stability analysis according to eigenvalues of the state transition matrix $\mathbf{A}_c$}

If a linear self-evolutionary system is stable, then its state $\mathbf{x}$ always converges to the equilibrium state 
\begin{align*}
\mathbf{x}_\mathrm{E} = \mathbf{0}
\end{align*}
no matter for what initial state $\mathbf{x}_0$. This is equivalent to the condition
\begin{align}  \label{eq:exp_Ac_t=0}
\lim_{t \to \infty} \mathrm{e}^{\mathbf{A}_c t} = 0.
\end{align}
In other words, the sufficient and necessary condition for the linear self-evolutionary system to be stable is given in (\ref{eq:exp_Ac_t=0}). 

For further analysis of (\ref{eq:exp_Ac_t=0}), we resort to \textit{Jordan canonical decomposition} \cite{Horn2012} of the state transition matrix $\mathbf{A}_c$. Suppose $\mathbf{A}_c$ is decomposed as
\begin{align}  \label{eq:Ac_jordan_decomposition}
\mathbf{A}_c = \mathbf{S} \begin{bmatrix} \mathbf{J}_{\lambda_1} & \mathbf{0} & \cdots & \mathbf{0} \\ \mathbf{0} & \mathbf{J}_{\lambda_2} &  \cdots & \mathbf{0} \\ \vdots & \vdots & \ddots & \vdots \\ \mathbf{0} & \mathbf{0} & \cdots & \mathbf{J}_{\lambda_q} \end{bmatrix} \mathbf{S}^{-1} \equiv \mathbf{S} \mathbf{J} \mathbf{S}^{-1},
\end{align}
where 
\begin{align*}
\mathbf{J}_{\lambda_1}, \quad \mathbf{J}_{\lambda_2}, \quad \cdots \quad, \quad \mathbf{J}_{\lambda_q}
\end{align*}
are \textit{Jordan blocks} with each $\mathbf{J}_{\lambda_i}$ ($i \in \{1, 2, \cdots, q\}$) corresponding to an \textit{eigenvalue}
\footnote{The German term ``eigen'' (in ``eigenwerte'') was introduced by \textit{Hilbert} in \cite{Hilbert1904}, which means ``own'', ``proprietary'' (sense extended from its French origin ``propre''), or ``characteristic''. This new ``English'' term has been adopted as mathematics convention ever since.}
$\lambda_i$ of $\mathbf{A}_c$ in the form as
\footnote{Each Jordan block can also be in the form as
\begin{align*}
\mathbf{J}_{\lambda_i} = \begin{bmatrix} \lambda_i & & & & & \\ 1 & \lambda_i & & & & \\ & 1 & \lambda_i & & & \\ & & & \ddots & & \\ & & & & \lambda_i & \\ & & & & 1 & \lambda_i \end{bmatrix},
\end{align*}
but this has no essential influence on following analysis.}
\begin{align*}
\mathbf{J}_{\lambda_i} = \begin{bmatrix} \lambda_i & 1 & & & & \\ & \lambda_i & 1 & & & \\ & & \lambda_i & & & \\ & & & \ddots & & \\ & & & & \lambda_i & 1 \\ & & & & & \lambda_i \end{bmatrix}.
\end{align*}

Since
\begin{align*}
\mathbf{A}_c^k &= (\mathbf{S} \mathbf{J} \mathbf{S}^{-1})^k = \begin{matrix} \underbrace{(\mathbf{S} \mathbf{J} \mathbf{S}^{-1}) (\mathbf{S} \mathbf{J} \mathbf{S}^{-1}) \cdots (\mathbf{S} \mathbf{J} \mathbf{S}^{-1})} \\ k \mbox{ times} \end{matrix} = \mathbf{S} \mathbf{J}^k \mathbf{S}^{-1} \\
  &= \mathbf{S} \begin{bmatrix} \mathbf{J}_{\lambda_1} & & \\ & \ddots & \\ & & \mathbf{J}_{\lambda_q} \end{bmatrix}^k \mathbf{S}^{-1} = \mathbf{S} \begin{bmatrix} \mathbf{J}_{\lambda_1}^k & & \\ & \ddots & \\ & & \mathbf{J}_{\lambda_q}^k \end{bmatrix} \mathbf{S}^{-1},
\end{align*}
we have
\begin{align}  \label{eq:exp_Ac_t}
\mathrm{e}^{\mathbf{A}_c t} &= \sum_{k=0}^{\infty} \frac{\mathbf{A}_c^k t^k}{k !} = \sum_{k=0}^{\infty} (\mathbf{S} \begin{bmatrix} \mathbf{J}_{\lambda_1}^k & & \\ & \ddots & \\ & & \mathbf{J}_{\lambda_q}^k \end{bmatrix} \frac{t^k}{k !} \mathbf{S}^{-1})  \nonumber \\
  &= \mathbf{S} \begin{bmatrix} \sum_{k=0}^{\infty} \mathbf{J}_{\lambda_1}^k \frac{t^k}{k !} & & \\ & \ddots & \\ & & \sum_{k=0}^{\infty} \mathbf{J}_{\lambda_q}^k \frac{t^k}{k !} \end{bmatrix} \mathbf{S}^{-1} = \mathbf{S} \begin{bmatrix} \mathrm{e}^{\mathbf{J}_{\lambda_1} t} & & \\ & \ddots & \\ & & \mathrm{e}^{\mathbf{J}_{\lambda_q} t} \end{bmatrix} \mathbf{S}^{-1}.
\end{align}

For each Jordan block $\mathbf{J}_{\lambda_i}$, we have
\begin{align*}
\mathbf{J}_{\lambda_i}^k = \begin{bmatrix} \lambda_i^k & C_k^1 \lambda_i^{k-1} & C_k^2 \lambda_i^{k-2} & \cdots & C_k^{d_i-2} \lambda_i^{k-d_i+2} & C_k^{d_i-1} \lambda_i^{k-d_i+1} \\ & \lambda_i^k & C_k^1 \lambda_i^{k-1} & \cdots & C_k^{d_i-3} \lambda_i^{k-d_i+3} & C_k^{d_i-2} \lambda_i^{k-d_i+2} \\ & & \lambda_i^k & \cdots & \ddots & C_k^{d_i-3} \lambda_i^{k-d_i+3} \\ & & & \ddots & \vdots & \vdots \\ & & & & \lambda_i^k & C_k^1 \lambda_i^{k-1} \\ & & & & & \lambda_i^k \end{bmatrix},
\end{align*}
which is an \textit{upper triangular matrix} having the following regularity: elements of the first diagonal line are the same to $\lambda_i^k$, elements of the second diagonal line are the same to $C_k^1 \lambda_i^{k-1}$, elements of the third diagonal line are the same to $C_k^2 \lambda_i^{k-2}$, $\cdots$, elements of the $(d_i-1)$-th diagonal line are the same to $C_k^{d_i-2} \lambda_i^{k-d_i+2}$, and the top-right element in the $d_i$-th diagonal line is $C_k^{d_i-1} \lambda_i^{k-d_i+1}$, where $d_i$ denotes the dimension of $\mathbf{J}_{\lambda_i}$. We further have
\begin{align*}
\mathrm{e}^{\mathbf{J}_{\lambda_i} t} = \sum_{k=0}^{\infty} \mathbf{J}_{\lambda_i}^k \frac{t^k}{k !} = \begin{bmatrix} D_{\lambda_i, t, 1} & D_{\lambda_i, t, 2} & D_{\lambda_i, t, 3} & \cdots & D_{\lambda_i, t, d_i-1} & D_{\lambda_i, t, d_i} \\ & D_{\lambda_i, t, 1} & D_{\lambda_i, t, 2} & \cdots & D_{\lambda_i, t, d_i-2} & D_{\lambda_i, t, d_i-1} \\ & & D_{\lambda_i, t, 1} & \cdots & \ddots & D_{\lambda_i, t, d_i-2} \\ & & & \ddots & \vdots & \vdots \\ & & & & D_{\lambda_i, t, 1} & D_{\lambda_i, t, 2} \\ & & & & & D_{\lambda_i, t, 1} \end{bmatrix},
\end{align*}
where
\begin{align*}
D_{\lambda_i, t, 1} &= \sum_{k=0}^{\infty} \lambda_i^k \frac{t^k}{k !} = \mathrm{e}^{\lambda_i t},  \\
D_{\lambda_i, t, 2} &= \sum_{k=1}^{\infty} C_k^1 \lambda_i^{k-1} \frac{t^k}{k !} = t \sum_{k=1}^{\infty} \lambda_i^{k-1} \frac{t^{k-1}}{(k-1) !} = t \mathrm{e}^{\lambda_i t}, \\
D_{\lambda_i, t, 3} &= \sum_{k=2}^{\infty} C_k^2 \lambda_i^{k-2} \frac{t^k}{k !} = \frac{t^2}{2 !} \sum_{k=2}^{\infty} \lambda_i^{k-2} \frac{t^{k-2}}{(k-2) !} = \frac{t^2}{2 !} \mathrm{e}^{\lambda_i t}, \\
\cdots &\quad \quad \cdots \\
D_{\lambda_i, t, d_i} &= \sum_{k=d_i-1}^{\infty} C_k^{d_i-1} \lambda_i^{k-d_i+1} \frac{t^k}{k !} = \frac{t^{d_i-1}}{(d_i-1) !} \mathrm{e}^{\lambda_i t}.
\end{align*}

Since
\begin{align*}
\mbox{Re}(\lambda_i) < 0 &\iff \lim_{t \to \infty} D_{\lambda_i, t, 1} = 0, \\
\mbox{Re}(\lambda_i) < 0 &\iff \lim_{t \to \infty} D_{\lambda_i, t, 2} = 0, \\
\mbox{Re}(\lambda_i) < 0 &\iff \lim_{t \to \infty} D_{\lambda_i, t, 3} = 0, \\
\cdots &\quad \qquad \cdots \\
\mbox{Re}(\lambda_i) < 0 &\iff \lim_{t \to \infty} D_{\lambda_i, t, d_i} = 0,
\end{align*}
we have
\begin{align}  \label{eq:exp_J_lambda_i=0}
\mbox{Re}(\lambda_i) < 0 &\iff \lim_{t \to \infty} \mathrm{e}^{\mathbf{J}_{\lambda_i} t} = 0.
\end{align}
From (\ref{eq:exp_Ac_t}) and (\ref{eq:exp_J_lambda_i=0}) we can infer
\begin{align}  \label{eq:linear_self_evolutionary_system_stability}
\mbox{Re}(\lambda_1) < 0, \mbox{Re}(\lambda_2) < 0, \cdots, \mbox{Re}(\lambda_q) < 0 &\iff \lim_{t \to \infty} \mathrm{e}^{\mathbf{A}_c t} = 0,
\end{align}
which implies that a linear self-evolutionary system is stable if and only if the real parts of all its eigenvalues (namely eigenvalues of its state transition matrix) are negative.

\begin{framed} 
\noindent \textbf{Control system stability criterion}: \textit{A linear self-evolutionary system is stable if and only if the real parts of all its eigenvalues are negative}.
\end{framed}

A control system that evolves only according to its initial condition is a typical kind of self-evolutionary system. A stable control system's initial response will always fade away no matter given what initial condition. So a stable control system that evolves only according to its initial condition is also a stable self-evolutionary system. On the other hand, if a control system that evolves only according to its initial condition is a stable self-evolutionary system, then the control system will always have convergent initial response no matter given what initial condition and hence is stable. Therefore, to judge whether a linear time-invariant control system especially closed-loop feedback control system is stable, we can use above stability criterion to judge whether the control system is a stable linear self-evolutionary system when it evolves only according to its initial condition.

\subsubsection*{Relationship between linear state-space modelling and system modelling via Laplace transform}

To facilitate understanding of mutual relationship between linear state-space modelling and system modelling via Laplace transform
\footnote{Readers may refer to the previous book \textit{Control Theory For Practical Applications} \cite{Li2024CTPA_Springer, Li2024CTPA_SJTU_1} for a systematic knowledge of system modelling via Laplace transform.}, 
we may consider a linear time-invariant control system that evolves only according to its initial condition and consider the mutual transform as follows: On one hand, suppose the control system adopts linear state-space modelling and dynamics of its initial response is modelled by (\ref{eq:closed-loop_feedback_SDE_linear}). Transform the linear state-space model into a transfer function model by performing the Laplace transform on both sides of (\ref{eq:closed-loop_feedback_SDE_linear}) and obtain
\begin{align} \label{eq:closed-loop_feedback_SDE_linear_LT}
s \mathbf{x}(s) - \mathbf{x}_0 &= \mathbf{A}_c \mathbf{x}(s) \nonumber \\
\mathbf{x}(s) &= (s \mathbf{I} - \mathbf{A}_c)^{-1} \mathbf{x}_0 = \frac{\mathbf{C}_{s \mathbf{I} - \mathbf{A}_c}}{\det (s \mathbf{I} - \mathbf{A}_c)} \mathbf{x}_0.
\end{align}
Derivation of (\ref{eq:closed-loop_feedback_SDE_linear_LT}) involves the \textit{Cramer rule} \cite{Golub1996}. The numerator part $\mathbf{C}_{s \mathbf{I} - \mathbf{A}_c}$ denotes the \textit{co-matrix} of $s \mathbf{I} - \mathbf{A}_c$ that consists of \textit{co-factors}, whereas the denominator part $\det (s \mathbf{I} - \mathbf{A}_c)$ denotes the determinant of $s \mathbf{I} - \mathbf{A}_c$. Note that the co-factors are polynomials in terms of the Laplace variable $s$, so (\ref{eq:closed-loop_feedback_SDE_linear_LT}) implies that $\mathbf{x}(s)$ consists of fractional polynomials that share a common denominator namely the polynomial $\det (s \mathbf{I} - \mathbf{A}_c)$ which is right the characteristic polynomial of the control system. By definition of eigenvalues and poles, we know that eigenvalues of the state transition matrix $\mathbf{A}_c$ and control system poles refer to the same thing namely roots of the characteristic polynomial $\det (s \mathbf{I} - \mathbf{A}_c)$.

For example, consider rotating disk dynamics illustrated in Figure \ref{fig:RDP_open_loop_initial_response} and the self-evolutionary system described by the linear state differential equation (\ref{eq:RDP_state_DE_initial_response})
\begin{align*}
\frac{\mathrm{d}}{\mathrm{d} t} \begin{bmatrix} p \\ v \end{bmatrix} = \begin{bmatrix} 0 & 1 \\ 0 & -\frac{b}{J} \end{bmatrix} \begin{bmatrix} p \\ v \end{bmatrix},
\end{align*}
which determines the rotating disk position open-loop initial response. Perform the Laplace transform on both sides of (\ref{eq:RDP_state_DE_initial_response}) and obtain
\begin{align*}
s \begin{bmatrix} p(s) \\ v(s) \end{bmatrix} - \begin{bmatrix} p_0 \\ v_0 \end{bmatrix} &= \begin{bmatrix} 0 & 1 \\ 0 & -\frac{b}{J} \end{bmatrix} \begin{bmatrix} p(s) \\ v(s) \end{bmatrix}  \\
(s \begin{bmatrix} 1 & 0 \\ 0 & 1 \end{bmatrix} - \begin{bmatrix} 0 & 1 \\ 0 & -\frac{b}{J} \end{bmatrix}) \begin{bmatrix} p(s) \\ v(s) \end{bmatrix} &= \begin{bmatrix} p_0 \\ v_0 \end{bmatrix},
\end{align*}
which gives
\begin{align*}
\begin{bmatrix} p(s) \\ v(s) \end{bmatrix} = \begin{bmatrix} s & -1 \\ 0 & s+\frac{b}{J} \end{bmatrix}^{-1} \begin{bmatrix} p_0 \\ v_0 \end{bmatrix} = \begin{bmatrix} \frac{1}{s} & \frac{J}{s (J s+b)} \\ 0 & \frac{J}{J s+b} \end{bmatrix} \begin{bmatrix} p_0 \\ v_0 \end{bmatrix} = \begin{bmatrix} \frac{(J s+b) p_0 + J v_0}{s (J s+b)} \\ \frac{J v_0}{J s+b} \end{bmatrix}.
\end{align*}
The terms
\begin{align*}
p(s) &= \frac{(J s+b) p_0 + J v_0}{s (J s+b)},  \\ 
v(s) &= \frac{J v_0}{J s+b} = \frac{J s v_0}{s (J s+b)}
\end{align*}
share a common denominator namely the polynomial
\begin{align*}
s (J s+b) \propto \det (s \begin{bmatrix} 1 & 0 \\ 0 & 1 \end{bmatrix} - \begin{bmatrix} 0 & 1 \\ 0 & -\frac{b}{J} \end{bmatrix}) = \det (\begin{bmatrix} s & -1 \\ 0 & s+\frac{b}{J} \end{bmatrix}) = \frac{1}{J} s (J s+b)
\end{align*}
which is right the characteristic polynomial of the rotating disk position open-loop control system. It is worth noting that for a control system, the characteristic polynomial scaled by a constant factor is still the characteristic polynomial.

On the other hand, suppose the control system adopts system modelling via Laplace transform and its characteristic polynomial is
\begin{align*}
C_{LT}(s) \equiv s^n + a_{n-1} s^{n-1} + \cdots + a_0.
\end{align*}
Dynamics of its initial response is modelled by
\begin{align}  \label{eq:closed-loop_feedback_LT_initial_response}
\frac{\mathrm{d}^n}{\mathrm{d} t^n} y + a_{n-1} \frac{\mathrm{d}^{n-1}}{\mathrm{d} t^{n-1}} y + \cdots + a_0 y &= 0, \\
(s^n + a_{n-1} s^{n-1} + \cdots + a_0) y(s) &= A_0(s), \nonumber
\end{align}
where $A_0(s)$ is a constant polynomial determined by initial conditions of the output $y$. 

For the control system, denote its state $\mathbf{x}$ as
\begin{align}  \label{eq:closed-loop_feedback_LT_initial_response_state}
\mathbf{x} \equiv \begin{bmatrix} y & \frac{\mathrm{d} y}{\mathrm{d} t} & \frac{\mathrm{d}^2 y}{\mathrm{d} t^2} & \cdots & \frac{\mathrm{d}^{n-1} y}{\mathrm{d} t^{n-1}} \end{bmatrix}^\mathrm{T}
\end{align}
and transform (\ref{eq:closed-loop_feedback_LT_initial_response}) into a linear state differential equation as
\begin{equation}  \label{eq:closed-loop_feedback_LT_initial_response_SDE}
\frac{\mathrm{d}}{\mathrm{d} t} \mathbf{x} = \begin{bmatrix}  & 1 &  & & & \\ & & 1 & & & \\ & & & \ddots & & \\ & & & & 1 & \\ & & & & & 1 \\ -a_0 & -a_1 & -a_2 & \cdots & -a_{n-2} & -a_{n-1} \end{bmatrix} \mathbf{x} \equiv \mathbf{A}_c \mathbf{x}.
\end{equation}
The characteristic polynomial of $\mathbf{A}_c$ is 
\begin{align*}
\det (s \mathbf{I} - \mathbf{A}_c) &= \begin{bmatrix} s & -1 &  & & \\ & s & -1 & & \\ & & s & \ddots & \\ & & & & -1 \\ a_0 & a_1 & a_2 & \cdots & s + a_{n-1} \end{bmatrix} = \begin{bmatrix} s & &  & & \\ & s & -1 & & \\ & & s & \ddots & \\ & & & & -1 \\ a_0 & a_1 + \frac{a_0}{s} & a_2 & \cdots & s + a_{n-1} \end{bmatrix} \\
  &= \cdots = \begin{bmatrix} s & & & \\ & \ddots & & \\ & & s & \\ a_0 & \cdots & * & s + a_{n-1} + \frac{a_{n-2}}{s} + \cdots + \frac{a_0}{s^{n-1}}   \end{bmatrix} = C_{LT}(s), \\
  &(\mbox{cancel -1 in each column gradually})
\end{align*}
which implies that poles of the control system whose characteristic polynomial is $C_{LT}(s)$ coincide with eigenvalues of the state transition matrix $\mathbf{A}_c$ whose characteristic polynomial $\det (s \mathbf{I} - \mathbf{A}_c)$ is right $C_{LT}(s)$.

\subsubsection*{Perspective of equivalently-constructed transfer function block diagram}

\begin{figure}[h!]
\begin{center}
\includegraphics[width=0.8\columnwidth]{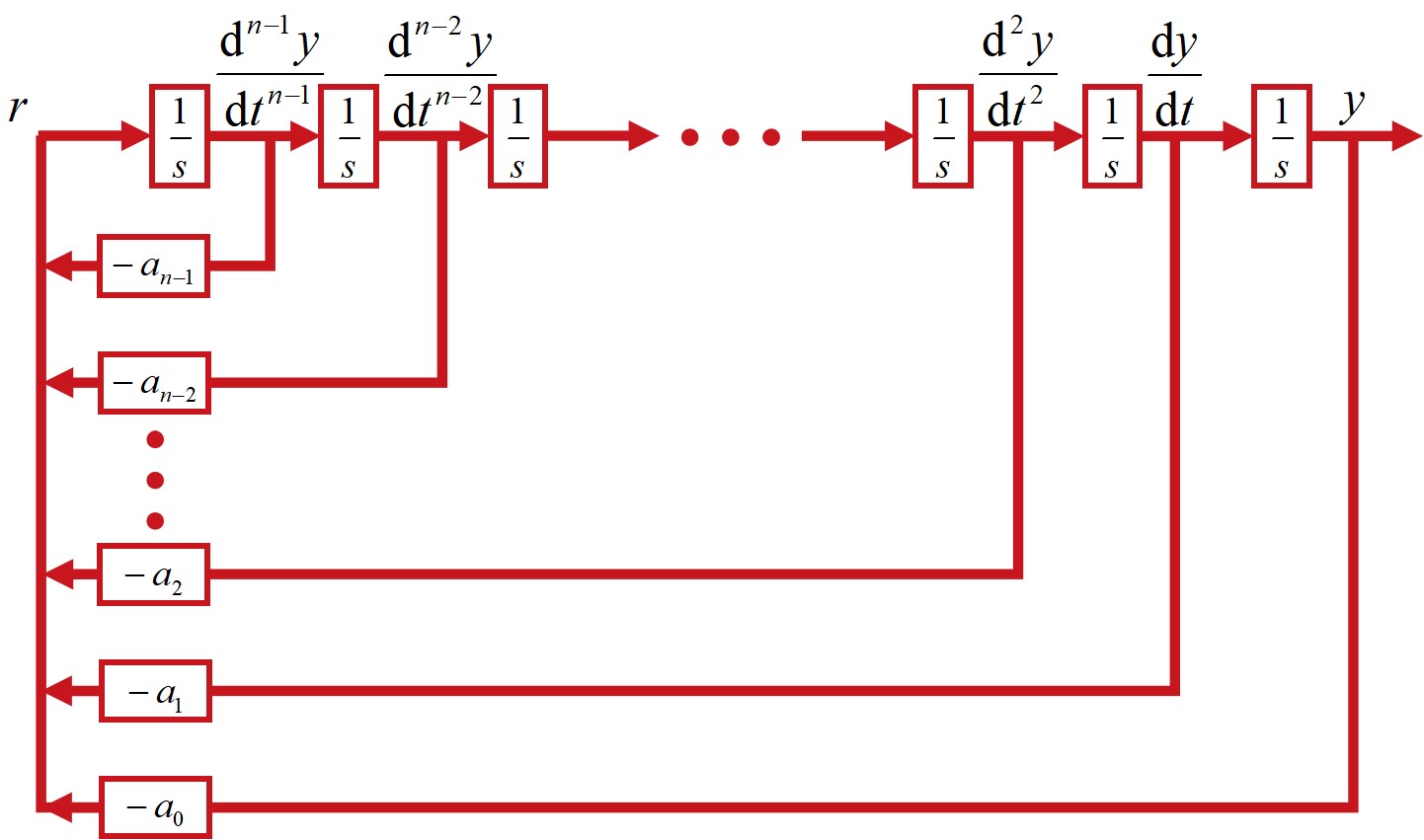}
\end{center}
\caption{Transfer function block diagram constructed according to linear state-space modelling}
\label{fig:constructed_TF_block_diagram}
\end{figure}

Construct an equivalent transfer function block diagram for the control system according to linear state-space modelling described by (\ref{eq:closed-loop_feedback_LT_initial_response_SDE}), as illustrated in Figure \ref{fig:constructed_TF_block_diagram}. Take $y$ as the control system output and add an imaginary control input $r$ to the transfer function block diagram.

For synthesis of the transfer function block diagram illustrated in Figure \ref{fig:constructed_TF_block_diagram}, resort to the \textit{Mason signal-flow gain formula} \cite{Mason1956}
\begin{equation}  \label{eq:mason_gain_formula}
T^{r,y}(s) = \frac{\sum_{P \in P_G^{r,y}} P(s) \Delta_{G/P}(s)}{\Delta_G(s)},
\end{equation}
where $T^{r,y}$ denotes the transfer function from the diagram input $r$ to the diagram output $y$ and $P_G^{r,y}$ denotes all possible open-loop paths from $r$ to $y$ in the graph $G$. $G/P$ denotes the imagined subgraph of $G$ obtained by removing the closed-loops that touch the path $P$. $\Delta$ denotes the determinant of the associated graph.

For a generic signal-flow graph $G$, its \textit{determinant} is computed via
\begin{align}  \label{eq:graph_determinant}
\Delta_G(s) &= 1 - \sum_{L_i \in L_G} L_i + \sum_{\substack{L_i, L_j \in L_G \\ non-touching}} L_i L_j - \sum_{\substack{L_i, L_j, L_k \in L_G \\ non-touching}} L_i L_j L_k \\ 
  & \quad + (-1)^m \sum_{\substack{L_{i_1}, L_{i_2}, \cdots, L_{i_m} \in L_G \\ non-touching}} L_{i_1} L_{i_2} \cdots L_{i_m} + \cdots , \nonumber
\end{align}
where $L_G$ denotes the set of all closed-loops in $G$ as well as their signed closed-loop gains. The second sum involves all pairs of non-touching closed-loops in $G$, the third sum involves all triplets of non-touching closed-loops in $G$, and so on. A simple proof of the Mason signal-flow gain formula via \textit{mathematical induction} is given in the previous book \textit{Control Theory For Practical Applications} \cite{Li2024CTPA_Springer, Li2024CTPA_SJTU_1}.

In the transfer function block diagram illustrated in Figure \ref{fig:constructed_TF_block_diagram}, there is only one open-loop path $P$ from $r$ to $y$, and its gain is
\begin{align*}
P(s) = \begin{matrix} \underbrace{\frac{1}{s} \cdot \frac{1}{s} \cdots \frac{1}{s} \cdot \frac{1}{s}} \\ n \mbox{ times} \end{matrix} = \frac{1}{s^n}.
\end{align*}
There are $n$ closed-loops 
\begin{align*}
L_0, \quad L_1, \quad L_2, \quad \cdots \quad, \quad L_{n-2}, \quad L_{n-1} 
\end{align*}
and their gains are respectively
\begin{align*}
L_{n-1}(s) &= \frac{1}{s} \cdot (-a_{n-1}) = -\frac{a_{n-1}}{s},  \\
L_{n-2}(s) &= \frac{1}{s} \cdot \frac{1}{s} \cdot (-a_{n-2}) = -\frac{a_{n-2}}{s^2},  \\
L_{n-3}(s) &= \frac{1}{s} \cdot \frac{1}{s} \cdot \frac{1}{s} \cdot (-a_{n-3}) = -\frac{a_{n-3}}{s^3},  \\
  &\vdots  \\
L_2(s) &= \begin{matrix} \underbrace{\frac{1}{s} \cdot \frac{1}{s} \cdots \frac{1}{s} \cdot \frac{1}{s}} \cdot (-a_2) \\ n-2 \mbox{ times} \end{matrix} = -\frac{a_2}{s^{n-2}},  \\
L_1(s) &= \begin{matrix} \underbrace{\frac{1}{s} \cdot \frac{1}{s} \cdots \frac{1}{s} \cdot \frac{1}{s}} \cdot (-a_1) \\ n-1 \mbox{ times} \end{matrix} = -\frac{a_1}{s^{n-1}},  \\
L_0(s) &= \begin{matrix} \underbrace{\frac{1}{s} \cdot \frac{1}{s} \cdots \frac{1}{s} \cdot \frac{1}{s}} \cdot (-a_0) \\ n \mbox{ times} \end{matrix} = -\frac{a_0}{s^n}.
\end{align*}
All the $n$ closed-loops are mutually touching. In other words, there is no pair of non-touching closed-loops. Besides, all the $n$ closed-loops touch the open-loop path $P$ as well.

Therefore, the determinant of the transfer function block diagram is
\begin{align*}
\Delta_G(s) &= 1 - \sum_{i=0}^{n-1} L_i = 1 + \frac{a_0}{s^n} + \frac{a_1}{s^{n-1}} + \frac{a_2}{s^{n-2}} + \cdots + \frac{a_{n-3}}{s^3} + \frac{a_{n-2}}{s^2} + \frac{a_{n-1}}{s}  \\
  &= \frac{s^n + a_{n-1} s^{n-1} + a_{n-2} s^{n-2} + a_{n-3} s^{n-3} + \cdots + a_2 s^2 + a_1 s + a_0}{s^n} 
\end{align*}
and the determinant of the imagined subgraph $G/P$ is
\begin{align*}
\Delta_{G/P}(s) = 1.
\end{align*}
Apply the Mason signal-flow gain formula (\ref{eq:mason_gain_formula}) and obtain
\begin{align*}
T^{r,y}(s) &= \frac{\frac{1}{s^n} \cdot 1}{\frac{s^n + a_{n-1} s^{n-1} + a_{n-2} s^{n-2} + a_{n-3} s^{n-3} + \cdots + a_2 s^2 + a_1 s + a_0}{s^n}}  \\
  &= \frac{1}{s^n + a_{n-1} s^{n-1} + a_{n-2} s^{n-2} + a_{n-3} s^{n-3} + \cdots + a_2 s^2 + a_1 s + a_0} = \frac{1}{C_{LT}(s)}.
\end{align*}

For a control system that adopts linear state-space modelling described by (\ref{eq:closed-loop_feedback_LT_initial_response_SDE}), from the perspective of an equivalently-constructed transfer function block diagram, above analysis also enables us to conclude that its characteristic polynomial is right $C_{LT}(s)$. The author believes above analysis from such perspective would even strengthen readers' understanding of mutual relationship between linear state-space modelling and system modelling via Laplace transform.

\subsubsection*{State transform}

Concerning the transform of (\ref{eq:closed-loop_feedback_LT_initial_response}) into (\ref{eq:closed-loop_feedback_LT_initial_response_SDE}), a question arises naturally: why set the state $\mathbf{x}$ as in (\ref{eq:closed-loop_feedback_LT_initial_response_state})
\begin{align*}
\mathbf{x} \equiv \begin{bmatrix} y & \frac{\mathrm{d} y}{\mathrm{d} t} & \frac{\mathrm{d}^2 y}{\mathrm{d} t^2} & \cdots & \frac{\mathrm{d}^{n-1} y}{\mathrm{d} t^{n-1}} \end{bmatrix}^\mathrm{T} ?
\end{align*}
In fact, unnecessarily so. We can fairly set the state in another way, denoted as $\bar{\mathbf{x}}$. 

No matter how $\bar{\mathbf{x}}$ is set, there must be an invertible mapping between $\mathbf{x}$ and $\bar{\mathbf{x}}$. Otherwise, system dynamics under consideration are altered, which is forbidden. Besides, since we focus on linear state-space modelling here, the invertible mapping between $\mathbf{x}$ and $\bar{\mathbf{x}}$ must be linear as well. Then suppose the \textbf{state transform} between $\mathbf{x}$ and $\bar{\mathbf{x}}$ is
\begin{equation}  \label{eq:state_transform_x_z}
\mathbf{x} = \mathbf{P} \bar{\mathbf{x}},
\end{equation}
where the invertible matrix $\mathbf{P}$ is the \textbf{state transform matrix}.

Substitute (\ref{eq:state_transform_x_z}) into (\ref{eq:closed-loop_feedback_LT_initial_response_SDE}) and obtain
\begin{equation}  \label{eq:CL_feedback_LT_initial_response_SDE_transform}
\frac{\mathrm{d}}{\mathrm{d} t} (\mathbf{P} \bar{\mathbf{x}}) = \mathbf{A}_c \mathbf{P} \bar{\mathbf{x}} \iff \frac{\mathrm{d}}{\mathrm{d} t} \bar{\mathbf{x}} = \mathbf{P}^{-1} \mathbf{A}_c \mathbf{P} \bar{\mathbf{x}} \equiv \bar{\mathbf{A}}_c \bar{\mathbf{x}}
\end{equation}
Note that
\begin{align*}
\det (s \mathbf{I} - \bar{\mathbf{A}}_c) &= \det (s \mathbf{I} - \mathbf{P}^{-1} \mathbf{A}_c \mathbf{P}) = \det (\mathbf{P}^{-1} (s \mathbf{I} - \mathbf{A}_c) \mathbf{P})  \\
  &= \det (\mathbf{P}^{-1}) \det (s \mathbf{I} - \mathbf{A}_c) \det (\mathbf{P}) = \det (\mathbf{P}^{-1}) \det (\mathbf{P}) \det (s \mathbf{I} - \mathbf{A}_c)  \\
  &= \det (\mathbf{I}) \det (s \mathbf{I} - \mathbf{A}_c) = \det (s \mathbf{I} - \mathbf{A}_c),
\end{align*}
which implies that the state transform (\ref{eq:state_transform_x_z}) does not change the characteristic polynomial of the control system. So stability analysis for the control system is essentially the same, be the state set as $\mathbf{x}$ or as $\bar{\mathbf{x}}$.

\subsection{Routh-Hurwitz criterion}  \label{sec:RH_criterion}

If we can explicitly compute eigenvalues of a linear self-evolutionary system, then we can directly take advantage of the \textit{control system stability criterion} to judge its stability, namely to check whether the real parts of all its eigenvalues are negative.

However, in practical applications where we cannot explicitly compute relevant eigenvalues, especially when parametrized characteristic polynomials are involved, we cannot directly take advantage of the \textit{control system stability criterion} for stability analysis. Instead, we can resort to the \textbf{Routh-Hurwitz criterion} method, a representative method of locating a generic polynomial's roots qualitatively according to the polynomial coefficients
\footnote{Studies on locating polynomial roots qualitatively can date back to Hermite's works in 1850s, in his paper \textit{Sur le nombre des racines d'une équation algébrique comprise entre des limites données}, originally published on \textit{Journal de Crelle} and also collected in \textit{Oeuvres de Charles Hermite} \cite{Hermite}.}.
The \textit{Routh-Hurwitz criterion} method was developed independently by E. Routh and A. Hurwitz in the late nineteenth century \cite{Dorf2008}.

The \textit{Routh-Hurwitz criterion} method is based on checking an array called \textbf{Routh array} that is completed iteratively from the ordered coefficients of the characteristic polynomial. Consider a generic characteristic polynomial
\begin{equation}  \label{eq:RH_criterion_CP}
a_n s^n + a_{n-1} s^{n-1} + \cdots + a_0
\end{equation}
with 
\begin{align*}
a_n > 0
\end{align*}
or even
\begin{align*}
a_n = 1.
\end{align*}
Order its coefficients into the first two rows of the following Routh array
\begin{align*}
\begin{matrix}
s^n & | & a_n & a_{n-2} & a_{n-4} & \cdots \\
s^{n-1} & | & a_{n-1} & a_{n-3} & a_{n-5} & \cdots \\
s^{n-2} & | & b_{n-2} & b_{n-4} & b_{n-6} & \cdots \\
s^{n-3} & | & c_{n-3} & c_{n-5} & c_{n-7} & \cdots \\
\cdots & | & \cdots & \cdots & \cdots \\
s^0 & | & z_0
\end{matrix}
\end{align*}
where rows after the second row, namely rows initiated by 
\begin{align*}
s^{n-2}, \quad s^{n-3}, \quad \cdots \quad, \quad s^1, \quad s^0, 
\end{align*}
are computed iteratively as
\begin{align*}
b_{n-2} &= -\frac{1}{a_{n-1}} \begin{vmatrix} a_n & a_{n-2} \\ a_{n-1} & a_{n-3} \end{vmatrix} = \frac{a_{n-1} a_{n-2} - a_n a_{n-3}}{a_{n-1}}, \\
b_{n-4} &= -\frac{1}{a_{n-1}} \begin{vmatrix} a_n & a_{n-4} \\ a_{n-1} & a_{n-5} \end{vmatrix}, \quad b_{n-6} = -\frac{1}{a_{n-1}} \begin{vmatrix} a_n & a_{n-6} \\ a_{n-1} & a_{n-7} \end{vmatrix}, \quad \cdots  \\
c_{n-3} &= -\frac{1}{b_{n-2}} \begin{vmatrix} a_{n-1} & a_{n-3} \\ b_{n-2} & b_{n-4} \end{vmatrix}, \quad c_{n-5} = -\frac{1}{b_{n-2}} \begin{vmatrix} a_{n-1} & a_{n-5} \\ b_{n-2} & b_{n-6} \end{vmatrix}, \quad \cdots
\end{align*}
Then we can conclude that \textit{the number of characteristic polynomial roots with positive real part is equal to the number of sign changes in the first column of the Routh array}
\footnote{Purely from mathematical perspective, the term ``characteristic'' can be omitted in above description. However, we keep the mathematically unnecessary term ``characteristic'' to emphasize the practical sense of such polynomials in the context of control systems.}.
A simplified version of proof of the \textit{Routh-Hurwitz criterion}, which catches the essential spirit of the proof given in \cite{Anagnost1991}, is given in the previous book \textit{Control Theory For Practical Applications} \cite{Li2024CTPA_Springer, Li2024CTPA_SJTU_1}.

\begin{framed} 
\noindent \textbf{Routh-Hurwitz criterion}: \textit{The number of characteristic polynomial roots with positive real part is equal to the number of sign changes in the first column of the Routh array}.
\end{framed}

According to the proof (refer to the previous book), a control system's characteristic polynomial has roots all with negative real part (namely the control system is stable) if and only if neither zero nor sign change exists in the first column of its associated Routh array. It is worth noting that just for stability analysis, people are actually not concerned with zeros in the first column of the Routh array. But in case of other concerns than stability analysis, readers may refer to the previous book for knowledge of handling zeros in the first column of the Routh array.

\begin{framed} 
\noindent \textbf{Routh-Hurwitz stability criterion}: \textit{A linear time-invariant control system is stable if and only if neither zero nor sign change exists in the first column of the Routh array associated with its characteristic polynomial}.
\end{framed}

\subsubsection*{Application: low-speed vehicle P-lateral control failure analysis}

Consider lateral control for low-speed autonomous vehicle navigation illustrated in Figure \ref{fig:low_speed_autonomous_vehicle}. Given that the vehicle lateral control system adopts linear state-space modelling described by (\ref{eq:vehicle_lateral_control_approximation})
\begin{align*}
\frac{\mathrm{d}}{\mathrm{d} t} \mathbf{x} = \begin{bmatrix} 0 & v & 0 \\ 0 & 0 & \frac{v}{L} \\ 0 & 0 & -\frac{1}{\tau_{\beta}} \end{bmatrix} \mathbf{x} + \begin{bmatrix} 0 \\ 0 \\ \frac{1}{\tau_{\beta}} \end{bmatrix} \beta_I \equiv \mathbf{A} \mathbf{x} + \mathbf{B} \beta_I,
\end{align*}
where the vehicle lateral state 
\begin{align*}
\mathbf{x} \equiv \begin{bmatrix} y & \theta & \beta \end{bmatrix}^\mathrm{T}
\end{align*}
consists of the vehicle lateral position $y$, the vehicle orientation $\theta$, and the vehicle steering angle $\beta$. The control input $\beta_I$ denotes the vehicle steering angle command.

If the P-lateral control method
\footnote{``P'' means ``proportional'', like ``P'' in the famous and popular family of \textit{proportional-integral-derivative (PID)} controllers \cite{Samad2017} that share the same control methodology of generating the control law by combining linearly a proportional term, an integral term, and a derivative term of the feedback error.}
is used, namely
\begin{equation}  \label{eq:P-lateral_control}
\beta_I = -P y = -\begin{bmatrix} P & 0 & 0 \end{bmatrix} \begin{bmatrix} y \\ \theta \\ \beta \end{bmatrix} = -\begin{bmatrix} P & 0 & 0 \end{bmatrix} \mathbf{x}.
\end{equation}
Substitute (\ref{eq:P-lateral_control}) into (\ref{eq:vehicle_lateral_control_approximation}) and obtain
\begin{equation}  \label{eq:P-lateral_control_closed-loop}
\frac{\mathrm{d}}{\mathrm{d} t} \mathbf{x} = \begin{bmatrix} 0 & v & 0 \\ 0 & 0 & \frac{v}{L} \\ 0 & 0 & -\frac{1}{\tau_{\beta}} \end{bmatrix} \mathbf{x} - \begin{bmatrix} 0 \\ 0 \\ \frac{1}{\tau_{\beta}} \end{bmatrix} \begin{bmatrix} P & 0 & 0 \end{bmatrix} \mathbf{x} = \begin{bmatrix} 0 & v & 0 \\ 0 & 0 & \frac{v}{L} \\ -\frac{P}{\tau_{\beta}} & 0 & -\frac{1}{\tau_{\beta}} \end{bmatrix} \mathbf{x},
\end{equation}
where the closed-loop state transition matrix is
\begin{equation}  \label{eq:P-lateral_control_Ac}
\mathbf{A}_c = \begin{bmatrix} 0 & v & 0 \\ 0 & 0 & \frac{v}{L} \\ -\frac{P}{\tau_{\beta}} & 0 & -\frac{1}{\tau_{\beta}} \end{bmatrix}
\end{equation}
and the parametrized characteristic polynomial is
\begin{equation}  \label{eq:P-lateral_control_CP}
\det (s \mathbf{I} - \mathbf{A}_c) = \begin{vmatrix} s & -v & 0 \\ 0 & s & -\frac{v}{L} \\ \frac{P}{\tau_{\beta}} & 0 & s+\frac{1}{\tau_{\beta}} \end{vmatrix} = s^3 + \frac{1}{\tau_{\beta}} s^2 + \frac{P v^2}{\tau_{\beta} L}.
\end{equation}

Establish a parametrized Routh array for the parametrized characteristic polynomial (\ref{eq:P-lateral_control_CP}) as
\begin{align*}
\begin{matrix}
s^3 & | & 1 & 0  \\
s^2 & | & \frac{1}{\tau_{\beta}} & \frac{P v^2}{\tau_{\beta} L}  \\
s^1 & | & -\frac{P v^2}{L} &   \\
s^0 & | & \frac{P v^2}{\tau_{\beta} L} &   
\end{matrix}
\end{align*}
No matter how the proportional coefficient $P$ is set, sign changes always exist in the first column of the Routh array. According to the \textit{Routh-Hurwitz stability criterion}, the closed-loop control system described by (\ref{eq:P-lateral_control_closed-loop}) is definitely unstable. Therefore, the P-lateral control method fails definitely, regardless of concrete configuration of $P$.

\subsection{Lyapunov stability criterion for nonlinear self-evolutionary systems}  \label{sec:Lyapunov_stability_criterion}

We have presented stability criteria for linear self-evolutionary systems. On the other hand, how to analyse stability of a nonlinear self-evolutionary system? Given a nonlinear control system that adopts generic state-space modelling described by (\ref{eq:state_differential_equation}) 
\footnote{Explicit expression of ``nonlinear'' is only to highlight consideration of systems that may not adopt linear system modelling, but by no means to imply that we exclude consideration of linear systems. So in this section, we treat linear systems as a special case of the so-called nonlinear systems.}
\begin{align*}
\frac{\mathrm{d}}{\mathrm{d} t} \mathbf{x} = f(\mathbf{x}, \mathbf{u}).
\end{align*}
As already clarified in Section \ref{sec:self_evo_systems}, the control input $\mathbf{u}$ is normally generated according to feedback of the state $\mathbf{x}$. Once the feedback control law 
\begin{align*}
\mathbf{u} = g(\mathbf{x})
\end{align*}
is determined, dynamics of the closed-loop feedback control system is equivalent to dynamics of a self-evolutionary system described by (\ref{eq:closed-loop_feedback_SDE})
\begin{align*}
\frac{\mathrm{d}}{\mathrm{d} t} \mathbf{x} = f(\mathbf{x}, g(\mathbf{x})) \equiv f_c(\mathbf{x})
\end{align*}
So we consider stability of a nonlinear self-evolutionary system that adopts state-space modelling described by (\ref{eq:closed-loop_feedback_SDE}). For presentation convenience, suppose the equilibrium or stable state of the nonlinear self-evolutionary system is 
\begin{align*}
\mathbf{x}_\mathrm{E} = \mathbf{0}
\end{align*}
by default.

Readers may expect certain general stability criterion for nonlinear self-evolutionary systems as those for linear self-evolutionary systems. Unfortunately, we do not have such general stability criterion. Instead, we have a strategy that may guide us to analyse stability of a nonlinear self-evolutionary system in \textit{ad hoc} way. The strategy consists in finding a scalar function $V(\mathbf{x})$ in terms of the state $\mathbf{x}$ such that
\begin{itemize}
\item The scalar function $V(\mathbf{x})$ is \textit{positive definite}.
\item The derivative $\frac{\mathrm{d}}{\mathrm{d} t} V(\mathbf{x})$ is \textit{negative definite} or at least \textit{negative semi-definite}
\footnote{If the derivative $\frac{\mathrm{d}}{\mathrm{d} t} V(\mathbf{x})$ is negative semi-definite, some further care is required, namely to verify that the stable state is the only fixed point or attractor for $\frac{\mathrm{d}}{\mathrm{d} t} V(\mathbf{x}) \equiv 0$. In other words, $\frac{\mathrm{d}}{\mathrm{d} t} V(\mathbf{x}) = 0$ may momentarily hold at some states other than the stable state, but will not continue to hold as the state evolves on. Here, ``$\equiv$'' is used instead of normal ``$=$'' to emphasize that $\frac{\mathrm{d}}{\mathrm{d} t} V(\mathbf{x})$ is ``stably'' zero.}.
\end{itemize}
The strategy is called the \textbf{Lyapunov strategy} \cite{Liapounoff1907}. The scalar function $V(\mathbf{x})$ that satisfies above two conditions is called a \textbf{Lyapunov function}. In one word, the Lyapunov strategy consists in finding a Lyapunov function for the nonlinear self-evolutionary system. \textit{If we do find a Lyapunov function for the nonlinear self-evolutionary system, then we can conclude that the nonlinear self-evolutionary system is stable} --- But attention that if we cannot find any Lyapunov function for the nonlinear self-evolutionary system, then we should refrain from concluding that the nonlinear self-evolutionary system is unstable. Perhaps the nonlinear self-evolutionary system is stable and proper Lyapunov functions exist for it, yet it is simply that we lack ability to find any of them.

\begin{framed} 
\noindent \textbf{Lyapunov stability criterion}: \textit{If we do find a Lyapunov function for the nonlinear self-evolutionary system, then the nonlinear self-evolutionary system is stable}.
\end{framed}

For example, consider a linear self-evolutionary system that adopts linear state-space modelling (\ref{eq:closed-loop_feedback_SDE_linear})
\begin{align*}
\frac{\mathrm{d}}{\mathrm{d} t} \mathbf{x} = \mathbf{A}_c \mathbf{x},
\end{align*}
where the state transition matrix $\mathbf{A}_c$ has eigenvalues all with negative real part. The linear self-evolutionary system is stable according to the \textit{control system stability criterion} presented in Section \ref{sec:stability_criterion_linear}. Here, apply the Lyapunov strategy to verify its stability. Consider the following Lyapunov equation
\begin{align*}
\mathbf{P} \mathbf{A}_c + \mathbf{A}_c^\mathrm{T} \mathbf{P} = -\mathbf{I}.
\end{align*}
Above Lyapunov equation has a unique positive definite solution of $\mathbf{P}$ (why will be explained in following sections). Then define the Lyapunov function as
\begin{align*}
V \equiv V(\mathbf{x}) = \mathbf{x}^\mathrm{T} \mathbf{P} \mathbf{x},
\end{align*}
which is positive definite. Besides, we can verify
\begin{align*}
\frac{\mathrm{d}}{\mathrm{d} t} V = \mathbf{x}^\mathrm{T} \mathbf{P} \frac{\mathrm{d} \mathbf{x}}{\mathrm{d} t} + \frac{\mathrm{d} \mathbf{x}^\mathrm{T}}{\mathrm{d} t} \mathbf{P} \mathbf{x} = \mathbf{x}^\mathrm{T} (\mathbf{P} \mathbf{A}_c + \mathbf{A}_c^\mathrm{T} \mathbf{P}) \mathbf{x} = -\mathbf{x}^\mathrm{T} \mathbf{x} < 0.
\end{align*}
So the defined function $V(\mathbf{x})$ is indeed a Lyapunov function for the linear self-evolutionary system whose stability is then verified.

For another example, consider a nonlinear self-evolutionary system that adopts state-space modelling as
\begin{align*}
\frac{\mathrm{d}}{\mathrm{d} t} \mathbf{x} = -(\mathbf{x}^\mathrm{T} \mathbf{x}) \mathbf{x}.
\end{align*}
For the nonlinear self-evolutionary system, we can find a Lyapunov function as 
\begin{align*}
V(\mathbf{x}) = \mathbf{x}^\mathrm{T} \mathbf{x}
\end{align*}
because such $V(\mathbf{x})$ is positive definite and its derivative
\begin{align*}
\frac{\mathrm{d}}{\mathrm{d} t} V = \mathbf{x}^\mathrm{T} \frac{\mathrm{d} \mathbf{x}}{\mathrm{d} t} + \frac{\mathrm{d} \mathbf{x}^\mathrm{T}}{\mathrm{d} t} \mathbf{x} = \mathbf{x}^\mathrm{T} (-\mathbf{x}^\mathrm{T} \mathbf{x}) \mathbf{x} + \mathbf{x}^\mathrm{T} (-\mathbf{x}^\mathrm{T} \mathbf{x}) \mathbf{x} = -2 (\mathbf{x}^\mathrm{T} \mathbf{x})^2 < 0.
\end{align*}
So the nonlinear self-evolutionary system is stable.

\subsubsection*{Application: damped pendulum stability analysis}

\begin{figure}[h!]
\begin{center}
\includegraphics[width=0.3\columnwidth]{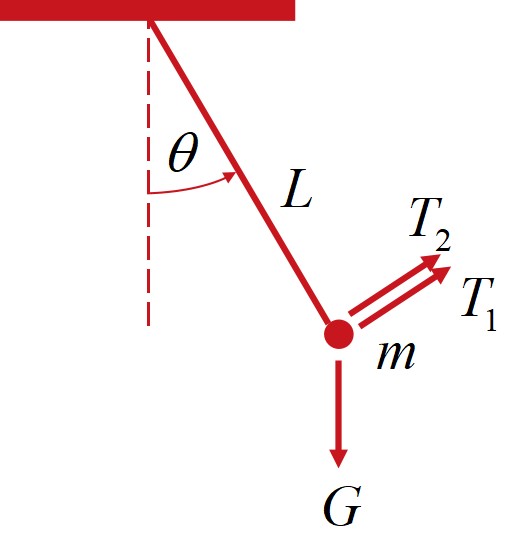}
\end{center}
\caption{Damped pendulum}
\label{fig:damped_pendulum}
\end{figure}

Consider a daily-life pendulum
\footnote{People's intuitive conception of pendulums may normally be attributed to pendulum clocks in daily life, yet pendulums have a much wider range of fascinating forms and extensions such as torsion pendulums, chaotic pendulums, coupled pendulums, quantum pendulums, and superconductivity-related pendulums \cite{Baker2005}.}
or in other words a \textit{damped pendulum} (in contrast with the \textit{simple pendulum} namely the imagined ideal pendulum that is under influence of gravity only but is exempt from any other influence especially that of damping factors such as friction), as illustrated in Figure \ref{fig:damped_pendulum}. Here, $m$ denotes the pendulum mass, $L$ denotes the pendulum length, $G$ denotes gravity, $T_1$ denotes the torque component contributed by gravity, and $T_2$ denotes the damping torque.

The rotating inertia of the damped pendulum is
\begin{align*}
J = m L^2
\end{align*}
and the two torques are computed as
\begin{align*}
T_1 &= -m g L \sin \theta,  \\ 
T_2 &= -b \frac{\mathrm{d} \theta}{\mathrm{d} t},
\end{align*}
where $b$ denotes the damping coefficient. 

So dynamics of the damped pendulum can be described by the following differential equation
\begin{align}  \label{eq:damped_pendulum_DE}
J \frac{\mathrm{d}^2 \theta}{\mathrm{d} t^2} = T_1 + T_2 &\iff m L^2 \frac{\mathrm{d}^2 \theta}{\mathrm{d} t^2} = -m g L \sin \theta - b \frac{\mathrm{d} \theta}{\mathrm{d} t}  \nonumber \\
  &\iff \frac{\mathrm{d}^2 \theta}{\mathrm{d} t^2} = -\frac{g}{L} \sin \theta - \frac{b}{m L^2} \frac{\mathrm{d} \theta}{\mathrm{d} t}
\end{align}
or expressed as a state differential equation
\begin{equation}  \label{eq:damped_pendulum_SDE}
\frac{\mathrm{d}}{\mathrm{d} t} \mathbf{x} \equiv \frac{\mathrm{d}}{\mathrm{d} t} \begin{bmatrix} \theta \\ \frac{\mathrm{d} \theta}{\mathrm{d} t} \end{bmatrix} = \begin{bmatrix} \frac{\mathrm{d} \theta}{\mathrm{d} t} \\ -\frac{g}{L} \sin \theta - \frac{b}{m L^2} \frac{\mathrm{d} \theta}{\mathrm{d} t} \end{bmatrix} \equiv f(\mathbf{x}),
\end{equation}
where the state 
\begin{align*}
\mathbf{x} \equiv \begin{bmatrix} \theta & \frac{\mathrm{d} \theta}{\mathrm{d} t} \end{bmatrix}^\mathrm{T}
\end{align*}
consists of the damped pendulum angle and angular velocity. Also denote variable derivatives as
\begin{align*}
\dot \theta \equiv \frac{\mathrm{d} \theta}{\mathrm{d} t}, \qquad \ddot \theta \equiv \frac{\mathrm{d}^2 \theta}{\mathrm{d} t^2}.
\end{align*}

\begin{figure}[h!]
\begin{center}
\includegraphics[width=0.6\columnwidth]{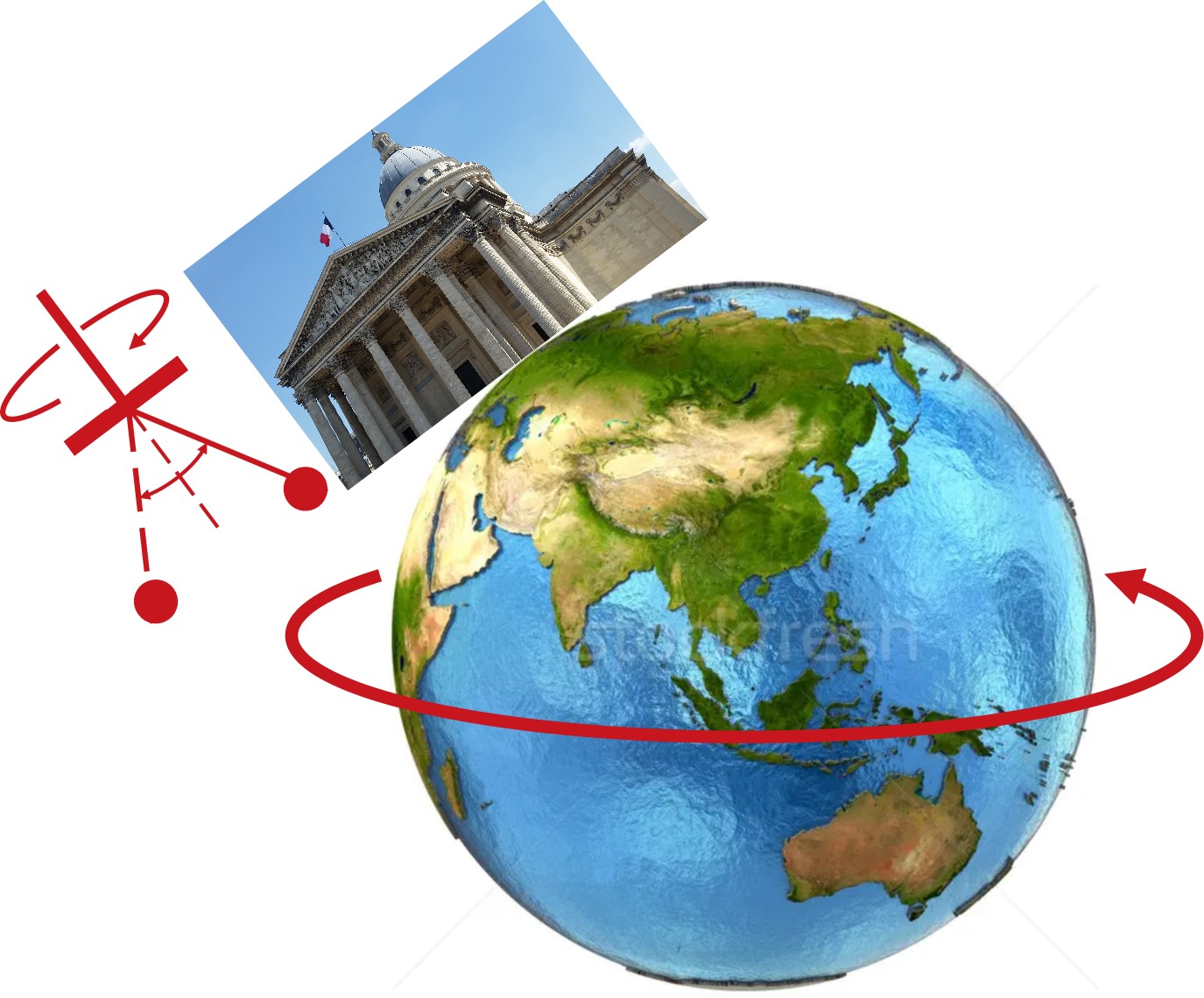}
\end{center}
\caption{Foucault pendulum}
\label{fig:Foucault_pendulum}
\end{figure}

It is worth noting that motion of the damped pendulum is assumed to be constrained in a plane intentionally, so the damped pendulum will not demonstrate the gyroscope-style dynamics like the famous \textit{Foucault pendulum} \cite{Foucault1851} that is installed on the dome inside \textit{Panthéon} in Paris and was conceived as an experiment to give simple and direct evidence of the Earth's rotation for the first time in history, as illustrated in Figure \ref{fig:Foucault_pendulum}.

Construct a candidate Lyapunov function as 
\begin{equation}  \label{eq:damped_pendulum_Lyapunov_func}
V(\mathbf{x}) = \frac{g}{L} (1 - \cos \theta) + \frac{1}{2} \dot \theta^2.
\end{equation}
First, the scalar function $V(\mathbf{x})$ is positive definite, because
\begin{align*}
1 - \cos \theta &\geq 0,  \\ 
\dot \theta^2 &\geq 0
\end{align*}
where both equalities hold if and only if 
\begin{align*}
\theta = 0. 
\end{align*}

Second, compute the derivative
\begin{align*}
\frac{\mathrm{d}}{\mathrm{d} t} V(\mathbf{x}) = \frac{g}{L} \sin \theta \dot \theta + \dot \theta \ddot \theta = \frac{g}{L} \sin \theta \dot \theta + \dot \theta (-\frac{g}{L} \sin \theta - \frac{b}{m L^2} \dot \theta) = - \frac{b}{m L^2} \dot \theta^2 \leq 0
\end{align*}
For 
\begin{align*}
\frac{\mathrm{d}}{\mathrm{d} t} V(\mathbf{x}) \equiv 0 \iff \dot \theta \equiv 0,
\end{align*}
the following condition
\begin{align*}
\ddot \theta = -\frac{g}{L} \sin \theta - \frac{b}{m L^2} \dot \theta \iff \frac{g}{L} \sin \theta = -\ddot \theta - \frac{b}{m L^2} \dot \theta \equiv 0 \iff \theta \equiv 0
\end{align*}
needs to hold, which implies that the stable state is the only fixed point or attractor for 
\begin{align*}
\frac{\mathrm{d}}{\mathrm{d} t} V(\mathbf{x}) \equiv 0.
\end{align*}
As already mentioned, ``$\equiv$'' emphasizes the status of being ``stably'' zero. 

Finally, we can conclude that the scalar function $V(\mathbf{x})$ given in (\ref{eq:damped_pendulum_Lyapunov_func}) is indeed a Lyapunov function for the damped pendulum and hence the damped pendulum is stable.

\section{Lyapunov and Riccati equations}

The Lyapunov equation and the Riccati equation (or more specifically, the matrix Riccati equation) are two important kinds of equations closely related to state-space analysis for control systems.

\subsection{Lyapunov equation}  \label{sec:Lyapunov_equation}

The generic formalism of \textbf{Lyapunov equation} is
\begin{equation}  \label{eq:Lyapunov_equation}
\mathbf{P} \mathbf{X} + \mathbf{X}^\mathrm{T} \mathbf{P} = \mathbf{Y},
\end{equation}
where $\mathbf{P}$ is the unknown square matrix to be solved with known square matrices $\mathbf{X}$ and $\mathbf{Y}$. By default, $\mathbf{X}$ is a real-value matrix and $\mathbf{Y}$ is a real-value symmetric matrix. Instead of generally discussing the Lyapunov equation described in (\ref{eq:Lyapunov_equation}), we focus on special cases of the Lyapunov equation that are closely related to control problems.

\subsubsection*{Special case 1: The matrix $\mathbf{X}$ is stable}

The matrix $\mathbf{X}$ being \textbf{stable} means that it has its eigenvalues all with negative real part. Decompose $\mathbf{X}$ via \textit{Schur decomposition} \cite{Horn2012} as
\begin{align*}
\mathbf{X} = \mathbf{U} \mathbf{\Sigma} \mathbf{U}^*
\end{align*}
such that $\mathbf{\Sigma}$ is an upper-triangular matrix and $\mathbf{U}$ is a unitary matrix satisfying
\begin{align*}
\mathbf{U} \mathbf{U}^* = \mathbf{U}^* \mathbf{U} = \mathbf{I}.
\end{align*}
Note that $\mathbf{X}$ is a real-value matrix and hence $\mathbf{X}^\mathrm{T} = \mathbf{X}^*$, transform (\ref{eq:Lyapunov_equation}) into an equivalent Lyapunov equation as
\begin{align*}
\mathbf{P} \mathbf{X} + \mathbf{X}^\mathrm{T} \mathbf{P} = \mathbf{Y} &\iff \mathbf{P} \mathbf{U} \mathbf{\Sigma} \mathbf{U}^* + \mathbf{U} \mathbf{\Sigma}^* \mathbf{U}^* \mathbf{P} = \mathbf{Y} \\
  &\iff \mathbf{U}^* \mathbf{P} \mathbf{U} \mathbf{\Sigma} + \mathbf{\Sigma}^* \mathbf{U}^* \mathbf{P} \mathbf{U} = \mathbf{U}^* \mathbf{Y} \mathbf{U}
\end{align*}
namely 
\begin{align}  \label{eq:Lyapunov_equation_equivalent}
\bar{\mathbf{P}} \mathbf{\Sigma} + \mathbf{\Sigma}^* \bar{\mathbf{P}} = \bar{\mathbf{Y}},
\end{align}
where 
\begin{align*}
\bar{\mathbf{P}} = \mathbf{U}^* \mathbf{P} \mathbf{U}, \quad \bar{\mathbf{Y}} = \mathbf{U}^* \mathbf{Y} \mathbf{U}.
\end{align*}

Perform \textit{matrix vectorization} on both sides of (\ref{eq:Lyapunov_equation_equivalent}) and obtain
\begin{align}  \label{eq:Lyapunov_equation_equivalent_M_vectorization}
\mbox{vec}(\bar{\mathbf{P}} \mathbf{\Sigma} + \mathbf{\Sigma}^* \bar{\mathbf{P}}) = \mbox{vec}(\bar{\mathbf{Y}}) \iff (\mathbf{\Sigma}^\mathrm{T} \otimes \mathbf{I} + \mathbf{I} \otimes \mathbf{\Sigma}^*) \mbox{vec}(\bar{\mathbf{P}}) = \mbox{vec}(\bar{\mathbf{Y}}),
\end{align}
where $\otimes$ denotes the \textit{Kronecker product} \cite{Horn1991} --- Matrix vectorization is defined as
\begin{align*}
\mbox{vec}(\begin{bmatrix} \mathbf{b}_1 & \mathbf{b}_2 & \cdots & \mathbf{b}_n \end{bmatrix}) \equiv \begin{bmatrix} \mathbf{b}_1 \\ \mathbf{b}_2 \\ \vdots \\ \mathbf{b}_n \end{bmatrix},
\end{align*}
where 
\begin{align*}
\mathbf{b}_1, \quad \mathbf{b}_2, \quad \cdots \quad, \quad \mathbf{b}_n 
\end{align*}
are vectors. Kronecker product is defined as
\begin{align*}
\mathbf{A} \otimes \mathbf{B} \equiv \begin{bmatrix} a_{11} \mathbf{B} & \cdots & a_{1n} \mathbf{B} \\ \vdots & \ddots & \vdots \\ a_{m1} \mathbf{B} & \cdots & a_{mn} \mathbf{B} \end{bmatrix}. 
\end{align*}
We have 
\begin{align*}
\mbox{vec}(\mathbf{A} \mathbf{B} \mathbf{C}) = (\mathbf{C}^\mathrm{T} \otimes \mathbf{A}) \mbox{vec}(\mathbf{B}).
\end{align*}

Denote the eigenvalues of $\mathbf{X}$ as
\begin{align*}
\lambda_1, \quad \lambda_2, \quad \cdots \quad, \quad \lambda_n, 
\end{align*}
which are all with negative real part as $\mathbf{X}$ is stable. Since
\begin{align*}
\mathbf{\Sigma}^\mathrm{T} \otimes \mathbf{I} + \mathbf{I} \otimes \mathbf{\Sigma}^* = \begin{bmatrix} \lambda_1 \mathbf{I} + \mathbf{\Sigma}^* & & \\  & \ddots & \\ * & & \lambda_n \mathbf{I} + \mathbf{\Sigma}^* \end{bmatrix}
\end{align*}
and each of its diagonal blocks is of the form
\begin{align*}
\lambda_i \mathbf{I} + \mathbf{\Sigma}^* = \begin{bmatrix} \lambda_i + \bar{\lambda}_1 & & \\ & \ddots & \\ * & & \lambda_i + \bar{\lambda}_n \end{bmatrix}
\end{align*}
whose diagonal elements are all with negative real part, the matrix
\begin{align*}
\mathbf{\Sigma}^\mathrm{T} \otimes \mathbf{I} + \mathbf{I} \otimes \mathbf{\Sigma}^*
\end{align*}
is a lower-triangular matrix with non-zero diagonal elements. Therefore, the equation described in (\ref{eq:Lyapunov_equation_equivalent_M_vectorization}) has a unique solution of $\mbox{vec}(\bar{\mathbf{P}})$, which implies that the original Lyapunov equation described in (\ref{eq:Lyapunov_equation}) has a unique solution of $\mathbf{P}$.

\begin{framed} 
\noindent \textbf{Lyapunov criterion I}: \textit{If the matrix $\mathbf{X}$ is stable, the Lyapunov equation described in (\ref{eq:Lyapunov_equation}) has a unique solution of $\mathbf{P}$.}
\end{framed}

The linear equation group described in (\ref{eq:Lyapunov_equation_equivalent_M_vectorization}) can be solved efficiently, thanks to the fact that the matrix
\begin{align*}
\mathbf{\Sigma}^\mathrm{T} \otimes \mathbf{I} + \mathbf{I} \otimes \mathbf{\Sigma}^*
\end{align*}
is a lower-triangular matrix. For a Lyapunov equation of large scale, we can take advantage of \textit{Schur decomposition} to transform it into an equivalent Lyapunov equation of the form described in (\ref{eq:Lyapunov_equation_equivalent_M_vectorization}) that can be solved efficiently. On the other hand, for a Lyapunov equation of moderate scale, we may perform matrix vectorization directly on the original Lyapunov equation described in (\ref{eq:Lyapunov_equation}) as
\begin{equation}  \label{eq:Lyapunov_equation_equivalent_M_vectorization_direct}
\mbox{vec}(\mathbf{P} \mathbf{X} + \mathbf{X}^\mathrm{T} \mathbf{P}) = \mbox{vec}(\mathbf{Y}) \iff (\mathbf{X}^\mathrm{T} \otimes \mathbf{I} + \mathbf{I} \otimes \mathbf{X}^\mathrm{T}) \mbox{vec}(\mathbf{P}) = \mbox{vec}(\mathbf{Y})
\end{equation}
and solve $\mathbf{P}$ via (\ref{eq:Lyapunov_equation_equivalent_M_vectorization_direct}). Matlab code for demonstration of Lyapunov equation solving is given as follows.

\begin{framed} 
\noindent \textbf{SolveLyapunovEquation.m} \\
\noindent \%\% Lyapunov equation: P X + X' P = Y \\
function P = SolveLyapunovEquation(X, Y) \\
$~~~~$ n = size(X,1); vecY = reshape(Y,[],1); \% [Y(:,1); Y(:,2); ...; Y(:,n)] \\
$~~~~$ M = zeros(n*n,n*n); k = 0; \\
$~~~~$ for c = 1:n \\
$~~~~$ $~~~~$ for r = 1:n \\
$~~~~$ $~~~~$ $~~~~$ k = k+1; \\
$~~~~$ $~~~~$ $~~~~$ M(k,(c*n-n+1):(c*n)) = M(k,(c*n-n+1):(c*n)) + X(:,r)'; \\
$~~~~$ $~~~~$ $~~~~$ M(k,r:n:end) = M(k,r:n:end) + X(:,c)'; \\
$~~~~$ $~~~~$ end \\
$~~~~$ end \\
$~~~~$ vecP = M$\backslash$vecY; \% [P(:,1); P(:,2); ...; P(:,n)] \\
$~~~~$ P = reshape(vecP,n,n); \\
end
\end{framed}

Matlab code for demonstration of symbolic operation oriented Lyapunov equation solving is given as follows.

\begin{framed} 
\noindent \textbf{SolveLyapunovEquationSym.m} \\
\noindent \%\% Lyapunov equation: P X + X' P = Y (symbolic operation) \\
function P = SolveLyapunovEquationSym(X, Y) \\
$~~~~$ n = size(X,1); vecY = reshape(Y,[],1); \% [Y(:,1); Y(:,2); ...; Y(:,n)] \\
$~~~~$ M = sym(zeros(n*n,n*n)); k = 0; \\
$~~~~$ for c = 1:n \\
$~~~~$ $~~~~$ for r = 1:n \\
$~~~~$ $~~~~$ $~~~~$ k = k+1; \\
$~~~~$ $~~~~$ $~~~~$ M(k,(c*n-n+1):(c*n)) = M(k,(c*n-n+1):(c*n)) + transpose(X(:,r)); \\
$~~~~$ $~~~~$ $~~~~$ M(k,r:n:end) = M(k,r:n:end) + transpose(X(:,c)); \\
$~~~~$ $~~~~$ end \\
$~~~~$ end \\
$~~~~$ vecP = M$\backslash$vecY; \% [P(:,1); P(:,2); ...; P(:,n)] \\
$~~~~$ P = reshape(vecP,n,n); \\
end
\end{framed}

\subsubsection*{Special case 2: The matrix $\mathbf{X}$ is stable and the matrix $\mathbf{Y}$ is negative definite}

According to the \textit{Lyapunov criterion I}, the Lyapunov equation described in (\ref{eq:Lyapunov_equation}) has a unique solution of $\mathbf{P}$. Since $-\mathbf{Y}$ is positive definite, i.e. 
\begin{align*}
-\mathbf{Y}>0, 
\end{align*}
we further have
\begin{align*}
\mathbf{P} &= \mathrm{e}^{\mathbf{X}^\mathrm{T} t} \mathbf{P} \mathrm{e}^{\mathbf{X} t} |_{t=0} - \mathrm{e}^{\mathbf{X}^\mathrm{T} t} \mathbf{P} \mathrm{e}^{\mathbf{X} t} |_{t \to \infty} = -\int_0^{\infty} \frac{\mathrm{d}}{\mathrm{d} t} [\mathrm{e}^{\mathbf{X}^\mathrm{T} t} \mathbf{P} \mathrm{e}^{\mathbf{X} t}] \mathrm{d} t \\
  &= -\int_0^{\infty} \mathrm{e}^{\mathbf{X}^\mathrm{T} t} (\mathbf{X}^\mathrm{T} \mathbf{P} + \mathbf{P} \mathbf{X}) \mathrm{e}^{\mathbf{X} t} \mathrm{d} t = \int_0^{\infty} \mathrm{e}^{\mathbf{X}^\mathrm{T} t} (-\mathbf{Y}) \mathrm{e}^{\mathbf{X} t} \mathrm{d} t > 0,
\end{align*}
which implies that the Lyapunov equation described in (\ref{eq:Lyapunov_equation}) has a unique solution of $\mathbf{P}$ that is positive definite.

\begin{framed} 
\noindent \textbf{Lyapunov criterion II}: \textit{If the matrix $\mathbf{X}$ is stable and the matrix $\mathbf{Y}$ is negative definite, the Lyapunov equation described in (\ref{eq:Lyapunov_equation}) has a unique solution of $\mathbf{P}$ that is positive definite.}
\end{framed}

The \textit{special case 2} is closely related to the following matrix inequality
\begin{equation}  \label{eq:Lyapunov_inequality}
\mathbf{P} \mathbf{X} + \mathbf{X}^\mathrm{T} \mathbf{P} < 0
\end{equation}
which is called the \textbf{Lyapunov inequality} \cite{Boyd1994} or \textbf{matrix Lyapunov inequality}. It is worth noting that the Lyapunov inequality can also refer to a probabilistic inequality
\begin{align*}
(E[|x|^a])^{\frac{1}{a}} \leq (E[|x|^b])^{\frac{1}{b}} \qquad (0 < a \leq b),
\end{align*}
which is generalized from Lyapunov's original works \cite{Liapounoff1900}. Set 
\begin{align*}
y \equiv |x|^a, \quad r \equiv b/a \geq 1 
\end{align*}
and obtain
\begin{align*}
(E[|x|^a])^{\frac{1}{a}} \leq (E[|x|^b])^{\frac{1}{b}} \iff (E[y])^r \leq E[y^r] 
\end{align*}
which can be verified by the \textit{Jensen inequality} \cite{Mitrinovic1970} (note that $y^r$ is a convex function). The two Lyapunov inequalities can be well distinguished from each other, if they are called \textit{matrix Lyapunov inequality} and \textit{probabilistic Lyapunov inequality} respectively. Yet throughout this book, the Lyapunov inequality refers to the matrix Lyapunov inequality by default.

Similar to derivation of the \textit{Lyapunov criterion II}, we have
\begin{align*}
\mathbf{P} &= \mathrm{e}^{\mathbf{X}^\mathrm{T} t} \mathbf{P} \mathrm{e}^{\mathbf{X} t} |_{t=0} - \mathrm{e}^{\mathbf{X}^\mathrm{T} t} \mathbf{P} \mathrm{e}^{\mathbf{X} t} |_{t \to \infty} = -\int_0^{\infty} \frac{\mathrm{d}}{\mathrm{d} t} [\mathrm{e}^{\mathbf{X}^\mathrm{T} t} \mathbf{P} \mathrm{e}^{\mathbf{X} t}] \mathrm{d} t \\
  &= -\int_0^{\infty} \mathrm{e}^{\mathbf{X}^\mathrm{T} t} (\mathbf{X}^\mathrm{T} \mathbf{P} + \mathbf{P} \mathbf{X}) \mathrm{e}^{\mathbf{X} t} \mathrm{d} t > 0,
\end{align*}
which implies that every solution $\mathbf{P}$ of the Lyapunov inequality described in (\ref{eq:Lyapunov_inequality}) is positive definite.

\begin{framed} 
\noindent \textbf{Lyapunov criterion II-B}: \textit{If the matrix $\mathbf{X}$ is stable, every solution $\mathbf{P}$ of the Lyapunov inequality described in (\ref{eq:Lyapunov_inequality}) is positive definite.}
\end{framed}

\subsubsection*{Special case 3: The matrix $\mathbf{Y}$ is negative definite and the solution $\mathbf{P}$ is positive definite}

In the spirit of the Lyapunov strategy presented in Section \ref{sec:Lyapunov_stability_criterion}, construct a dynamic system with state $\mathbf{z}$ that follows
\begin{align*}
\frac{\mathrm{d}}{\mathrm{d} t} \mathbf{z} = \mathbf{X} \mathbf{z}.
\end{align*}
The state 
\begin{align*}
\mathbf{z} = \mathbf{0}
\end{align*}
is obviously the stable or equilibrium state for the constructed dynamic system. 

Define a positive definite function $V$ as
\begin{align*}
V \equiv V(\mathbf{z}) = \mathbf{z}^\mathrm{T} \mathbf{P} \mathbf{z}.
\end{align*}
When 
\begin{align*}
\mathbf{z} \not = \mathbf{0}, 
\end{align*}
we have
\begin{align*}
\frac{\mathrm{d}}{\mathrm{d} t} V = \frac{\mathrm{d} \mathbf{z}^\mathrm{T}}{\mathrm{d} t} \mathbf{P} \mathbf{z} + \mathbf{z}^\mathrm{T} \mathbf{P} \frac{\mathrm{d} \mathbf{z}}{\mathrm{d} t} = \mathbf{z}^\mathrm{T} (\mathbf{X}^\mathrm{T} \mathbf{P} + \mathbf{P} \mathbf{X}) \mathbf{z} = \mathbf{z}^\mathrm{T} \mathbf{Y} \mathbf{z} < 0,
\end{align*}
which implies that the positive definite function $V$ decreases monotonically and definitely converges to a limit with 
\begin{align*}
\frac{\mathrm{d}}{\mathrm{d} t} V = 0. 
\end{align*}
Note that
\begin{align*}
\frac{\mathrm{d}}{\mathrm{d} t} V = \mathbf{z}^\mathrm{T} \mathbf{Y} \mathbf{z} = 0 \iff \mathbf{z} = 0
\end{align*} 
and all above derivation holds regardless of the initial state $\mathbf{z}_0$, so the constructed dynamic system always converges to the stable state and hence the matrix $\mathbf{X}$ is stable. 

In fact, the positive definiteness of the scalar function $V(\mathbf{z})$ and the negative definiteness of $\frac{\mathrm{d}}{\mathrm{d} t} V(\mathbf{z})$ imply that $V(\mathbf{z})$ is a Lyapunov function for the constructed dynamic system and hence the constructed dynamic system is stable (according to the \textit{Lyapunov stability criterion} presented in Section \ref{sec:Lyapunov_stability_criterion}). From this perspective we can also conclude that the matrix $\mathbf{X}$ is stable. Therefore, for the Lyapunov equation described in (\ref{eq:Lyapunov_equation}), if the matrix $\mathbf{Y}$ is negative definite and the solution $\mathbf{P}$ is positive definite, the matrix $\mathbf{X}$ is stable.

\begin{framed} 
\noindent \textbf{Lyapunov criterion III}: \textit{For the Lyapunov equation described in (\ref{eq:Lyapunov_equation}), if the matrix $\mathbf{Y}$ is negative definite and the solution $\mathbf{P}$ is positive definite, the matrix $\mathbf{X}$ is stable.}
\end{framed}

Suppose there is a positive definite matrix $\mathbf{P}$ such that
\begin{align*}
\mathbf{P} \mathbf{X} + \mathbf{X}^\mathrm{T} \mathbf{P} < 0.
\end{align*}
In other words, suppose the Lyapunov inequality described in (\ref{eq:Lyapunov_inequality}) has a positive definite solution $\mathbf{P}$. Similar to above analysis, construct a dynamic system with state $\mathbf{z}$ that follows
\begin{align*}
\frac{\mathrm{d}}{\mathrm{d} t} \mathbf{z} = \mathbf{X} \mathbf{z}.
\end{align*}
Define a positive definite function $V$ as
\begin{align*}
V \equiv V(\mathbf{z}) = \mathbf{z}^\mathrm{T} \mathbf{P} \mathbf{z}.
\end{align*}
We have
\begin{align*}
\frac{\mathrm{d}}{\mathrm{d} t} V = \frac{\mathrm{d} \mathbf{z}^\mathrm{T}}{\mathrm{d} t} \mathbf{P} \mathbf{z} + \mathbf{z}^\mathrm{T} \mathbf{P} \frac{\mathrm{d} \mathbf{z}}{\mathrm{d} t} = \mathbf{z}^\mathrm{T} (\mathbf{X}^\mathrm{T} \mathbf{P} + \mathbf{P} \mathbf{X}) \mathbf{z}
\end{align*}
which is a negative definite function in terms of the state $\mathbf{z}$.

So $V(\mathbf{z})$ is a Lyapunov function for the constructed dynamic system and hence the constructed dynamic system is stable (according to the \textit{Lyapunov stability criterion}), which further implies that the matrix $\mathbf{X}$ is stable. Therefore, if the Lyapunov inequality described in (\ref{eq:Lyapunov_inequality}) has a positive definite solution $\mathbf{P}$, the matrix $\mathbf{X}$ is stable.

\begin{framed} 
\noindent \textbf{Lyapunov criterion III-B}: \textit{If the Lyapunov inequality described in (\ref{eq:Lyapunov_inequality}) has a positive definite solution $\mathbf{P}$, the matrix $\mathbf{X}$ is stable.}
\end{framed}

\subsubsection*{Lyapunov stability criterion for linear self-evolutionary systems}

For the Lyapunov inequality described in (\ref{eq:Lyapunov_inequality})
\begin{align*}
\mathbf{P} \mathbf{X} + \mathbf{X}^\mathrm{T} \mathbf{P} < 0,
\end{align*}
we say that the Lyapunov inequality is \textit{characterized} by the matrix $\mathbf{X}$, or the matrix $\mathbf{X}$ \textit{characterizes} the Lyapunov inequality.

For linear self-evolutionary systems, the \textit{Lyapunov criterion II-B} and the \textit{Lyapunov criterion III-B} lead to another stability criterion besides the \textit{control system stability criterion} presented in Section \ref{sec:stability_criterion_linear} and the \textit{Routh-Hurwitz stability criterion} presented in Section \ref{sec:RH_criterion}.

\begin{framed} 
\noindent \textbf{Lyapunov stability criterion II}: \textit{A linear self-evolutionary system is stable if and only if the Lyapunov inequality characterized by its state transition matrix has a positive definite solution}.
\end{framed}

In practice, to determine whether the Lyapunov inequality
\begin{align*}
\mathbf{P} \mathbf{X} + \mathbf{X}^\mathrm{T} \mathbf{P} < 0
\end{align*}
has a positive definite solution or not, it is unnecessary to handle the problem purely from the inequality perspective. According to the \textit{Lyapunov criterion II} and the \textit{Lyapunov criterion III}, we can turn the problem into an equivalent one handled from the equality perspective. More specifically, we can choose an arbitrary negative definite matrix $\mathbf{Y}$. For example, simply choose 
\begin{align*}
\mathbf{Y} = -\mathbf{I}, 
\end{align*}
then we just determine whether the Lyapunov equation
\begin{align*}
\mathbf{P} \mathbf{X} + \mathbf{X}^\mathrm{T} \mathbf{P} = \mathbf{Y} = -\mathbf{I}
\end{align*}
has a positive definite solution or not.

\subsubsection*{Application: rotating disk position P-control stability analysis}

Consider the rotating disk position control system illustrated in Figure \ref{fig:RDP_open_loop_initial_response}, state dynamics of which is modelled by the linear state differential equation (\ref{eq:RDP_state_DE_linear})
\begin{align*}
\frac{\mathrm{d}}{\mathrm{d} t} \mathbf{x} \equiv \frac{\mathrm{d}}{\mathrm{d} t} \begin{bmatrix} p \\ v \end{bmatrix} = \begin{bmatrix} 0 & 1 \\ 0 & -\frac{b}{J} \end{bmatrix} \begin{bmatrix} p \\ v \end{bmatrix} + \begin{bmatrix} 0 \\ \frac{1}{J} \end{bmatrix} T \equiv \mathbf{A} \mathbf{x} + \mathbf{B} T.
\end{align*}
Let the expected rotating disk position be zero for simplicity. 

Suppose the P-control method namely
\begin{align*}
T = -P p = -\begin{bmatrix} P & 0 \end{bmatrix} \mathbf{x}
\end{align*}
is adopted. Then we have the linear closed-loop feedback state differential equation
\begin{align*}
\frac{\mathrm{d}}{\mathrm{d} t} \mathbf{x} = (\begin{bmatrix} 0 & 1 \\ 0 & -\frac{b}{J} \end{bmatrix} - \begin{bmatrix} 0 \\ \frac{1}{J} \end{bmatrix} \begin{bmatrix} P & 0 \end{bmatrix}) \mathbf{x} = \begin{bmatrix} 0 & 1 \\ -\frac{P}{J} & -\frac{b}{J} \end{bmatrix} \mathbf{x} \equiv \mathbf{A}_c \mathbf{x},
\end{align*}
which describes dynamics of the closed-loop feedback control system namely dynamics of a linear self-evolutionary system.

Construct a candidate matrix $\mathbf{P}$ of the form
\begin{align*}
\mathbf{P} = \begin{bmatrix} r & s \\ s & 1 \end{bmatrix}
\end{align*}
and compute
\begin{align*}
\mathbf{P} \mathbf{A}_c + \mathbf{A}_c^\mathrm{T} \mathbf{P} = -\frac{1}{J} \begin{bmatrix} 2 P s & P + b s - J r \\ P + b s - J r & 2 (b - J s) \end{bmatrix}.
\end{align*}
Tentatively set
\begin{align*}
r = \frac{P}{J} + \frac{b}{J} s
\end{align*}
and reduce the matrix
\begin{align*}
\mathbf{P} \mathbf{A}_c + \mathbf{A}_c^\mathrm{T} \mathbf{P}
\end{align*}
to a diagonal matrix
\begin{align*}
\mathbf{P} \mathbf{A}_c + \mathbf{A}_c^\mathrm{T} \mathbf{P} = -\frac{1}{J} \begin{bmatrix} 2 P s & 0 \\ 0 & 2 (b - J s) \end{bmatrix}.
\end{align*}

For the Lyapunov inequality 
\begin{align*}
\mathbf{P} \mathbf{A}_c + \mathbf{A}_c^\mathrm{T} \mathbf{P} < 0
\end{align*}
to hold, the necessary and sufficient condition is
\begin{align*}
0 < s < \frac{b}{J}.
\end{align*}
Under above condition of $s$, we have
\begin{align*}
r = \frac{P}{J} + \frac{b}{J} s > \frac{P}{J} + s^2 > s^2,
\end{align*}
so the condition
\begin{align*}
s > 0, \quad r > s^2
\end{align*}
that guarantees the positive definiteness of $\mathbf{P}$ is naturally satisfied. 

In other words, any matrix $\mathbf{P}$ of the form
\begin{align*}
\mathbf{P} = \begin{bmatrix} \frac{P}{J} + \frac{b}{J} s & s \\ s & 1 \end{bmatrix}
\end{align*}
with
\begin{align*}
0 < s < \frac{b}{J}
\end{align*}
is a positive definite solution of the Lyapunov inequality characterized by the closed-loop state transition matrix $\mathbf{A}_c$. According to the \textit{Lyapunov stability criterion II}, $\mathbf{A}_c$ is stable, and so is the closed-loop feedback control system when the P-control method is adopted.

We can also use the second way to determine whether the Lyapunov inequality 
\begin{align*}
\mathbf{P} \mathbf{A}_c + \mathbf{A}_c^\mathrm{T} \mathbf{P} < 0
\end{align*}
has a positive definite solution $\mathbf{P}$ or not. Solve the Lyapunov equation
\begin{align*}
\mathbf{P} \mathbf{A}_c + \mathbf{A}_c^\mathrm{T} \mathbf{P} = -\mathbf{I}
\end{align*}
and obtain
\begin{align*}
\mathbf{P} = \frac{1}{2 P b} \begin{bmatrix} P^2 + J P + b^2 & J b \\ J b & J^2 + J P \end{bmatrix}
\end{align*}
which is indeed positive definite, because
\begin{align*}
P^2 + J P + b^2 &> 0,  \\
(P^2 + J P + b^2) (J^2 + J P) - (J b)^2 &> (b^2) (J^2) - (J b)^2 = 0.
\end{align*}
Then we can also conclude that $\mathbf{A}_c$ is stable. Matlab code for solving above Lyapunov equation via symbolic operation is given as follows.

\begin{framed} 
\noindent syms b J P \\
A = [0, 1; 0, -b/J]; B = [0; 1/J]; Ac = A-B*[P, 0]; \\
Pm = SolveLyapunovEquationSym(Ac, sym(-eye(2)));
\end{framed}

\subsection{First formalism of Riccati equation}  \label{sec:first_formalism_Riccati_equation}

The first generic formalism of \textbf{matrix Riccati equation} or simply \textbf{Riccati equation} is
\begin{equation}  \label{eq:Riccati_equation_form1} 
\mathbf{P} \mathbf{A} + \mathbf{A}^\mathrm{T} \mathbf{P} - \mathbf{P} \mathbf{B} \mathbf{R}^{-1} \mathbf{B}^\mathrm{T} \mathbf{P} + \mathbf{Q} = \mathbf{0},
\end{equation}
where $\mathbf{P}$ is the unknown square matrix to be solved with known matrices $\mathbf{A}$, $\mathbf{B}$, $\mathbf{R}$, and $\mathbf{Q}$. By default, $\mathbf{A}$ is a real-value square matrix, $\mathbf{R}$ and $\mathbf{Q}$ are real-value positive definite matrices, and $\mathbf{B}$ is a real-value matrix whose dimensions are consistent with both $\mathbf{A}$ and $\mathbf{R}$. 

The formalism of Riccati equation described in (\ref{eq:Riccati_equation_form1}) seems to have redundant notations $\mathbf{B}$ and $\mathbf{R}$. On one hand, the notations are indeed redundant purely from mathematics perspective. On the other hand, the formalism of (\ref{eq:Riccati_equation_form1}) is deeply rooted in the kingdom of control theory especially in the context of optimal control, where the notations $\mathbf{A}$, $\mathbf{B}$, $\mathbf{R}$, and $\mathbf{Q}$ all have concrete senses in practical applications. More specifically, $\mathbf{A}$ denotes the state transition matrix, $\mathbf{B}$ denotes the control input matrix, and $\mathbf{R}$ and $\mathbf{Q}$ denote cost matrices. Details will be presented in Section 6.1 in Chapter 6.
\footnote{Namely Chapter 6 of the author's works \cite{Li2026ACTPA_SJTU_2, Li2026ACTPA_SJTU_1}. Note that this article is Chapter 1 of the works.}

The Riccati equation described in (\ref{eq:Riccati_equation_form1}), especially in the context of optimal control, can be solved in iterative way as follows.

\begin{framed}
\noindent \textbf{Riccati equation iterative solving}\cite{Kleinman1968} \\
\noindent Initialization: \\
$~~~~$ Find $\mathbf{K}_0$ such that $\mathbf{A} - \mathbf{B} \mathbf{K}_0^\mathrm{T}$ is stable. \\
\noindent Iteration: \\
$~~~~$ Compute $\mathbf{A}_i = \mathbf{A} - \mathbf{B} \mathbf{K}_i^\mathrm{T}$ and $\mathbf{Q}_i =  -\mathbf{K}_i \mathbf{R} \mathbf{K}_i^\mathrm{T} - \mathbf{Q}$. \\
$~~~~$ Solve the \textit{Lyapunov equation} $\mathbf{P}_i \mathbf{A}_i + \mathbf{A}_i^\mathrm{T} \mathbf{P}_i = \mathbf{Q}_i$. \\
$~~~~$ Update $\mathbf{K}_{i+1}^\mathrm{T} = \mathbf{R}^{-1} \mathbf{B}^\mathrm{T} \mathbf{P}_i \iff \mathbf{K}_{i+1} = \mathbf{P}_i \mathbf{B} \mathbf{R}^{-1}$.
\end{framed}

By default, only consider the following matrix set
\begin{align}  \label{eq:FSFC_stable_gain_matrix_set}
\mathbf{K}_{\Omega} = \{\mathbf{K} \mbox{ } | \mbox{ } \mathbf{A} - \mathbf{B} \mathbf{K}^\mathrm{T} \mbox{ is stable}\},
\end{align}
namely the set of gain matrices $\mathbf{K}$ that can stabilize the control system. $\mathbf{K}_{\Omega}$ is called the \textbf{stabilizing gain matrix set} of the control system. Suppose the control system's target process is controllable and hence the stabilizing gain matrix set $\mathbf{K}_{\Omega}$ is non-empty. A systematic presentation of control system controllability, of the role that the gain matrix $\mathbf{K}$ can play, of why $\mathbf{K}_{\Omega}$ is non-empty given a controllable target process, and of how to find 
\begin{align*}
\mathbf{K}_0 \in \mathbf{K}_{\Omega}
\end{align*}
will be postponed to Section \ref{sec:controllability} and Chapter 2. 
\footnote{Namely Chapter 2 of the author's works \cite{Li2026ACTPA_SJTU_2, Li2026ACTPA_SJTU_1}. Note that this article is Chapter 1 of the works.}
For the moment, readers only need to be aware that $\mathbf{K}_{\Omega}$ does be non-empty in relevant discussions and we do have certain way to obtain such $\mathbf{K}_0$ that serves as a proper initial value for the Riccati equation iterative solving algorithm. 

Once initialization is done, then in each iteration of the Riccati equation iterative solving algorithm, apply the method presented in Section \ref{sec:Lyapunov_equation} to solve the following Lyapunov equation
\begin{equation}  \label{eq:Lyapunov_equation_iterative}
\mathbf{P}_i (\mathbf{A} - \mathbf{B} \mathbf{K}_i^\mathrm{T}) + (\mathbf{A} - \mathbf{B} \mathbf{K}_i^\mathrm{T})^\mathrm{T} \mathbf{P}_i = - \mathbf{K}_i \mathbf{R} \mathbf{K}_i^\mathrm{T} - \mathbf{Q}.
\end{equation}
Three points need to be clarified for the Riccati equation iterative solving algorithm. 

\subsubsection*{Point 1: The matrices $\mathbf{A}_i = \mathbf{A} - \mathbf{B} \mathbf{K}_i^\mathrm{T}$ are stable}

The matrices 
\begin{align*}
\mathbf{A}_i = \mathbf{A} - \mathbf{B} \mathbf{K}_i^\mathrm{T}
\end{align*}
encountered during iteration of the Riccati equation iterative solving algorithm are stable. This can be proved via \textit{mathematical induction}: Initially, the matrix 
\begin{align*}
\mathbf{A}_0 = \mathbf{A} - \mathbf{B} \mathbf{K}_0^\mathrm{T}
\end{align*}
is stable. Suppose the matrix $\mathbf{A}_i$ is stable and consider 
\begin{align*}
\mathbf{A}_{i+1} = \mathbf{A} - \mathbf{B} \mathbf{K}_{i+1}^\mathrm{T}.
\end{align*}

Since $\mathbf{A}_i$ is stable and 
\begin{align*}
\mathbf{Q}_i =  -\mathbf{K}_i \mathbf{R} \mathbf{K}_i^\mathrm{T} - \mathbf{Q} < 0, 
\end{align*}
the matrix $\mathbf{P}_i$, which is the solution of the Lyapunov equation 
\begin{align*}
\mathbf{P}_i \mathbf{A}_i + \mathbf{A}_i^\mathrm{T} \mathbf{P}_i = \mathbf{Q}_i, 
\end{align*}
is positive definite (according to the \textit{Lyapunov criterion II}). Further note that
\begin{align*}
&\mathbf{P}_i \mathbf{A}_{i+1} + \mathbf{A}_{i+1}^\mathrm{T} \mathbf{P}_i \\
=& \mathbf{P}_i [\mathbf{A}_i + \mathbf{B} (\mathbf{K}_i - \mathbf{K}_{i+1})^\mathrm{T}] + [\mathbf{A}_i + \mathbf{B} (\mathbf{K}_i - \mathbf{K}_{i+1})^\mathrm{T}]^\mathrm{T} \mathbf{P}_i \\
=& \mathbf{P}_i \mathbf{A}_i + \mathbf{A}_i^\mathrm{T} \mathbf{P}_i + \mathbf{P}_i \mathbf{B} \mathbf{R}^{-1} \mathbf{R} (\mathbf{K}_i - \mathbf{K}_{i+1})^\mathrm{T} + (\mathbf{K}_i - \mathbf{K}_{i+1}) \mathbf{R} \mathbf{R}^{-1} \mathbf{B}^\mathrm{T} \mathbf{P}_i \\
=& - \mathbf{Q} - \mathbf{K}_i \mathbf{R} \mathbf{K}_i^\mathrm{T} + \mathbf{K}_{i+1} \mathbf{R} (\mathbf{K}_i - \mathbf{K}_{i+1})^\mathrm{T} + (\mathbf{K}_i - \mathbf{K}_{i+1}) \mathbf{R} \mathbf{K}_{i+1}^\mathrm{T} \\
=& - \mathbf{Q} - (\mathbf{K}_i - \mathbf{K}_{i+1}) \mathbf{R} (\mathbf{K}_i - \mathbf{K}_{i+1})^\mathrm{T} - \mathbf{K}_{i+1} \mathbf{R} \mathbf{K}_{i+1}^\mathrm{T} < 0,
\end{align*}
which implies that $\mathbf{A}_{i+1}$ is stable (according to the \textit{Lyapunov criterion III}). Therefore, all matrices $\mathbf{A}_i$ ($i \in \{0, 1, 2, \cdots\}$) are stable.

\subsubsection*{Point 2: The matrices $\mathbf{P}_i$ are positive definite and converge to the solution}

The matrices $\mathbf{P}_i$ encountered during iteration of the Riccati equation iterative solving algorithm are positive definite and converge to the positive definite solution of the Riccati equation described in (\ref{eq:Riccati_equation_form1}). Their positive definiteness is already clarified above during clarification of the first point, so we focus on their convergence. Denote 
\begin{align*}
\Delta \mathbf{K}_i = \mathbf{K}_i - \mathbf{K}_{i+1}.
\end{align*}
During clarification of the first point, we have already computed
\begin{align*}
\mathbf{P}_i \mathbf{A}_{i+1} + \mathbf{A}_{i+1}^\mathrm{T} \mathbf{P}_i = - \mathbf{Q} - \Delta \mathbf{K}_i \mathbf{R} \Delta \mathbf{K}_i^\mathrm{T} - \mathbf{K}_{i+1} \mathbf{R} \mathbf{K}_{i+1}^\mathrm{T}.
\end{align*}
Then we have
\begin{align*}
&(\mathbf{P}_i - \mathbf{P}_{i+1}) \mathbf{A}_{i+1} + \mathbf{A}_{i+1}^\mathrm{T} (\mathbf{P}_i - \mathbf{P}_{i+1}) \\
=& (\mathbf{P}_i \mathbf{A}_{i+1} + \mathbf{A}_{i+1}^\mathrm{T} \mathbf{P}_i) - (\mathbf{P}_{i+1} \mathbf{A}_{i+1} + \mathbf{A}_{i+1}^\mathrm{T} \mathbf{P}_{i+1}) \\
=& - \mathbf{Q} - \Delta \mathbf{K}_i \mathbf{R} \Delta \mathbf{K}_i^\mathrm{T} - \mathbf{K}_{i+1} \mathbf{R} \mathbf{K}_{i+1}^\mathrm{T} - (- \mathbf{Q} - \mathbf{K}_{i+1} \mathbf{R} \mathbf{K}_{i+1}^\mathrm{T}) = - \Delta \mathbf{K}_i \mathbf{R} \Delta \mathbf{K}_i^\mathrm{T} \leq 0
\end{align*}
and
\begin{align*}
\mathbf{P}_i - \mathbf{P}_{i+1} &= \mathrm{e}^{\mathbf{A}_{i+1}^\mathrm{T} t} (\mathbf{P}_i - \mathbf{P}_{i+1}) \mathrm{e}^{\mathbf{A}_{i+1} t} |_{t=0} - \mathrm{e}^{\mathbf{A}_{i+1}^\mathrm{T} t} (\mathbf{P}_i - \mathbf{P}_{i+1}) \mathrm{e}^{\mathbf{A}_{i+1} t} |_{t \to \infty} \\
  &= -\int_0^{\infty} \frac{\mathrm{d}}{\mathrm{d} t} [\mathrm{e}^{\mathbf{A}_{i+1}^\mathrm{T} t} (\mathbf{P}_i - \mathbf{P}_{i+1}) \mathrm{e}^{\mathbf{A}_{i+1} t}] \mathrm{d} t \\
  &= -\int_0^{\infty} \mathrm{e}^{\mathbf{A}_{i+1}^\mathrm{T} t} [(\mathbf{P}_i - \mathbf{P}_{i+1}) \mathbf{A}_{i+1} + \mathbf{A}_{i+1}^\mathrm{T} (\mathbf{P}_i - \mathbf{P}_{i+1})] \mathrm{e}^{\mathbf{A}_{i+1} t} \mathrm{d} t \\
  &= \int_0^{\infty} \mathrm{e}^{\mathbf{A}_{i+1}^\mathrm{T} t} \Delta \mathbf{K}_i \mathbf{R} \Delta \mathbf{K}_i^\mathrm{T} \mathrm{e}^{\mathbf{A}_{i+1} t} \mathrm{d} t \geq 0,
\end{align*}
which implies that the positive definite matrices $\mathbf{P}_i$ ($i \in \{0, 1, 2, \cdots\}$) decrease monotonically in terms of positive definiteness. Therefore, they definitely converge to a limit $\mathbf{P}$ with a convergent 
\begin{align*}
\mathbf{K} = \mathbf{P} \mathbf{B} \mathbf{R}^{-1}
\end{align*}
as well. Consider limits on both sides of (\ref{eq:Lyapunov_equation_iterative}) and obtain
\begin{align*}
&\mathbf{P} (\mathbf{A} - \mathbf{B} \mathbf{K}^\mathrm{T}) + (\mathbf{A} - \mathbf{B} \mathbf{K}^\mathrm{T})^\mathrm{T} \mathbf{P} = - \mathbf{K} \mathbf{R} \mathbf{K}^\mathrm{T} - \mathbf{Q} \\
\iff & \mathbf{P} \mathbf{A} + \mathbf{A}^\mathrm{T} \mathbf{P} - \mathbf{P} \mathbf{B} \mathbf{R}^{-1} \mathbf{B}^\mathrm{T} \mathbf{P} + \mathbf{Q} = \mathbf{0},
\end{align*}
which is of the Riccati equation form described in (\ref{eq:Riccati_equation_form1}). In other words, the convergent $\mathbf{P}$ is right the positive definite solution of the Riccati equation described in (\ref{eq:Riccati_equation_form1}). 

\subsubsection*{Point 3: The convergent $\mathbf{P}$ is the unique solution}

The convergent $\mathbf{P}$ is the unique positive definite solution of the Riccati equation described in (\ref{eq:Riccati_equation_form1}). Suppose there is another positive definite solution $\bar{\mathbf{P}}$. Denote 
\begin{align*}
\bar{\mathbf{K}} &= \bar{\mathbf{P}} \mathbf{B} \mathbf{R}^{-1}, \quad \Delta \mathbf{K} = \mathbf{K} - \bar{\mathbf{K}},  \\ 
\mathbf{A}_{\mathbf{P}} &= \mathbf{A} - \mathbf{B} \mathbf{K}^\mathrm{T}, \quad \mathbf{A}_{\bar{\mathbf{P}}} = \mathbf{A} - \mathbf{B} \bar{\mathbf{K}}^\mathrm{T}. 
\end{align*}
We have
\begin{align*}
\bar{\mathbf{P}} \mathbf{A}_{\bar{\mathbf{P}}} + \mathbf{A}_{\bar{\mathbf{P}}}^\mathrm{T} \bar{\mathbf{P}} = \bar{\mathbf{P}} (\mathbf{A} - \mathbf{B} \bar{\mathbf{K}}^\mathrm{T}) + (\mathbf{A} - \mathbf{B} \bar{\mathbf{K}}^\mathrm{T})^\mathrm{T} \bar{\mathbf{P}} = - \mathbf{Q} - \bar{\mathbf{K}} \mathbf{R} \bar{\mathbf{K}}^\mathrm{T},
\end{align*}
which implies that $\mathbf{A}_{\bar{\mathbf{P}}}$ is stable (according to the \textit{Lyapunov criterion III}). We further have
\begin{align*}
&(\mathbf{P} - \bar{\mathbf{P}}) \mathbf{A}_{\bar{\mathbf{P}}} + \mathbf{A}_{\bar{\mathbf{P}}}^\mathrm{T} (\mathbf{P} - \bar{\mathbf{P}}) \\
=& (\mathbf{P} \mathbf{A}_{\mathbf{P}} + \mathbf{A}_{\mathbf{P}}^\mathrm{T} \mathbf{P}) + \mathbf{P} \mathbf{B} \Delta \mathbf{K}^\mathrm{T} + \Delta \mathbf{K} \mathbf{B}^\mathrm{T} \mathbf{P} - (\bar{\mathbf{P}} \mathbf{A}_{\bar{\mathbf{P}}} + \mathbf{A}_{\bar{\mathbf{P}}}^\mathrm{T} \bar{\mathbf{P}}) \\
=& (- \mathbf{Q} - \mathbf{K} \mathbf{R} \mathbf{K}^\mathrm{T}) + \mathbf{K} \mathbf{R} \Delta \mathbf{K}^\mathrm{T} + \Delta \mathbf{K} \mathbf{R} \mathbf{K}^\mathrm{T} -(- \mathbf{Q} - \bar{\mathbf{K}} \mathbf{R} \bar{\mathbf{K}}^\mathrm{T}) \\
=& \Delta \mathbf{K} \mathbf{R} \Delta \mathbf{K}^\mathrm{T} \geq 0
\end{align*}
and similarly by symmetry of above derivation we have
\begin{align*}
&(\mathbf{P} - \bar{\mathbf{P}}) \mathbf{A}_{\mathbf{P}} + \mathbf{A}_{\mathbf{P}}^\mathrm{T} (\mathbf{P} - \bar{\mathbf{P}}) = -[(\bar{\mathbf{P}} - \mathbf{P}) \mathbf{A}_{\mathbf{P}} + \mathbf{A}_{\mathbf{P}}^\mathrm{T} (\bar{\mathbf{P}} - \mathbf{P})] \\
=& - (-\Delta \mathbf{K}) \mathbf{R} (-\Delta \mathbf{K})^\mathrm{T} = - \Delta \mathbf{K} \mathbf{R} \Delta \mathbf{K}^\mathrm{T} \leq 0.
\end{align*}
On one hand
\begin{align*}
\mathbf{P} - \bar{\mathbf{P}} &= -\int_0^{\infty} \frac{\mathrm{d}}{\mathrm{d} t} [\mathrm{e}^{\mathbf{A}_{\bar{\mathbf{P}}}^\mathrm{T} t} (\mathbf{P} - \bar{\mathbf{P}}) \mathrm{e}^{\mathbf{A}_{\bar{\mathbf{P}}} t}] \mathrm{d} t \\
  &= -\int_0^{\infty} \mathrm{e}^{\mathbf{A}_{\bar{\mathbf{P}}}^\mathrm{T} t} [(\mathbf{P} - \bar{\mathbf{P}}) \mathbf{A}_{\bar{\mathbf{P}}} + \mathbf{A}_{\bar{\mathbf{P}}}^\mathrm{T} (\mathbf{P} - \bar{\mathbf{P}})] \mathrm{e}^{\mathbf{A}_{\bar{\mathbf{P}}} t} \mathrm{d} t \\
  &= -\int_0^{\infty} \mathrm{e}^{\mathbf{A}_{\bar{\mathbf{P}}}^\mathrm{T} t} \Delta \mathbf{K} \mathbf{R} \Delta \mathbf{K}^\mathrm{T} \mathrm{e}^{\mathbf{A}_{\bar{\mathbf{P}}} t} \mathrm{d} t \leq 0,
\end{align*}
and on the other hand
\begin{align*}
\mathbf{P} - \bar{\mathbf{P}} &= -\int_0^{\infty} \frac{\mathrm{d}}{\mathrm{d} t} [\mathrm{e}^{\mathbf{A}_{\mathbf{P}}^\mathrm{T} t} (\mathbf{P} - \bar{\mathbf{P}}) \mathrm{e}^{\mathbf{A}_{\mathbf{P}} t}] \mathrm{d} t \\
  &= -\int_0^{\infty} \mathrm{e}^{\mathbf{A}_{\mathbf{P}}^\mathrm{T} t} [(\mathbf{P} - \bar{\mathbf{P}}) \mathbf{A}_{\mathbf{P}} + \mathbf{A}_{\mathbf{P}}^\mathrm{T} (\mathbf{P} - \bar{\mathbf{P}})] \mathrm{e}^{\mathbf{A}_{\mathbf{P}} t} \mathrm{d} t \\
  &= \int_0^{\infty} \mathrm{e}^{\mathbf{A}_{\mathbf{P}}^\mathrm{T} t} \Delta \mathbf{K} \mathbf{R} \Delta \mathbf{K}^\mathrm{T} \mathrm{e}^{\mathbf{A}_{\mathbf{P}} t} \mathrm{d} t \geq 0.
\end{align*}
Therefore, we have
\begin{align*}
\mathbf{P} - \bar{\mathbf{P}} = \mathbf{0} \iff \mathbf{P} = \bar{\mathbf{P}}
\end{align*}
and the uniqueness of the positive definite solution of the Riccati equation described in (\ref{eq:Riccati_equation_form1}) is proved. 

Matlab code for demonstration of Riccati equation iterative solving is given as follows.

\begin{framed} 
\noindent \textbf{SolveRiccatiEquationForm1.m} \\
\noindent \%\% Riccati equation Form 1: P A + A' P - P B inv(R) B' P + Q = 0  \\
\%\% Solve the symmetric matrix P and obtain Kc = (inv(R)*B'*P)' \\
function [P, Kc] = SolveRiccatiEquationForm1(A, B, Q, R, Kinit) \\
$~~~~$ Kc = Kinit; errK = 10000; \\
$~~~~$ while (errK$>$0.00001) \\
$~~~~$ $~~~~$ P = SolveLyapunovEquation(A-B*Kc', -Kc*R*Kc'-Q); \\
$~~~~$ $~~~~$ Kold = Kc; Kc = (inv(R)*B'*P)'; \% Transpose \\
$~~~~$ $~~~~$ errK = trace((Kc-Kold)'*(Kc-Kold)); \\
$~~~~$ end \\
end
\end{framed}

\subsection{Second formalism of Riccati equation}

The second generic formalism of \textbf{Riccati equation} is
\begin{equation}  \label{eq:Riccati_equation_form2} 
\mathbf{P} \mathbf{A} + \mathbf{A}^\mathrm{T} \mathbf{P} - \mathbf{P} \mathbf{B} \mathbf{B}^\mathrm{T} \mathbf{P} + \mathbf{Q} = \mathbf{0},
\end{equation}
where $\mathbf{P}$ is the unknown square matrix to be solved with known matrices $\mathbf{A}$, $\mathbf{B}$, and $\mathbf{Q}$. By default, $\mathbf{A}$ is a real-value square matrix, $\mathbf{Q}$ is a real-value positive definite matrix, and $\mathbf{B}$ is a real-value matrix whose row dimension is consistent with $\mathbf{A}$. 

In fact, the Riccati equation described in (\ref{eq:Riccati_equation_form2}) and the Riccati equation described in (\ref{eq:Riccati_equation_form1}) can be mutually transformed into each other and hence are equivalent. On one hand, the Riccati equation described in (\ref{eq:Riccati_equation_form2}) can be regarded as
\begin{align*}
\mathbf{P} \mathbf{A} + \mathbf{A}^\mathrm{T} \mathbf{P} - \mathbf{P} \mathbf{B} \mathbf{I}^{-1} \mathbf{B}^\mathrm{T} \mathbf{P} + \mathbf{Q} = \mathbf{0}
\end{align*}
which is consistent with the formalism of the Riccati equation described in (\ref{eq:Riccati_equation_form1}) --- The matrix $\mathbf{I}$ is the identity matrix --- In other words, the Riccati equation described in (\ref{eq:Riccati_equation_form2}) can be transformed into the Riccati equation described in (\ref{eq:Riccati_equation_form1}).

On the other hand, for the Riccati equation described in (\ref{eq:Riccati_equation_form1}), the positive definite matrix $\mathbf{R}^{-1}$ can always be decomposed into the product of a matrix and its transpose as
\begin{align*}
\mathbf{R}^{-1} = \bar{\mathbf{R}} \bar{\mathbf{R}}^{\mathrm{T}}.
\end{align*}
Matlab code for such kind of matrix decomposition is given as follows.

\begin{framed} 
\noindent \textbf{PDMtoBBT.m} \\
\noindent \%\% Decompose positive definite matrix (PDM) as A = B * B' (BBT) \\
function B = PDMtoBBT(A, mt) \\
$~~~~$ if (nargin$<$2) mt = 'eig'; end \\
$~~~~$ if (strcmp(mt,'eig')) \\
$~~~~$ $~~~~$ r = rank(A); [U, E] = eig(A); E = diag(E); \\
$~~~~$ $~~~~$ B = U(:,end-r+1:end)*diag(sqrt(E(end-r+1:end))); \\
$~~~~$ elseif (strcmp(mt,'svd')) \\
$~~~~$ $~~~~$ r = rank(A); [U,S,V] = svd(A); S = diag(S); \\
$~~~~$ $~~~~$ B = U(:,1:r)*diag(sqrt(S(1:r))); \\
$~~~~$ else \\
$~~~~$ $~~~~$ B = cholcov(A)'; \% built-in Cholesky-like covariance decomposition \\
$~~~~$ end \\
end
\end{framed}

With $\mathbf{R}^{-1}$ decomposed in above way, denote
\begin{align*}
\bar{\mathbf{B}} \equiv \mathbf{B} \bar{\mathbf{R}}.
\end{align*}
Then the Riccati equation described in (\ref{eq:Riccati_equation_form1}) becomes
\begin{align*}
&\mathbf{P} \mathbf{A} + \mathbf{A}^\mathrm{T} \mathbf{P} - \mathbf{P} \mathbf{B} \bar{\mathbf{R}} \bar{\mathbf{R}}^{\mathrm{T}} \mathbf{B}^\mathrm{T} \mathbf{P} + \mathbf{Q} = \mathbf{0}  \\
\iff &\mathbf{P} \mathbf{A} + \mathbf{A}^\mathrm{T} \mathbf{P} - \mathbf{P} \bar{\mathbf{B}} \bar{\mathbf{B}}^\mathrm{T} \mathbf{P} + \mathbf{Q} = \mathbf{0}
\end{align*}
which is consistent with the formalism of the Riccati equation described in (\ref{eq:Riccati_equation_form2}). In other words, the Riccati equation described in (\ref{eq:Riccati_equation_form1}) can be transformed into the Riccati equation described in (\ref{eq:Riccati_equation_form2}).

For the Riccati equation described in (\ref{eq:Riccati_equation_form2}), if $\mathbf{B} \mathbf{B}^\mathrm{T}$ is treated holistically as a positive definite matrix, we can first decompose the holistic $\mathbf{B} \mathbf{B}^\mathrm{T}$ into the product of $\mathbf{B}$ and its transpose $\mathbf{B}^\mathrm{T}$ --- such decomposition is not unique but this does not influence the solution of the Riccati equation --- Then simply set 
\begin{align*}
\mathbf{R} = \mathbf{I}
\end{align*}
and take advantage of the Riccati equation iterative solving algorithm to compute the solution $\mathbf{P}$. Matlab code for implementing such idea of solving the Riccati equation described in (\ref{eq:Riccati_equation_form2}) is given as follows.

\begin{framed} 
\noindent \textbf{SolveRiccatiEquationForm2.m} \\
\noindent \%\% Riccati equation Form 2: P A + A' P - P B B' P + Q = 0  \\
\%\% Solve the symmetric matrix P \\
function P = SolveRiccatiEquationForm2(A, BBT, Q) \\
$~~~~$ B = PDMtoBBT(BBT); \\
$~~~~$ Kinit = DesignGainMatrix(A, B, repmat(-1,size(A,1),1)); \\
$~~~~$ P = SolveRiccatiEquationForm1(A, B, Q, eye(size(B,2)), Kinit); \\
end
\end{framed}

Here, the code \textbf{DesignGainMatrix.m} is to provide a valid 
\begin{align*}
\mathbf{K}_0 \in \mathbf{K}_{\Omega}
\end{align*}
and will be presented in Chapter 2.
\footnote{Namely Chapter 2 of the author's works \cite{Li2026ACTPA_SJTU_2, Li2026ACTPA_SJTU_1}. Note that this article is Chapter 1 of the works.}

\subsection{Third formalism of Riccati equation}  \label{sec:third_formalism_Riccati_equation}

The third generic formalism of \textbf{Riccati equation} is
\begin{equation}  \label{eq:Riccati_equation_form3} 
\mathbf{P} \mathbf{A} + \mathbf{A}^\mathrm{T} \mathbf{P} - \mathbf{P} \mathbf{M} \mathbf{P} + \mathbf{Q} = \mathbf{0},
\end{equation}
where $\mathbf{P}$ is the unknown square matrix to be solved with known matrices $\mathbf{A}$, $\mathbf{M}$, and $\mathbf{Q}$. By default, $\mathbf{A}$ is a real-value square matrix, $\mathbf{M}$ is a real-value symmetric matrix, and $\mathbf{Q}$ is a real-value positive definite matrix. Both $\mathbf{M}$ and $\mathbf{Q}$ are of the same dimension as $\mathbf{A}$.

The Riccati equation described in (\ref{eq:Riccati_equation_form2}) is apparently a special case of the Riccati equation described in (\ref{eq:Riccati_equation_form3}), if $\mathbf{B} \mathbf{B}^\mathrm{T}$ is treated holistically as a positive definite matrix $\mathbf{M}$. On the other hand, given any positive definite matrix $\mathbf{M}$, it can always be decomposed into the product of a matrix and its transpose as
\begin{align*}
\mathbf{M} = \mathbf{B} \mathbf{B}^\mathrm{T}.
\end{align*}
So the Riccati equation described in (\ref{eq:Riccati_equation_form2}) is the special case of the Riccati equation described in (\ref{eq:Riccati_equation_form3}) where $\mathbf{M}$ is positive definite. Besides, we have explained that the Riccati equation described in (\ref{eq:Riccati_equation_form2}) and the Riccati equation described in (\ref{eq:Riccati_equation_form1}) are equivalent. So the Riccati equation described in (\ref{eq:Riccati_equation_form1}) is also the same special case of the Riccati equation described in (\ref{eq:Riccati_equation_form3}).

If the matrix $\mathbf{M}$ in (\ref{eq:Riccati_equation_form3}) is positive definite, the method of Riccati equation solving presented previously can be applied. Yet in practical applications especially in the context of robust control, the matrix $\mathbf{M}$ in (\ref{eq:Riccati_equation_form3}) is not necessarily positive definite. To solve the Riccati equation, we may resort to another method: Suppose the \textbf{Hamiltonian matrix} $\mathbf{H}$ is diagonalizable with half of its eigenvalues having negative real part (corresponding to the diagonal block $\mathbf{\Sigma}_{-}$) as
\begin{equation}  \label{eq:Hamiltonian_matrix}
\mathbf{H} \equiv \begin{bmatrix} \mathbf{A} & -\mathbf{M} \\ -\mathbf{Q} & -\mathbf{A}^\mathrm{T} \end{bmatrix} = \begin{bmatrix} \mathbf{U}_{11} & \mathbf{U}_{12} \\ \mathbf{U}_{21} & \mathbf{U}_{22} \end{bmatrix} \begin{bmatrix} \mathbf{\Sigma}_{-} & \\ & \mathbf{\Sigma}_{+} \end{bmatrix} \begin{bmatrix} \mathbf{U}_{11} & \mathbf{U}_{12} \\ \mathbf{U}_{21} & \mathbf{U}_{22} \end{bmatrix}^{-1}
\end{equation}
and suppose the block $\mathbf{U}_{11}$ is invertible, then the Riccati equation described in (\ref{eq:Riccati_equation_form3}) is solved as 
\begin{equation}  \label{eq:Riccati_equation_form3_Hamiltonian_solution} 
\mathbf{P} = \mathbf{U}_{21} \mathbf{U}_{11}^{-1}. 
\end{equation}
Three points need to be clarified for the solution.

\subsubsection*{Point 1: $\mathbf{P} = \mathbf{U}_{21} \mathbf{U}_{11}^{-1}$ is indeed a solution}

The matrix $\mathbf{P}$ specified in (\ref{eq:Riccati_equation_form3_Hamiltonian_solution}) is indeed a solution of the Riccati equation described in (\ref{eq:Riccati_equation_form3}). Transform (\ref{eq:Hamiltonian_matrix}) into
\begin{align*}
&\begin{bmatrix} \mathbf{A} & -\mathbf{M} \\ -\mathbf{Q} & -\mathbf{A}^\mathrm{T} \end{bmatrix} \begin{bmatrix} \mathbf{U}_{11} & \mathbf{U}_{12} \\ \mathbf{U}_{21} & \mathbf{U}_{22} \end{bmatrix} = \begin{bmatrix} \mathbf{U}_{11} & \mathbf{U}_{12} \\ \mathbf{U}_{21} & \mathbf{U}_{22} \end{bmatrix} \begin{bmatrix} \mathbf{\Sigma}_{-} & \\ & \mathbf{\Sigma}_{+} \end{bmatrix}  \\
\implies &\begin{bmatrix} \mathbf{A} & -\mathbf{M} \\ -\mathbf{Q} & -\mathbf{A}^\mathrm{T} \end{bmatrix} \begin{bmatrix} \mathbf{U}_{11} \\ \mathbf{U}_{21} \end{bmatrix} = \begin{bmatrix} \mathbf{U}_{11} \\ \mathbf{U}_{21} \end{bmatrix} \mathbf{\Sigma}_{-}  
  \iff  \left\{
\begin{array}{l l}
\mathbf{A} \mathbf{U}_{11} -\mathbf{M} \mathbf{U}_{21} &= \mathbf{U}_{11} \mathbf{\Sigma}_{-}  \\
-\mathbf{Q} \mathbf{U}_{11} -\mathbf{A}^\mathrm{T} \mathbf{U}_{21} &= \mathbf{U}_{21} \mathbf{\Sigma}_{-}
\end{array}
\right.
\end{align*}
The first equation gives
\begin{align*}
\mathbf{\Sigma}_{-} = \mathbf{U}_{11}^{-1} (\mathbf{A} \mathbf{U}_{11} -\mathbf{M} \mathbf{U}_{21}).
\end{align*}
Substitute it into the second equation and obtain
\begin{align*}
&-\mathbf{Q} \mathbf{U}_{11} -\mathbf{A}^\mathrm{T} \mathbf{U}_{21} = \mathbf{U}_{21} \mathbf{U}_{11}^{-1} (\mathbf{A} \mathbf{U}_{11} -\mathbf{M} \mathbf{U}_{21})  \\
\iff &\mathbf{U}_{21} \mathbf{U}_{11}^{-1} \mathbf{A} \mathbf{U}_{11} + \mathbf{A}^\mathrm{T} \mathbf{U}_{21} - \mathbf{U}_{21} \mathbf{U}_{11}^{-1} \mathbf{M} \mathbf{U}_{21} + \mathbf{Q} \mathbf{U}_{11} = \mathbf{0}  \\
\iff &\mathbf{U}_{21} \mathbf{U}_{11}^{-1} \mathbf{A} + \mathbf{A}^\mathrm{T} \mathbf{U}_{21} \mathbf{U}_{11}^{-1} - \mathbf{U}_{21} \mathbf{U}_{11}^{-1} \mathbf{M} \mathbf{U}_{21} \mathbf{U}_{11}^{-1} + \mathbf{Q} = \mathbf{0},
\end{align*}
which is right in the Riccati equation form described in (\ref{eq:Riccati_equation_form3}) and implies that the matrix $\mathbf{P}$ specified in (\ref{eq:Riccati_equation_form3_Hamiltonian_solution}) is a solution.

It is worth noting that the supposed preliminary condition that the Hamiltonian matrix $\mathbf{H}$ is diagonalizable with half of its eigenvalues having negative real part is important. Readers had better not take it for granted that the preliminary condition tends to hold. For example, given
\begin{align*}
\mathbf{A} = \begin{bmatrix} 0 & 1 & 0 \\ 9 & 0 & 0 \\ 0 & 0 & 0 \end{bmatrix}, \quad \mathbf{M} = \begin{bmatrix} -15 & 0 & 0 \\ 0 & -11 & -6 \\ 0 & -6 & -5 \end{bmatrix}, \quad \mathbf{Q} = \begin{bmatrix} 101 & 0 & 0 \\ 0 & 1 & 0 \\ 0 & 0 & 1 \end{bmatrix},
\end{align*}
the Hamiltonian matrix
\begin{align*}
\mathbf{H} \equiv \begin{bmatrix} \mathbf{A} & -\mathbf{M} \\ -\mathbf{Q} & -\mathbf{A}^\mathrm{T} \end{bmatrix} = \begin{bmatrix} 0 & 1 & 0 & 15 & 0 & 0 \\ 9 & 0 & 0 & 0 & 11 & 6 \\ 0 & 0 & 0 & 0 & 6 & 5 \\ -101 & 0 & 0 & 0 & -9 & 0 \\ 0 & -1 & 0 & -1 & 0 & 0 \\ 0 & 0 & -1 & 0 & 0 & 0 \end{bmatrix}
\end{align*}
does not satisfy the preliminary condition. When the supposed preliminary condition does not hold, the Riccati equation described in (\ref{eq:Riccati_equation_form3}) cannot be solved via (\ref{eq:Riccati_equation_form3_Hamiltonian_solution}).

\subsubsection*{Point 2: $\mathbf{P} = \mathbf{U}_{21} \mathbf{U}_{11}^{-1}$ is a real-value symmetric matrix}

During clarification of the first point, we know that
\begin{align*}
\mathbf{A} \mathbf{U}_{11} -\mathbf{M} \mathbf{U}_{21} = \mathbf{U}_{11} \mathbf{\Sigma}_{-}. 
\end{align*}
The equality leads to
\begin{align*}
\mathbf{A} -\mathbf{M} \mathbf{U}_{21} \mathbf{U}_{11}^{-1} = \mathbf{U}_{11} \mathbf{\Sigma}_{-} \mathbf{U}_{11}^{-1} \iff \mathbf{A} -\mathbf{M} \mathbf{P} = \mathbf{U}_{11} \mathbf{\Sigma}_{-} \mathbf{U}_{11}^{-1},
\end{align*}
which implies that the matrix
\begin{align*}
\mathbf{A}_c \equiv \mathbf{A} -\mathbf{M} \mathbf{P}
\end{align*}
is stable.

The solution $\mathbf{P}$ specified in (\ref{eq:Riccati_equation_form3_Hamiltonian_solution}) satisfies the Riccati equation described in (\ref{eq:Riccati_equation_form3})
\begin{align*}
\mathbf{P} \mathbf{A} + \mathbf{A}^\mathrm{T} \mathbf{P} - \mathbf{P} \mathbf{M} \mathbf{P} + \mathbf{Q} = \mathbf{0}.
\end{align*}
Transpose the Riccati equation and obtain
\begin{align*}
(\mathbf{P} \mathbf{A} + \mathbf{A}^\mathrm{T} \mathbf{P} - \mathbf{P} \mathbf{M} \mathbf{P} + \mathbf{Q})^\mathrm{T} &= \mathbf{0}  \\
\iff \mathbf{P}^\mathrm{T} \mathbf{A} + \mathbf{A}^\mathrm{T} \mathbf{P}^\mathrm{T} - \mathbf{P}^\mathrm{T} \mathbf{M} \mathbf{P}^\mathrm{T} + \mathbf{Q} &= \mathbf{0}.
\end{align*}
Compare the original Riccati equation and the transposed Riccati equation as follows
\begin{align*}
&\mathbf{P} \mathbf{A} + \mathbf{A}^\mathrm{T} \mathbf{P} - \mathbf{P} \mathbf{M} \mathbf{P} + \mathbf{Q} = \mathbf{P}^\mathrm{T} \mathbf{A} + \mathbf{A}^\mathrm{T} \mathbf{P}^\mathrm{T} - \mathbf{P}^\mathrm{T} \mathbf{M} \mathbf{P}^\mathrm{T} + \mathbf{Q}  \\
\iff &\mathbf{P} \mathbf{A} + \mathbf{A}^\mathrm{T} \mathbf{P} - \mathbf{P} \mathbf{M} \mathbf{P} - \mathbf{P}^\mathrm{T} \mathbf{A} - \mathbf{A}^\mathrm{T} \mathbf{P}^\mathrm{T} + \mathbf{P}^\mathrm{T} \mathbf{M} \mathbf{P}^\mathrm{T} = \mathbf{0}  \\
\iff &(\mathbf{P} - \mathbf{P}^\mathrm{T}) \mathbf{A} + \mathbf{A}^\mathrm{T} (\mathbf{P} - \mathbf{P}^\mathrm{T}) - (\mathbf{P} - \mathbf{P}^\mathrm{T}) \mathbf{M} \mathbf{P} - \mathbf{P}^\mathrm{T} \mathbf{M} (\mathbf{P} - \mathbf{P}^\mathrm{T}) = \mathbf{0}  \\
\iff &(\mathbf{P} - \mathbf{P}^\mathrm{T}) (\mathbf{A} -\mathbf{M} \mathbf{P}) + (\mathbf{A} -\mathbf{M} \mathbf{P})^\mathrm{T} (\mathbf{P} - \mathbf{P}^\mathrm{T}) = \mathbf{0}
\end{align*}
or
\begin{align*}
(\mathbf{P} - \mathbf{P}^\mathrm{T}) \mathbf{A}_c + \mathbf{A}_c^\mathrm{T} (\mathbf{P} - \mathbf{P}^\mathrm{T}) = \mathbf{0}.
\end{align*}
Since the matrix $\mathbf{A}_c$ is stable, according to the \textit{Lyapunov criterion I} presented in Section \ref{sec:Lyapunov_equation}, above Lyapunov equation in terms of 
\begin{align*}
\mathbf{P} - \mathbf{P}^\mathrm{T}
\end{align*}
has a unique solution
\begin{align*}
\mathbf{P} - \mathbf{P}^\mathrm{T} = \mathbf{0} \iff \mathbf{P} = \mathbf{P}^\mathrm{T},
\end{align*}
namely that $\mathbf{P}$ is a real-value symmetric matrix.

\subsubsection*{Point 3: $\mathbf{M} > 0$ implies that $\mathbf{P} = \mathbf{U}_{21} \mathbf{U}_{11}^{-1}$ is the positive definite solution}

We have known that the matrix $\mathbf{P}$ specified in (\ref{eq:Riccati_equation_form3_Hamiltonian_solution}) is a real-value symmetric solution of the Riccati equation described in (\ref{eq:Riccati_equation_form3}) and that the matrix $\mathbf{A}_c$ is stable. If the matrix $\mathbf{M}$ is positive definite, then we have
\begin{align*}
&\mathbf{P} \mathbf{A} + \mathbf{A}^\mathrm{T} \mathbf{P} - \mathbf{P} \mathbf{M} \mathbf{P} + \mathbf{Q} = \mathbf{0}  \\
\iff &\mathbf{P} (\mathbf{A} -\mathbf{M} \mathbf{P}) + (\mathbf{A} -\mathbf{M} \mathbf{P})^\mathrm{T} \mathbf{P} + \mathbf{P} \mathbf{M} \mathbf{P} + \mathbf{Q} = \mathbf{0}  \\
\iff &\mathbf{P} (\mathbf{A} -\mathbf{M} \mathbf{P}) + (\mathbf{A} -\mathbf{M} \mathbf{P})^\mathrm{T} \mathbf{P} = - \mathbf{P} \mathbf{M} \mathbf{P} - \mathbf{Q} < 0
\end{align*}
or
\begin{align*}
\mathbf{P} \mathbf{A}_c + \mathbf{A}_c^\mathrm{T} \mathbf{P} = - \mathbf{P} \mathbf{M} \mathbf{P} - \mathbf{Q} < 0,
\end{align*}
which implies that the matrix $\mathbf{P}$ specified in (\ref{eq:Riccati_equation_form3_Hamiltonian_solution}) is positive definite, according to the \textit{Lyapunov criterion II} or the \textit{Lyapunov criterion II-B} presented in Section \ref{sec:Lyapunov_equation}. So the matrix $\mathbf{P}$ specified in (\ref{eq:Riccati_equation_form3_Hamiltonian_solution}) is the positive definite solution of the Riccati equation described in (\ref{eq:Riccati_equation_form3}).  

Some explanations hover over uniqueness of the solution $\mathbf{P}$. In fact, as mentioned above, given any positive definite matrix $\mathbf{M}$, it can always be decomposed into the product of a matrix and its transpose as
\begin{align*}
\mathbf{M} = \mathbf{B} \mathbf{B}^\mathrm{T}.
\end{align*}
So the Riccati equation described in (\ref{eq:Riccati_equation_form3}) is reduced to the Riccati equations described in (\ref{eq:Riccati_equation_form2}) and (\ref{eq:Riccati_equation_form1}), uniqueness of the positive definite solution for which has already been verified. 

It is worth noting that the positive definiteness of $\mathbf{M}$, i.e.
\begin{align*}
\mathbf{M} > 0
\end{align*}
is a sufficient condition for the matrix $\mathbf{P}$ specified in (\ref{eq:Riccati_equation_form3_Hamiltonian_solution}) to be the positive definite solution, but is not a necessary condition for so. Sometimes even when the positive definiteness of $\mathbf{M}$ does not hold, the matrix $\mathbf{P}$ specified in (\ref{eq:Riccati_equation_form3_Hamiltonian_solution}) is still the positive definite solution
\footnote{For example, in the context of robust control.}. Matlab code for Riccati equation solving is given as follows.

\begin{framed} 
\noindent \textbf{SolveRiccatiEquation.m} \\
\noindent \%\% Riccati equation Form 3: P A + A' P - P M P + Q = 0  \\
\%\% Solve the symmetric matrix P \\
function P = SolveRiccatiEquation(A, M, Q, mt) \\
$~~~~$ if (nargin$<$4) mt = 'iterative'; end \\
$~~~~$ n = size(A,1); \\
$~~~~$ if (strcmp(mt,'hamilton')) \\
$~~~~$ $~~~~$ \% If H is diagonalizable with half eigs having negative real part \\
$~~~~$ $~~~~$ H = [A, -M; -Q, -A']; [U, E] = eig(H); E = diag(E); \\
$~~~~$ $~~~~$ [Er, idx] = sort(real(E)); E = E(idx); U = U(:,idx); \\
$~~~~$ $~~~~$ U11 = U(1:n,1:n); U21 = U(n+1:end,1:n); P = real(U21*inv(U11)); \\
$~~~~$ elseif (strcmp(mt,'iterative')) \\
$~~~~$ $~~~~$ P = SolveRiccatiEquationForm2(A, M, Q); \\
$~~~~$ end \\
end
\end{framed}

\section{Controllability}  \label{sec:controllability}

Fundamentals of state-space analysis especially the stability criteria presented in previous sections provide valuable theoretical guide for handling difficult control problems such as double inverted pendulum control, though more knowledge of modern control theory is still needed. For a control problem, before any tentative design of a control method, a preliminary question arises naturally: Is the target process \textit{controllable}? In other words, is it ever possible to design a control system that can control the target process as desired? Suppose the target process is characterized by certain state, then above preliminary question may be posed in another way: Is it ever possible to design a control system that enables the state to evolve as desired especially to achieve any expected state?

Take double inverted pendulum control illustrated in Figure \ref{fig:double_inverted_pendulum_control} as example, the difficulty of such control problem is easily understood if we imagine that we use our own hand instead of the moving cart to perform double inverted pendulum control. This control problem is so difficult that we might even doubt whether it would ever be possible to succeed in double inverted pendulum control. To avoid blind trials, it is worth making a preliminary theoretical judgement of whether the double inverted pendulum is \textit{controllable}: If the double inverted pendulum is \textit{proved} to be \textit{controllable}, then the intention to design a double inverted pendulum control system tends to make sense. In contrast, if the double inverted pendulum is \textit{proved} to be \textit{uncontrollable}, then we should waste no time on designing a double inverted pendulum control system. 

To design a control system, \textit{it is worth making a preliminary theoretical judgement (if possible) of controllability of the target process}. It is difficult to have a general method to theoretically determine controllability of any arbitrary target process especially severely nonlinear target process. But fortunately, many target processes encountered in practical applications can fairly adopt linear state-space modelling. For example, dynamics of the double inverted pendulum state can be fairly modelled by the linear state differential equation described in (\ref{eq:DIP_state_DE_linear}) if both inverted pendulum angles $\theta_1$ and $\theta_2$ are close to zero. For a target process that can fairly adopt linear state-space modelling, we have a systematic method to theoretically determine its controllability.

\subsection{Solution of linear state differential equation}

Given a control system with its state denoted as $\mathbf{x}$ and its control input to the target process denoted as $\mathbf{u}$, suppose dynamics of the state $\mathbf{x}$ is modelled generically by a linear state differential equation described in (\ref{eq:state_differential_equation_linear})
\begin{align*}
\frac{\mathrm{d}}{\mathrm{d} t} \mathbf{x} = \mathbf{A} \mathbf{x} + \mathbf{B} \mathbf{u}.
\end{align*}
Its homogeneous counterpart equation is
\begin{align*}
\frac{\mathrm{d}}{\mathrm{d} t} \mathbf{x} = \mathbf{A} \mathbf{x}.
\end{align*} 

According to analysis presented in Section \ref{sec:stability_criterion_linear}, the homogeneous counterpart equation has the solution as
\begin{align*}
\mathbf{x} = \mathrm{e}^{\mathbf{A} t} \mathbf{x}_0.
\end{align*}
Definition of the matrix exponential function is already clarified in (\ref{eq:closed-loop_feedback_SDE_linear_solution}). Replace 
\begin{align*}
\mathbf{x}_0 \equiv \mathbf{x}(0)
\end{align*}
by an unknown function $\mathbf{y}$, let 
\begin{align*}
\mathbf{x} = \mathrm{e}^{\mathbf{A} t} \mathbf{y}  
\end{align*}
with the initial condition apparently satisfying
\begin{align*}
\mathbf{y}_0 = \mathbf{x}_0. 
\end{align*}
Substitute it into (\ref{eq:state_differential_equation_linear}) and obtain
\begin{align*}
&\frac{\mathrm{d}}{\mathrm{d} t} (\mathrm{e}^{\mathbf{A} t} \mathbf{y}) = \mathbf{A} \mathrm{e}^{\mathbf{A} t} \mathbf{y} + \mathbf{B} \mathbf{u} \iff \mathbf{A} \mathrm{e}^{\mathbf{A} t} \mathbf{y} + \mathrm{e}^{\mathbf{A} t} \frac{\mathrm{d}}{\mathrm{d} t} \mathbf{y} = \mathbf{A} \mathrm{e}^{\mathbf{A} t} \mathbf{y} + \mathbf{B} \mathbf{u} \\
&\iff \frac{\mathrm{d}}{\mathrm{d} t} \mathbf{y} = \mathrm{e}^{-\mathbf{A} t} \mathbf{B} \mathbf{u} \iff \mathbf{y} = \mathbf{y}_0 + \int_0^t \mathrm{e}^{-\mathbf{A} \tau} \mathbf{B} \mathbf{u}(\tau) \mathrm{d} \tau.
\end{align*}
Then we have
\begin{align}  \label{eq:state_differential_equation_linear_solution}
\mathbf{x} = \mathrm{e}^{\mathbf{A} t} \mathbf{y}_0 + \mathrm{e}^{\mathbf{A} t} \int_0^t \mathrm{e}^{-\mathbf{A} \tau} \mathbf{B} \mathbf{u}(\tau) \mathrm{d} \tau = \mathrm{e}^{\mathbf{A} t} \mathbf{x}_0 + \int_0^t \mathrm{e}^{\mathbf{A} (t - \tau)} \mathbf{B} \mathbf{u}(\tau) \mathrm{d} \tau.
\end{align}

In derivation of (\ref{eq:state_differential_equation_linear_solution}), the \textit{special associative law} of matrix exponential function operation namely 
\begin{align*}
\mathrm{e}^{\mathbf{A} t_1} \mathrm{e}^{\mathbf{A} t_2} = \mathrm{e}^{\mathbf{A} (t_1 + t_2)}
\end{align*}
is used --- Readers had better not take this \textit{special associative law} for granted. It needs to be proved: $\forall t_1, t_2$, define the function 
\begin{align*}
\phi (t) = \mathrm{e}^{\mathbf{A} (t_1 - t)} \mathrm{e}^{\mathbf{A} (t + t_2)}. 
\end{align*}
We have 
\begin{align*}
\frac{\mathrm{d}}{\mathrm{d} t} \phi (t) = \frac{\mathrm{d}}{\mathrm{d} t} \mathrm{e}^{\mathbf{A} (t_1 - t)} \mathrm{e}^{\mathbf{A} (t + t_2)} + \mathrm{e}^{\mathbf{A} (t_1 - t)} \frac{\mathrm{d}}{\mathrm{d} t} \mathrm{e}^{\mathbf{A} (t + t_2)} = -\mathbf{A} \phi (t) + \mathbf{A} \phi (t) = \mathbf{0},
\end{align*}
which implies that the function $\phi (t)$ is a constant matrix regardless of $t$. So 
\begin{align*}
\mathrm{e}^{\mathbf{A} t_1} \mathrm{e}^{\mathbf{A} t_2} = \phi (0) = \phi (t_1) = \mathrm{e}^{\mathbf{A} (t_1 + t_2)}.
\end{align*}
 
Rearrange (\ref{eq:state_differential_equation_linear_solution}) as
\begin{align}  \label{eq:SDE_linear_FE_u}
\int_0^t \mathrm{e}^{\mathbf{A} (t - \tau)} \mathbf{B} \mathbf{u}(\tau) \mathrm{d} \tau = \mathbf{x} - \mathrm{e}^{\mathbf{A} t} \mathbf{x}_0,
\end{align}
which can be treated as a \textit{functional equation} in terms of the function $\mathbf{u}$ once $t$ is determined. If the target process is controllable, then for arbitrary state 
\begin{align*}
\bar{\mathbf{x}}_t \equiv \mathbf{x} - \mathrm{e}^{\mathbf{A} t} \mathbf{x}_0, 
\end{align*}
the functional equation described in (\ref{eq:SDE_linear_FE_u}) must have a solution of $\mathbf{u}$ for some $t$, and \textit{vice versa}.

\subsection{Primitive controllability matrix}

Transform the left side of the functional equation described in (\ref{eq:SDE_linear_FE_u}) as
\begin{align*}
\int_0^t \mathrm{e}^{\mathbf{A} (t - \tau)} \mathbf{B} \mathbf{u}(\tau) \mathrm{d} \tau = \int_0^t \sum_{k=0}^{\infty} \mathbf{A}^k \frac{(t - \tau)^k}{k !} \mathbf{B} \mathbf{u}(\tau) \mathrm{d} \tau & \\
  = \sum_{k=0}^{\infty} \mathbf{A}^k \mathbf{B} \int_0^t \frac{(t - \tau)^k}{k !} \mathbf{u}(\tau) \mathrm{d} \tau = \mathbf{C}_{\mathbf{A}, \mathbf{B}} \mathbf{U}_t &,
\end{align*}
where
\begin{align*}
\mathbf{C}_{\mathbf{A}, \mathbf{B}} &= \begin{bmatrix} \mathbf{B} & \mathbf{A} \mathbf{B} & \mathbf{A}^2 \mathbf{B} & \cdots & \mathbf{A}^{n-1} \mathbf{B} & \cdots \end{bmatrix}, \\
\mathbf{U}_t &= \begin{bmatrix} \int_0^t \mathbf{u}(\tau) \mathrm{d} \tau  \\  \int_0^t (t - \tau) \mathbf{u}(\tau) \mathrm{d} \tau  \\  \int_0^t \frac{(t - \tau)^2}{2 !} \mathbf{u}(\tau) \mathrm{d} \tau  \\  \cdots  \\  \int_0^t \frac{(t - \tau)^{n-1}}{(n-1) !} \mathbf{u}(\tau) \mathrm{d} \tau  \\  \cdots \end{bmatrix}.
\end{align*}
Thus the functional equation described in (\ref{eq:SDE_linear_FE_u}) can be transformed into a matrix equation form as
\begin{align}  \label{eq:SDE_linear_FE2_u}
\mathbf{C}_{\mathbf{A}, \mathbf{B}} \mathbf{U}_t = \bar{\mathbf{x}}_t,
\end{align}
where $\mathbf{C}_{\mathbf{A}, \mathbf{B}}$ is called the \textbf{controllability matrix} of the control system.

The \textit{necessary and sufficient condition} for the matrix equation described in (\ref{eq:SDE_linear_FE2_u}) to always have a solution of $\mathbf{U}_t$ no matter given what $\bar{\mathbf{x}}_t$ is that \textit{the controllability matrix $\mathbf{C}_{\mathbf{A}, \mathbf{B}}$ is of full rank (by default in terms of column vectors)} or in other words \textit{the rank of $\mathbf{C}_{\mathbf{A}, \mathbf{B}}$ equals the state dimension}. If the controllability matrix $\mathbf{C}_{\mathbf{A}, \mathbf{B}}$ is rank-deficient, there must be $\bar{\mathbf{x}}_t$ such that the matrix equation described in (\ref{eq:SDE_linear_FE2_u}) has no solution of $\mathbf{U}_t$, which implies that the target process is uncontrollable. If the controllability matrix $\mathbf{C}_{\mathbf{A}, \mathbf{B}}$ is of full rank, the matrix equation described in (\ref{eq:SDE_linear_FE2_u}) has infinite solutions of $\mathbf{U}_t$ no matter given what $\bar{\mathbf{x}}_t$. For each specific solution of $\mathbf{U}_t$, there is $\mathbf{u}$ from which the specific solution of $\mathbf{U}_t$ can be obtained 
\footnote{Given a specific $\mathbf{U}_t$, since 
\begin{align*}
1, \quad t - \tau, \quad (t - \tau)^2, \quad \cdots \quad, \quad (t - \tau)^{n-1}, \quad \cdots 
\end{align*}
are \textit{linearly independent}, the following group of linear functional equations in terms of $\mathbf{u}$
\begin{align*}
\begin{bmatrix} \int_0^t \mathbf{u}(\tau) \mathrm{d} \tau  \\  \int_0^t (t - \tau) \mathbf{u}(\tau) \mathrm{d} \tau  \\  \int_0^t \frac{(t - \tau)^2}{2 !} \mathbf{u}(\tau) \mathrm{d} \tau  \\  \cdots \end{bmatrix} = \mathbf{U}_t
\end{align*}
are also linearly independent. Besides, the group of equations have \textit{countable} equations, whereas $\mathbf{u}$ has \textit{uncountable} dimensions or degrees of freedom \cite{Tao2010, Tao2011}. So the group of equations definitely have solutions of $\mathbf{u}$ (in fact, infinite solutions of $\mathbf{u}$).}, 
which implies that the target process is controllable. 

\subsection{Refined controllability matrix}  \label{sec:refined_controllability_matrix}

Denote the characteristic polynomial of $\mathbf{A}$ namely $\det (s \mathbf{I} - \mathbf{A})$ as 
\begin{align*}
C_{\mathbf{A}}(s) = s^n + a_{n-1} s^{n-1} + \cdots + a_0,
\end{align*}
where $n$ is the dimension of the state $\mathbf{x}$. The state transition matrix $\mathbf{A}$ satisfies
\begin{align}  \label{eq:characteristic_polynomial_A=0}
C_{\mathbf{A}}(\mathbf{A}) = \mathbf{A}^n + a_{n-1} \mathbf{A}^{n-1} + \cdots + a_0 \mathbf{I} = \mathbf{0},
\end{align}
which is called the \textit{Hamilton-Cayley theorem}.

\begin{proof}
By \textit{Jordan canonical decomposition}, suppose $\mathbf{A}$ is decomposed as
\begin{align*}
\mathbf{A} = \mathbf{S} \begin{bmatrix} \mathbf{J}_{\lambda_1} & & \\  & \ddots &  \\ & & \mathbf{J}_{\lambda_q} \end{bmatrix} \mathbf{S}^{-1} \equiv \mathbf{S} \mathbf{J} \mathbf{S}^{-1},
\end{align*}
where Jordan blocks $\mathbf{J}_{\lambda_i}$ have dimensions $d_i$ respectively ($i \in \{1, 2, \cdots, q\}$). The eigenvalues $\lambda_i$ are allowed to be the same. The characteristic polynomial $C_{\mathbf{A}}(s)$ is factorized as
\begin{align*}
C_{\mathbf{A}}(s) = (s - \lambda_1)^{d_1} \cdots (s - \lambda_q)^{d_q},
\end{align*}
so
\begin{align*}
C_{\mathbf{A}}(\mathbf{A}) &= (\mathbf{S} \mathbf{J} \mathbf{S}^{-1} - \lambda_1 \mathbf{I})^{d_1} \cdots (\mathbf{S} \mathbf{J} \mathbf{S}^{-1} - \lambda_q \mathbf{I})^{d_q} \\
  &= \mathbf{S} (\mathbf{J} - \lambda_1 \mathbf{I})^{d_1} \cdots (\mathbf{J} - \lambda_q \mathbf{I})^{d_q} \mathbf{S}^{-1} \\
  &= \mathbf{S} \begin{bmatrix} (\mathbf{J}_{\lambda_1} - \lambda_1 \mathbf{I})^{d_1} & \\ & \ddots \end{bmatrix} \cdots \begin{bmatrix} \ddots & \\ & (\mathbf{J}_{\lambda_q} - \lambda_q \mathbf{I})^{d_q} \end{bmatrix}  \mathbf{S}^{-1} \\
  &= \mathbf{S} \begin{bmatrix} \mathbf{0} & \\ & \ddots \end{bmatrix} \cdots \begin{bmatrix} \ddots & \\ & \mathbf{0} \end{bmatrix}  \mathbf{S}^{-1} = \mathbf{S} \quad \mathbf{0} \quad \mathbf{S}^{-1} = \mathbf{0}.
\end{align*}
The proof is done.
\end{proof}

The Hamilton-Cayley theorem described by (\ref{eq:characteristic_polynomial_A=0}) implies that each element after $\mathbf{A}^{n-1} \mathbf{B}$ in the controllability matrix $\mathbf{C}_{\mathbf{A}, \mathbf{B}}$ namely each $\mathbf{A}^k \mathbf{B}$ for $k \geq n$ can be transformed into a linear combination of 
\begin{align*}
\mathbf{B}, \quad \mathbf{A} \mathbf{B}, \quad \mathbf{A}^2 \mathbf{B}, \quad \cdots \quad, \quad \mathbf{A}^{n-1} \mathbf{B}. 
\end{align*}
So we have
\begin{align*}
\mbox{rank of } &\begin{bmatrix} \mathbf{B} & \mathbf{A} \mathbf{B} & \mathbf{A}^2 \mathbf{B} & \cdots & \mathbf{A}^{n-1} \mathbf{B} & \cdots \end{bmatrix} \\
= \mbox{rank of } &\begin{bmatrix} \mathbf{B} & \mathbf{A} \mathbf{B} & \mathbf{A}^2 \mathbf{B} & \cdots & \mathbf{A}^{n-1} \mathbf{B} \end{bmatrix}.
\end{align*}
Then we can define the \textbf{controllability matrix} simply as
\begin{equation}  \label{eq:controllability_matrix}
\mathbf{C}_{\mathbf{A}, \mathbf{B}} = \begin{bmatrix} \mathbf{B} & \mathbf{A} \mathbf{B} & \mathbf{A}^2 \mathbf{B} & \cdots & \mathbf{A}^{n-1} \mathbf{B} \end{bmatrix}.
\end{equation}
If the controllability matrix $\mathbf{C}_{\mathbf{A}, \mathbf{B}}$ described in (\ref{eq:controllability_matrix}) is rank-deficient, then the target process is uncontrollable. If the controllability matrix $\mathbf{C}_{\mathbf{A}, \mathbf{B}}$ described in (\ref{eq:controllability_matrix}) is of full rank, then the target process is controllable.

\begin{framed} 
\noindent \textbf{Control system controllability criterion}: \textit{For a linear control system, if its controllability matrix is rank-deficient, then its target process is uncontrollable. If its controllability matrix is of full rank, then its target process is controllable}.
\end{framed}

\subsubsection*{Application: double inverted pendulum controllability analysis}

Apply the controllability criterion to determine controllability of the double inverted pendulum that adopts linear state-space modelling described by (\ref{eq:DIP_state_DE_linear}). The state transition matrix $\mathbf{A}$ and the control input matrix $\mathbf{B}$ are respectively
\begin{align*}
\mathbf{A} = \begin{bmatrix} 0 & 1 & 0 & 0 & 0 & 0 \\
(1 + \frac{m_2}{m_1}) \frac{g}{L_1} & 0 & -\frac{m_2}{m_1} \frac{g}{L_1} & 0 & 0 & 0 \\
0 & 0 & 0 & 1 & 0 & 0 \\
-(1 + \frac{m_2}{m_1}) \frac{g}{L_2} & 0 & (1 + \frac{m_2}{m_1}) \frac{g}{L_2} & 0 & 0 & 0 \\
0 & 0 & 0 & 0 & 0 & 1  \\ 0 & 0 & 0 & 0 & 0 & 0 \end{bmatrix}, \quad \mathbf{B} = \begin{bmatrix} 0 \\ -\frac{1}{L_1} \\ 0 \\ 0 \\ 0 \\ 1 \end{bmatrix}.
\end{align*}
Compute the controllability matrix via (\ref{eq:controllability_matrix}) as
\begin{align*}
\mathbf{C}_{\mathbf{A}, \mathbf{B}} &= \begin{bmatrix} \mathbf{B} & \mathbf{A} \mathbf{B} & \mathbf{A}^2 \mathbf{B} & \mathbf{A}^3 \mathbf{B} & \mathbf{A}^4 \mathbf{B} & \mathbf{A}^5 \mathbf{B} \end{bmatrix}  \\
  &= \begin{bmatrix}
0 & -\frac{1}{L_1} & 0 & -\frac{c_1}{L_1} & 0 & -\frac{c_1^2}{L_1} - \frac{c_1 c_2}{L_2} \\ 
-\frac{1}{L_1} & 0 & -\frac{c_1}{L_1} & 0 & -\frac{c_1^2}{L_1} - \frac{c_1 c_2}{L_2} & 0 \\ 
0 & 0 & 0 & \frac{c_1}{L_2} & 0 & \frac{c_3^2}{L_1} + \frac{c_1 c_3}{L_1} \\ 
0 & 0 & \frac{c_1}{L_2} & 0 & \frac{c_3^2}{L_1} + \frac{c_1 c_3}{L_1} & 0 \\ 
0 & 1 & 0 & 0 & 0 & 0 \\ 
1 & 0 & 0 & 0 & 0 & 0
\end{bmatrix}
\end{align*}
where
\begin{align*}
c_1 &\equiv (1 + \frac{m_2}{m_1}) \frac{g}{L_1},  \\ 
c_2 &\equiv \frac{m_2 g}{m_1 L_1},  \\ 
c_3 &\equiv (1 + \frac{m_2}{m_1}) \frac{g}{L_2}.
\end{align*}
Matlab code for computing the controllability matrix via symbolic operation is given as follows.

\begin{framed} 
\noindent \textbf{ControllabilityMatrixSymDIP.m} \\
\noindent \%\% Double inverted pendulum parameters \\
syms m1  m2  L1  L2  g \\
A = [0, 1, 0, 0, 0, 0; ... \\
$~~~~$ (m1+m2)*g/(m1*L1), 0, -m2*g/(m1*L1), 0, 0, 0; ... \\
$~~~~$ 0, 0, 0, 1, 0, 0; ... \\
$~~~~$ -(m1+m2)*g/(m1*L2), 0, (m1+m2)*g/(m1*L2), 0, 0, 0; ... \\
$~~~~$ 0, 0, 0, 0, 0, 1; ... \\
$~~~~$ 0, 0, 0, 0, 0, 0]; \\
B = [0; -1/L1; 0; 0; 0; 1]; \\
n = size(A,1); \% State dimension \\
 \\
CM = sym(zeros(n)); CM(:,1) = B; \\
for k = 2:n \\
$~~~~$ CM(:,k) = A*CM(:,k-1); \\
end \% CM = [B, A*B, A\^{}2*B, A\^{}3*B, A\^{}4*B, A\^{}5*B]; \\
fprintf('Controllability matrix: '); CM \\
 \\
\%\% Check if the controllability matrix is of full rank \\
if (rank(CM) == n) \\
$~~~~$ fprintf('The double inverted pendulum is controllable$\backslash$n'); \\
else \\
$~~~~$ fprintf('The double inverted pendulum is uncontrollable$\backslash$n'); \\
end
\end{framed}

After trying the Matlab code, readers will find that the double inverted pendulum is indeed controllable, at least when both inverted pendulum angles $\theta_1$ and $\theta_2$ are close to zero so that linear state-space modelling can be fairly adopted. Based on this positive preliminary theoretical judgement, we can move forward with confidence to handle the double inverted pendulum control problem.

\subsubsection*{Application: low-speed autonomous vehicle controllability analysis}

Apply the controllability criterion to determine controllability of the low-speed autonomous vehicle lateral control system that adopts linear state-space modelling described by (\ref{eq:vehicle_lateral_control_approximation}). The state transition matrix $\mathbf{A}$ and the control input matrix $\mathbf{B}$ are respectively
\begin{align*}
\mathbf{A} = \begin{bmatrix} 0 & v & 0 \\ 0 & 0 & \frac{v}{L} \\ 0 & 0 & -\frac{1}{\tau_{\beta}} \end{bmatrix}, \quad \mathbf{B} = \begin{bmatrix} 0 \\ 0 \\ \frac{1}{\tau_{\beta}} \end{bmatrix}.
\end{align*}
Compute the controllability matrix via (\ref{eq:controllability_matrix}) as
\begin{align*}
\mathbf{C}_{\mathbf{A}, \mathbf{B}} &= \begin{bmatrix} \mathbf{B} & \mathbf{A} \mathbf{B} & \mathbf{A}^2 \mathbf{B} \end{bmatrix} = \begin{bmatrix} 0 & 0 & \frac{v^2}{\tau_{\beta} L} \\ 0 & \frac{v}{\tau_{\beta} L} & -\frac{v}{\tau_{\beta}^2 L} \\ \frac{1}{\tau_{\beta}} & -\frac{1}{\tau_{\beta}^2} & \frac{1}{\tau_{\beta}^3} \end{bmatrix}
\end{align*}
which is of full rank and hence the vehicle lateral control system is indeed controllable --- According to daily-life experiences, people of course know that the vehicle is controllable in terms of steering or lateral control. Above analysis is not to tell people this evident fact, but to demonstrate with another concrete example how to apply the methodology of controllability analysis. This is especially valuable to complicated control tasks for which people usually do not have much or even any daily-life experience. For example, the double inverted pendulum control task just demonstrated belongs to such complicated cases. In fact, the author once made a survey among his students, asking them whether they think the double inverted pendulum is controllable. Most students gave negative opinions by intuition and were finally surprised by its controllability. From this we could clearly see that theoretical analysis tends to be indispensable and even more important than pure daily-life experiences based intuition.

\appendix

\section{System Dynamics}  \label{app:system_dynamics}

For the large variety of control systems involved in this book, Appendix \ref{app:system_dynamics} focuses on clarification of system dynamics models only for those comparatively complicated ones among them. Fundamentals of physicals especially mechanics are necessary for readers to digest knowledge presented throughout Appendix \ref{app:system_dynamics}. Books worth recommendation are \textit{The Feynman Lectures on Physics} \cite{Feynman2004} and \textit{Mathematical Methods of Classical Mechanics} \cite{Arnold1989}. 

\subsection{Inverted pendulum dynamics}

\subsubsection{Single inverted pendulum dynamics}  \label{sec:SIP_dynamics}

\begin{figure}[h!]
\begin{center}
\includegraphics[width=0.6\columnwidth]{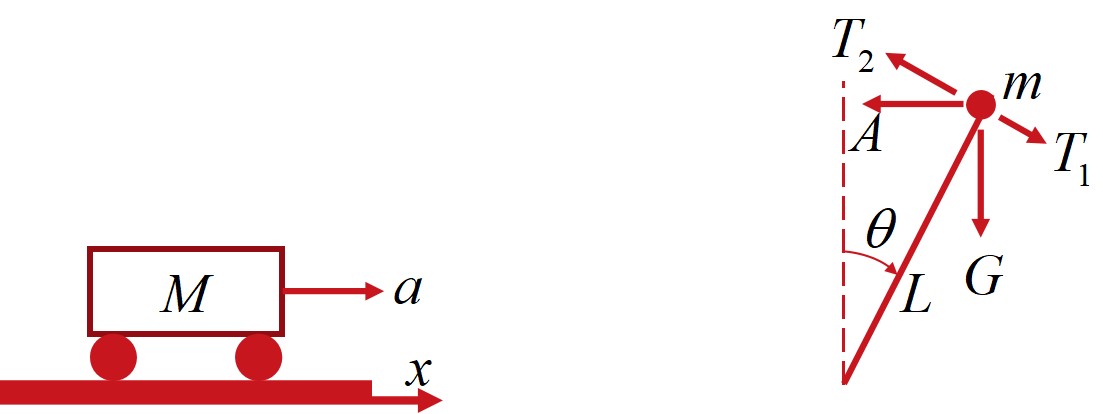}
\end{center}
\caption{Single inverted pendulum dynamics: (left) cart dynamics; (right) inverted pendulum body dynamics}
\label{fig:SIP_dynamics}
\end{figure}

Dynamics of the single inverted pendulum is separated into two parts, namely that of the cart and that of the inverted pendulum body. As illustrated in the left sub-figure of Figure \ref{fig:SIP_dynamics}, dynamics of the cart can be easily described as
\begin{subequations}  \label{eq:SIP_cart_dynamics}
\begin{align}
\frac{\mathrm{d}}{\mathrm{d} t} x &= \dot x,  \\
\frac{\mathrm{d}}{\mathrm{d} t} \dot x &= a,
\end{align}
\end{subequations}
where $x$ denotes the cart position and 
\begin{align*}
\dot x \equiv \frac{\mathrm{d} x}{\mathrm{d} t}
\end{align*}
denotes the cart speed or velocity. The first equation of (\ref{eq:SIP_cart_dynamics}) is a trivial equation.

As illustrated in the right sub-figure of Figure \ref{fig:SIP_dynamics}, dynamics of the inverted pendulum body involves a bit more analysis. Here, $m$ denotes the inverted pendulum mass, $L$ denotes the inverted pendulum length, $G$ denotes gravity, $A$ denotes the cart acceleration force, $T_1$ denotes the torque component contributed by gravity, and $T_2$ denotes the torque component contributed by the cart acceleration force. The rotating inertia of the inverted pendulum body is
\begin{align*}
J = m L^2,
\end{align*}
the two forces are
\begin{align*}
G &= m g,  \\
A &= -m a,
\end{align*} 
and the two torques are computed as
\begin{align*}
T_1 &= m g L \sin \theta,  \\
T_2 &= -m a L \cos \theta.
\end{align*}
So dynamics of the inverted pendulum body can be described by the following differential equation
\begin{align*}
J \frac{\mathrm{d}^2 \theta}{\mathrm{d} t^2} = T_1 + T_2 &\iff m L^2 \frac{\mathrm{d}^2 \theta}{\mathrm{d} t^2} = m g L \sin \theta - m a L \cos \theta  \\
  &\iff \frac{\mathrm{d}^2 \theta}{\mathrm{d} t^2} = \frac{\sin \theta}{L} g - \frac{\cos \theta}{L} a
\end{align*}
which can be further decomposed into two equations
\begin{subequations}  \label{eq:SIP_body_dynamics}
\begin{align}
\frac{\mathrm{d}}{\mathrm{d} t} \theta &= \dot \theta,  \\
\frac{\mathrm{d}}{\mathrm{d} t} \dot \theta &= \frac{\sin \theta}{L} g - \frac{\cos \theta}{L} a,
\end{align}
\end{subequations}
where $\theta$ denotes the inverted pendulum angle and 
\begin{align*}
\dot \theta \equiv \frac{\mathrm{d} \theta}{\mathrm{d} t}
\end{align*}
denotes the inverted pendulum angular speed. The first equation of (\ref{eq:SIP_body_dynamics}) is also a trivial equation.

Combine (\ref{eq:SIP_cart_dynamics}) and (\ref{eq:SIP_body_dynamics}) to obtain the state differential equation (\ref{eq:SIP_state_DE})
\begin{align*}
\frac{\mathrm{d}}{\mathrm{d} t} \mathbf{x} \equiv \frac{\mathrm{d}}{\mathrm{d} t} \begin{bmatrix} \theta \\ \frac{\mathrm{d} \theta}{\mathrm{d} t} \\ x \\ \frac{\mathrm{d} x}{\mathrm{d} t} \end{bmatrix} = \begin{bmatrix} \frac{\mathrm{d} \theta}{\mathrm{d} t} \\ \frac{\sin \theta}{L} g - \frac{\cos \theta}{L} a \\ \frac{\mathrm{d} x}{\mathrm{d} t} \\ a \end{bmatrix} \equiv f(\mathbf{x}, a),
\end{align*}
where the state 
\begin{align*}
\mathbf{x} \equiv \begin{bmatrix} \theta & \dot \theta & x & \dot x \end{bmatrix}^\mathrm{T}
\end{align*}
namely 
\begin{align*}
\mathbf{x} \equiv \begin{bmatrix} \theta & \frac{\mathrm{d} \theta}{\mathrm{d} t} & x & \frac{\mathrm{d} x}{\mathrm{d} t} \end{bmatrix}^\mathrm{T}
\end{align*}
consists of the inverted pendulum angle and angular velocity, and the cart position and velocity.

If the inverted pendulum angle $\theta$ is close to zero, then $\sin \theta$ and $\cos \theta$ can be approximated respectively as
\begin{align*}
\sin \theta \approx \theta, \qquad \cos \theta \approx 1
\end{align*}
and hence the state differential equation (\ref{eq:SIP_state_DE}) can be fairly linearized about the equilibrium state and simplified into the linear state differential equation (\ref{eq:SIP_state_DE_linear})
\begin{align*}
\frac{\mathrm{d}}{\mathrm{d} t} \mathbf{x} = \begin{bmatrix} \frac{\mathrm{d} \theta}{\mathrm{d} t} \\ \frac{\theta}{L} g - \frac{1}{L} a \\ \frac{\mathrm{d} x}{\mathrm{d} t} \\ a \end{bmatrix} = \begin{bmatrix} 0 & 1 & 0 & 0 \\ \frac{g}{L} & 0 & 0 & 0 \\ 0 & 0 & 0 & 1  \\ 0 & 0 & 0 & 0 \end{bmatrix} \mathbf{x} + \begin{bmatrix} 0 \\ -\frac{1}{L} \\ 0 \\ 1 \end{bmatrix} a \equiv \mathbf{A} \mathbf{x} + \mathbf{B} a.
\end{align*}

\subsubsection{Double inverted pendulum dynamics}  \label{sec:DIP_dynamics}

\begin{figure}[h!]
\begin{center}
\includegraphics[width=0.8\columnwidth]{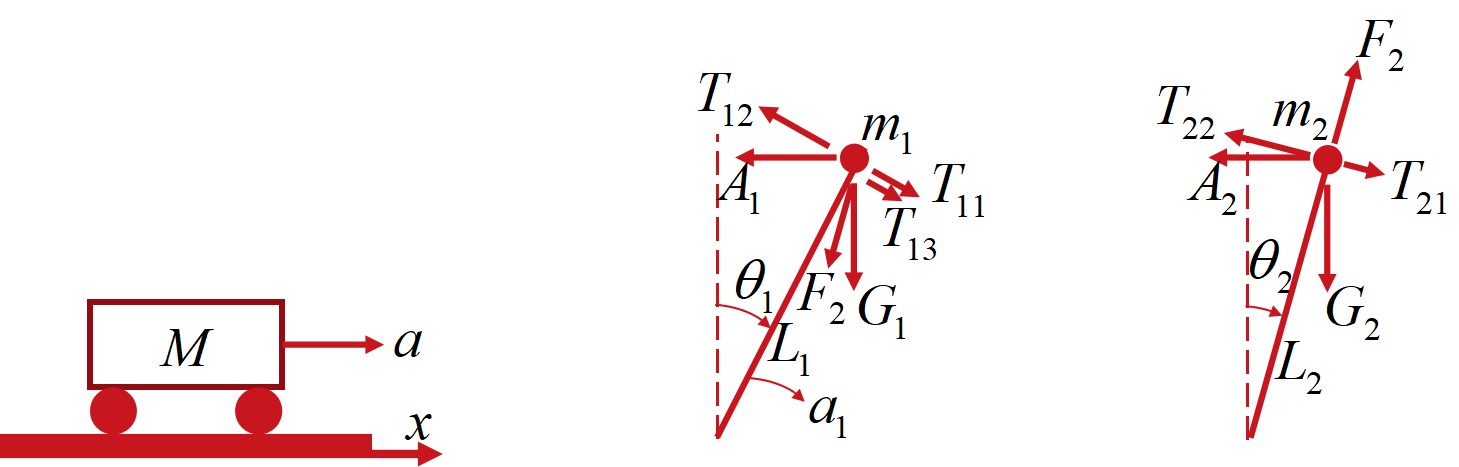}
\end{center}
\caption{Double inverted pendulum dynamics: (left) cart dynamics; (middle) first inverted pendulum body dynamics; (right) second inverted pendulum body dynamics}
\label{fig:DIP_dynamics}
\end{figure}

Consider dynamics of the double inverted pendulum variant illustrated in Figure \ref{fig:double_inverted_pendulum_variant_control} which is compatible with that of the original double inverted pendulum illustrated in Figure \ref{fig:double_inverted_pendulum_control}. Dynamics of the double inverted pendulum variant is separated into three parts, namely that of the cart, that of the first inverted pendulum body, and that of the second inverted pendulum body. As illustrated in the left sub-figure of Figure \ref{fig:DIP_dynamics}, dynamics of the cart can be easily described as in (\ref{eq:SIP_cart_dynamics})
\begin{align*}
\frac{\mathrm{d}}{\mathrm{d} t} x &= \dot x  \\
\frac{\mathrm{d}}{\mathrm{d} t} \dot x &= a
\end{align*}
where $x$ denotes the cart position and $\dot x$ denotes the cart speed or velocity.

Factors that influence dynamics of the first inverted pendulum body are illustrated in the middle sub-figure of Figure \ref{fig:DIP_dynamics}. Here, $m_1$ denotes the first inverted pendulum mass, $L_1$ denotes the first inverted pendulum length, $G_1$ denotes gravity of the first inverted pendulum body, $A_1$ denotes the cart acceleration force exerted on the first inverted pendulum body, $F_2$ denotes the force exerted on the first inverted pendulum body along the second inverted pendulum link, $T_{11}$ denotes the torque component contributed by gravity, $T_{12}$ denotes the torque component contributed by the cart acceleration force, $T_{13}$ denotes the torque component contributed by the second inverted pendulum link force, and $a_1$ denotes the control input of first inverted pendulum angular acceleration. 

Factors that influence dynamics of the second inverted pendulum body are illustrated in the right sub-figure of Figure \ref{fig:DIP_dynamics}. Here, $m_2$ denotes the second inverted pendulum mass, $L_2$ denotes the second inverted pendulum length, $G_2$ denotes gravity of the second inverted pendulum body, $A_2$ denotes the cart acceleration force exerted on the second inverted pendulum body, $F_2$ denotes the force exerted on the second inverted pendulum body along the second inverted pendulum link (in the opposite direction of that exerted on the first inverted pendulum body), $T_{21}$ denotes the torque component contributed by gravity, and $T_{22}$ denotes the torque component contributed by the cart acceleration force.

The forces $G_1$, $A_1$, $G_2$, and $A_2$ are known easily as
\footnote{Horizontal forces take the right as the positive direction. This is why there are minus signs before $A_1$ and $A_2$. In following derivations, do not mistake the actual positive direction with the intuitive illustration.}
\begin{align*}
G_1 &= m_1 g, \qquad A_1 = -m_1 a,  \\
G_2 &= m_2 g, \qquad A_2 = -m_2 a.
\end{align*}
Computation of the force $F_2$ needs some derivation. Consider acceleration of $m_2$ along the second inverted pendulum link, which is actually contributed by acceleration of $m_1$. Treat the cart as the stationary physical reference, or simply speaking, in the cart reference, the acceleration of $m_1$ is
\begin{align*}
a_{m_1} = L_1 \ddot \theta_1
\end{align*}
and the acceleration of $m_2$ along the second inverted pendulum link is
\begin{align*}
a_{m_2}^{L_2} = a_{m_1} \sin \Delta \theta = L_1 \ddot \theta_1 \sin \Delta \theta,
\end{align*}
where 
\begin{align*}
\Delta \theta \equiv \theta_1 - \theta_2
\end{align*}
and the acceleration direction points downside. Then we have
\begin{align*}
& G_2 \cos \theta_2 + A_2 \cos (\theta_2 + \pi/2) - F_2 = m_2 a_{m_2}^{L_2}  \\
\iff& m_2 g \cos \theta_2 + m_2 a \sin \theta_2 - F_2 = m_2 L_1 \ddot \theta_1 \sin \Delta \theta
\end{align*}
and
\begin{equation}  \label{eq:DIP_dynamics_F2}
F_2 = m_2 g \cos \theta_2 + m_2 a \sin \theta_2 - m_2 L_1 \ddot \theta_1 \sin \Delta \theta.
\end{equation}
Despite $F_2$ is still unknown by so far because $\ddot \theta_1$ is unknown yet, (\ref{eq:DIP_dynamics_F2}) at least conveys the relationship between $F_2$ and other physical quantities.

For the first inverted pendulum body, the rotating inertia is
\begin{align*}
J_1 = m_1 L_1^2
\end{align*}
and the three torques are computed as
\begin{align*}
T_{11} &= m_1 g L_1 \sin \theta_1,  \\
T_{12} &= -m_1 a L_1 \cos \theta_1,  \\
T_{13} &= F_2 L_1 \sin \Delta \theta.
\end{align*}
So dynamics of the first inverted pendulum body can be described by the following differential equation --- Substitute (\ref{eq:DIP_dynamics_F2}) into following derivation 
\begin{align*}
J_1 \ddot \theta_1 &= T_{11} + T_{12} + T_{13} + J_1 a_1 = m_1 g L_1 \sin \theta_1 - m_1 a L_1 \cos \theta_1 + F_2 L_1 \sin \Delta \theta + m_1 L_1^2 a_1  \\
  &= m_1 g L_1 \sin \theta_1 - m_1 a L_1 \cos \theta_1 + m_1 L_1^2 a_1  \\
  &\qquad + L_1 \sin \Delta \theta (m_2 g \cos \theta_2 + m_2 a \sin \theta_2 - m_2 L_1 \ddot \theta_1 \sin \Delta \theta)  \\
  &= L_1 (m_1 \sin \theta_1 + m_2 \sin \Delta \theta \cos \theta_2) g + L_1 (- m_1 \cos \theta_1 + m_2 \sin \Delta \theta \sin \theta_2) a  \\
  &\qquad + m_1 L_1^2 a_1 - m_2 L_1^2 (\sin \Delta \theta)^2 \ddot \theta_1
\end{align*}
from which we can solve $\ddot \theta_1$ as
\begin{align*}
\ddot \theta_1 = \frac{(m_1 \sin \theta_1 + m_2 \sin \Delta \theta \cos \theta_2) g + (- m_1 \cos \theta_1 + m_2 \sin \Delta \theta \sin \theta_2) a + m_1 L_1 a_1}{m_1 L_1 + m_2 L_1 (\sin \Delta \theta)^2}.
\end{align*}
Above equation can be further decomposed into two equations
\begin{subequations}  \label{eq:DIP_body1_dynamics}
\begin{align}
\frac{\mathrm{d}}{\mathrm{d} t} \theta_1 &= \dot \theta_1,  \\
\frac{\mathrm{d}}{\mathrm{d} t} \dot \theta_1 &= \frac{(\sin \theta_1 + \frac{m_2}{m_1} \sin \Delta \theta \cos \theta_2)}{1 + \frac{m_2}{m_1} (\sin \Delta \theta)^2} \frac{g}{L_1}  \nonumber  \\
  &\quad - \frac{(\cos \theta_1 - \frac{m_2}{m_1} \sin \Delta \theta \sin \theta_2)}{1 + \frac{m_2}{m_1} (\sin \Delta \theta)^2} \frac{a}{L_1} + \frac{1}{1 + \frac{m_2}{m_1} (\sin \Delta \theta)^2} a_1,
\end{align}
\end{subequations}
where $\theta_1$ denotes the first inverted pendulum angle and $\dot \theta_1$ denotes the first inverted pendulum angular speed.

The rotating inertia of the second inverted pendulum body is
\begin{align*}
J_2 = m_2 L_2^2
\end{align*}
and the two torques are computed as
\begin{align*}
T_{21} &= m_2 g L_2 \sin \theta_2,  \\
T_{22} &= -m_2 a L_2 \cos \theta_2.
\end{align*}
Note that the superposed angular acceleration of the second inverted pendulum is
\begin{align*}
\ddot \theta_2 + \frac{L_1 \ddot \theta_1 \cos \Delta \theta}{L_2}
\end{align*}
So dynamics of the second inverted pendulum body can be described by the following differential equation
\begin{align*}
& J_2 (\ddot \theta_2 + \frac{L_1 \ddot \theta_1 \cos \Delta \theta}{L_2}) = T_{21} + T_{22}    \\
\iff& m_2 L_2^2 \ddot \theta_2 + m_2 L_2 L_1 \ddot \theta_1 \cos \Delta \theta = m_2 g L_2 \sin \theta_2 - m_2 a L_2 \cos \theta_2  \\
\iff& \ddot \theta_2 = \sin \theta_2 \frac{g}{L_2} - \cos \theta_2 \frac{a}{L_2} - \cos \Delta \theta \frac{L_1}{L_2} \ddot \theta_1.
\end{align*}
Substitute the second equation of (\ref{eq:DIP_body1_dynamics}) into above equation and obtain
\begin{align*}
\ddot \theta_2 &= \sin \theta_2 \frac{g}{L_2} - \cos \theta_2 \frac{a}{L_2}    \\
  &\quad - \cos \Delta \theta \frac{(\sin \theta_1 + \frac{m_2}{m_1} \sin \Delta \theta \cos \theta_2) \frac{g}{L_2} + (- \cos \theta_1 + \frac{m_2}{m_1} \sin \Delta \theta \sin \theta_2) \frac{a}{L_2} + \frac{L_1}{L_2} a_1}{1 + \frac{m_2}{m_1} (\sin \Delta \theta)^2}  \\
  &= - \frac{\sin \Delta \theta \cos \theta_1}{1 + \frac{m_2}{m_1} (\sin \Delta \theta)^2} (1 + \frac{m_2}{m_1}) \frac{g}{L_2} - \frac{\sin \Delta \theta \sin \theta_1}{1 + \frac{m_2}{m_1} (\sin \Delta \theta)^2} (1 + \frac{m_2}{m_1}) \frac{a}{L_2} - \frac{\cos \Delta \theta \frac{L_1}{L_2}}{1 + \frac{m_2}{m_1} (\sin \Delta \theta)^2} a_1
\end{align*}
which can be further decomposed into two equations
\begin{subequations}  \label{eq:DIP_body2_dynamics}
\begin{align}
\frac{\mathrm{d}}{\mathrm{d} t} \theta_2 &= \dot \theta_2,  \\
\frac{\mathrm{d}}{\mathrm{d} t} \dot \theta_2 &= - \frac{\sin \Delta \theta \cos \theta_1}{1 + \frac{m_2}{m_1} (\sin \Delta \theta)^2} (1 + \frac{m_2}{m_1}) \frac{g}{L_2}   \nonumber  \\
  & \quad - \frac{\sin \Delta \theta \sin \theta_1}{1 + \frac{m_2}{m_1} (\sin \Delta \theta)^2} (1 + \frac{m_2}{m_1}) \frac{a}{L_2} - \frac{\cos \Delta \theta \frac{L_1}{L_2}}{1 + \frac{m_2}{m_1} (\sin \Delta \theta)^2} a_1,   
\end{align}
\end{subequations}
where $\theta_2$ denotes the second inverted pendulum angle and $\dot \theta_2$ denotes the second inverted pendulum angular speed.

Combine (\ref{eq:SIP_cart_dynamics}), (\ref{eq:DIP_body1_dynamics}), and (\ref{eq:DIP_body2_dynamics}) to obtain the state differential equation (\ref{eq:DIP_state_DE_MIMO})
\begin{align*}
\frac{\mathrm{d}}{\mathrm{d} t} \mathbf{x} &= \begin{bmatrix} \frac{\mathrm{d} \theta_1}{\mathrm{d} t} \\ 
\frac{(\sin \theta_1 + \frac{m_2}{m_1} \sin \Delta \theta \cos \theta_2)}{1 + \frac{m_2}{m_1} (\sin \Delta \theta)^2} \frac{g}{L_1} - \frac{(\cos \theta_1 - \frac{m_2}{m_1} \sin \Delta \theta \sin \theta_2)}{1 + \frac{m_2}{m_1} (\sin \Delta \theta)^2} \frac{a}{L_1} + \frac{a_1}{1 + \frac{m_2}{m_1} (\sin \Delta \theta)^2} \\ 
\frac{\mathrm{d} \theta_2}{\mathrm{d} t} \\ 
- \frac{\sin \Delta \theta \cos \theta_1}{1 + \frac{m_2}{m_1} (\sin \Delta \theta)^2} (1 + \frac{m_2}{m_1}) \frac{g}{L_2} - \frac{\sin \Delta \theta \sin \theta_1}{1 + \frac{m_2}{m_1} (\sin \Delta \theta)^2} (1 + \frac{m_2}{m_1}) \frac{a}{L_2} - \frac{\cos \Delta \theta \frac{L_1}{L_2} a_1}{1 + \frac{m_2}{m_1} (\sin \Delta \theta)^2}  \\ 
\frac{\mathrm{d} x}{\mathrm{d} t} \\ a \end{bmatrix}  \nonumber  \\
  &= \begin{bmatrix} \frac{\mathrm{d} \theta_1}{\mathrm{d} t} \\ 
\frac{(\sin \theta_1 + \frac{m_2}{m_1} \sin \Delta \theta \cos \theta_2)}{1 + \frac{m_2}{m_1} (\sin \Delta \theta)^2} \frac{g}{L_1} \\ 
\frac{\mathrm{d} \theta_2}{\mathrm{d} t} \\ 
- \frac{\sin \Delta \theta \cos \theta_1}{1 + \frac{m_2}{m_1} (\sin \Delta \theta)^2} (1 + \frac{m_2}{m_1}) \frac{g}{L_2}  \\ 
\frac{\mathrm{d} x}{\mathrm{d} t} \\ 0 \end{bmatrix} 
  - \begin{bmatrix} 0 \\ 
\frac{(\cos \theta_1 - \frac{m_2}{m_1} \sin \Delta \theta \sin \theta_2)}{1 + \frac{m_2}{m_1} (\sin \Delta \theta)^2} \frac{1}{L_1} \\ 
0 \\ 
\frac{\sin \Delta \theta \sin \theta_1}{1 + \frac{m_2}{m_1} (\sin \Delta \theta)^2} (1 + \frac{m_2}{m_1}) \frac{1}{L_2}  \\ 
0 \\ - 1 \end{bmatrix} a 
  + \begin{bmatrix} 0 \\ \frac{1}{1 + \frac{m_2}{m_1} (\sin \Delta \theta)^2} \\ 0 \\ - \frac{\cos \Delta \theta \frac{L_1}{L_2}}{1 + \frac{m_2}{m_1} (\sin \Delta \theta)^2} \\ 0 \\ 0 \end{bmatrix} a_1   \nonumber  \\
  &\equiv f(\mathbf{x}) - g(\mathbf{x}) a + g_1(\mathbf{x}) a_1,
\end{align*}
where the state 
\begin{align*}
\mathbf{x} \equiv \begin{bmatrix} \theta_1 & \dot \theta_1 & \theta_2 & \dot \theta_2 & x & \dot x \end{bmatrix}^\mathrm{T}
\end{align*}
namely 
\begin{align*}
\mathbf{x} \equiv \begin{bmatrix} \theta_1 & \frac{\mathrm{d} \theta_1}{\mathrm{d} t} & \theta_2 & \frac{\mathrm{d} \theta_2}{\mathrm{d} t} & x & \frac{\mathrm{d} x}{\mathrm{d} t} \end{bmatrix}^\mathrm{T}
\end{align*}
consists of the first inverted pendulum angle and angular velocity, the second inverted pendulum angle and angular velocity, and the cart position and velocity. Simply substitute 
\begin{align*}
a_1 = 0
\end{align*}
into (\ref{eq:DIP_state_DE_MIMO}) and obtain the state differential equation (\ref{eq:DIP_state_DE})
\begin{align*}
\frac{\mathrm{d}}{\mathrm{d} t} \mathbf{x} = \begin{bmatrix} \frac{\mathrm{d} \theta_1}{\mathrm{d} t} \\
\frac{(\sin \theta_1 + \frac{m_2}{m_1} \sin \Delta \theta \cos \theta_2)}{1 + \frac{m_2}{m_1} (\sin \Delta \theta)^2} \frac{g}{L_1} - \frac{(\cos \theta_1 - \frac{m_2}{m_1} \sin \Delta \theta \sin \theta_2)}{1 + \frac{m_2}{m_1} (\sin \Delta \theta)^2} \frac{a}{L_1} \\
\frac{\mathrm{d} \theta_2}{\mathrm{d} t} \\
- \frac{\sin \Delta \theta \cos \theta_1}{1 + \frac{m_2}{m_1} (\sin \Delta \theta)^2} (1 + \frac{m_2}{m_1}) \frac{g}{L_2} - \frac{\sin \Delta \theta \sin \theta_1}{1 + \frac{m_2}{m_1} (\sin \Delta \theta)^2} (1 + \frac{m_2}{m_1}) \frac{a}{L_2} \\
\frac{\mathrm{d} x}{\mathrm{d} t} \\ a \end{bmatrix} \equiv f(\mathbf{x}, a).
\end{align*}

If both inverted pendulum angles $\theta_1$ and $\theta_2$ are close to zero, then following approximations
\begin{align*}
\sin \theta_1 \approx \theta_1, & \qquad \cos \theta_1 \approx 1,  \\
\sin \theta_2 \approx \theta_2, & \qquad \cos \theta_2 \approx 1,  \\
\sin \Delta \theta \approx \theta_1 - \theta_2, & \qquad \cos \Delta \theta \approx 1
\end{align*}
can be taken respectively and hence the state differential equation (\ref{eq:DIP_state_DE_MIMO}) can be fairly linearized about the equilibrium state and simplified into the linear state differential equation (\ref{eq:DIP_state_DE_linear_MIMO})
\begin{align*}
\frac{\mathrm{d}}{\mathrm{d} t} \mathbf{x} &= \begin{bmatrix} \frac{\mathrm{d} \theta_1}{\mathrm{d} t} \\ 
\frac{\theta_1 + \frac{m_2}{m_1} (\theta_1 - \theta_2)}{1 + \frac{m_2}{m_1} (\theta_1 - \theta_2)^2} \frac{g}{L_1} - \frac{1 - \frac{m_2}{m_1} (\theta_1 - \theta_2) \theta_2}{1 + \frac{m_2}{m_1} (\theta_1 - \theta_2)^2} \frac{a}{L_1} + \frac{a_1}{1 + \frac{m_2}{m_1} (\theta_1 - \theta_2)^2} \\ 
\frac{\mathrm{d} \theta_2}{\mathrm{d} t} \\ 
- \frac{(\theta_1 - \theta_2)}{1 + \frac{m_2}{m_1} (\theta_1 - \theta_2)^2} (1 + \frac{m_2}{m_1}) \frac{g}{L_2} - \frac{(\theta_1 - \theta_2) \theta_1}{1 + \frac{m_2}{m_1} (\theta_1 - \theta_2)^2} (1 + \frac{m_2}{m_1}) \frac{a}{L_2} - \frac{\frac{L_1}{L_2} a_1}{1 + \frac{m_2}{m_1} (\theta_1 - \theta_2)^2}  \\ 
\frac{\mathrm{d} x}{\mathrm{d} t} \\ a \end{bmatrix}  \\
  &= \begin{bmatrix} \frac{\mathrm{d} \theta_1}{\mathrm{d} t} \\ 
(1 + \frac{m_2}{m_1}) \frac{g}{L_1} \theta_1 - \frac{m_2}{m_1} \frac{g}{L_1} \theta_2 - \frac{a}{L_1} + a_1 \\ 
\frac{\mathrm{d} \theta_2}{\mathrm{d} t} \\ 
- (1 + \frac{m_2}{m_1}) \frac{g}{L_2} \theta_1 + (1 + \frac{m_2}{m_1}) \frac{g}{L_2} \theta_2  - \frac{L_1}{L_2} a_1  \\ 
\frac{\mathrm{d} x}{\mathrm{d} t} \\ a \end{bmatrix} \equiv \mathbf{A} \begin{bmatrix} \theta_1 \\ \dot \theta_1 \\ \theta_2 \\ \dot \theta_2 \\ x \\ \dot x \end{bmatrix} + \mathbf{B} \begin{bmatrix} a \\ a_1 \end{bmatrix},
\end{align*}
where
\begin{align*}
\mathbf{A} \equiv \begin{bmatrix} 0 & 1 & 0 & 0 & 0 & 0 \\
(1 + \frac{m_2}{m_1}) \frac{g}{L_1} & 0 & -\frac{m_2}{m_1} \frac{g}{L_1} & 0 & 0 & 0 \\
0 & 0 & 0 & 1 & 0 & 0 \\
-(1 + \frac{m_2}{m_1}) \frac{g}{L_2} & 0 & (1 + \frac{m_2}{m_1}) \frac{g}{L_2} & 0 & 0 & 0 \\
0 & 0 & 0 & 0 & 0 & 1  \\ 0 & 0 & 0 & 0 & 0 & 0 \end{bmatrix}, \qquad \mathbf{B} \equiv \begin{bmatrix} 0 & 0 \\ -\frac{1}{L_1} & 1 \\ 0 & 0 \\ 0 & -\frac{L_1}{L_2} \\ 0 & 0 \\ 1 & 0 \end{bmatrix}.
\end{align*}
Simply substitute 
\begin{align*}
a_1 = 0
\end{align*}
into (\ref{eq:DIP_state_DE_linear_MIMO}) and obtain the linear state differential equation (\ref{eq:DIP_state_DE_linear})
\begin{align*}
\frac{\mathrm{d}}{\mathrm{d} t} \mathbf{x} = \begin{bmatrix} 0 & 1 & 0 & 0 & 0 & 0 \\
(1 + \frac{m_2}{m_1}) \frac{g}{L_1} & 0 & -\frac{m_2}{m_1} \frac{g}{L_1} & 0 & 0 & 0 \\
0 & 0 & 0 & 1 & 0 & 0 \\
-(1 + \frac{m_2}{m_1}) \frac{g}{L_2} & 0 & (1 + \frac{m_2}{m_1}) \frac{g}{L_2} & 0 & 0 & 0 \\
0 & 0 & 0 & 0 & 0 & 1  \\ 0 & 0 & 0 & 0 & 0 & 0 \end{bmatrix} \begin{bmatrix} \theta_1 \\ \dot \theta_1 \\ \theta_2 \\ \dot \theta_2 \\ x \\ \dot x \end{bmatrix} + \begin{bmatrix} 0 \\ -\frac{1}{L_1} \\ 0 \\ 0 \\ 0 \\ 1 \end{bmatrix} a.
\end{align*}

\subsubsection*{Note}

In the previous book \textit{Control Theory For Practical Applications} \cite{Li2024CTPA_Springer, Li2024CTPA_SJTU_1}, nonlinear modelling of double inverted pendulum dynamics follows certain formalism approximation that facilitates derivation as well as computation. So relevant approximated formalisms in the previous book are different from those described in (\ref{eq:DIP_state_DE}) and (\ref{eq:DIP_state_DE_MIMO}). Such approximated formalisms have no essential influence on the double inverted pendulum involved practices presented in the previous book, because the practices presented there after all involve double inverted pendulum dynamics only about the equilibrium state especially about the equilibrium angles. On the other hand, in this book, the author refrains from any formalism approximation adopted in the previous book, but provides nonlinear modelling of genuine dynamics of the double inverted pendulum.

\subsection{Vehicle dynamics}  \label{sec:vehicle_dynamics}

\subsubsection{Negligence of tyre side-slip angles}  \label{sec:vehicle_dynamics_no_slip}

In Section \ref{sec:vehicle_dynamics_no_slip}, we analyse vehicle dynamics without considering tyre side-slip angles. In many practical applications such as low-speed autonomous vehicle navigation illustrated in Figure \ref{fig:low_speed_autonomous_vehicle}, tyre side-slip angles can be fairly neglected. Specification of tyre side-slip angles and analysis of how they are taken into account in vehicle dynamics modelling will be postponed to Section \ref{sec:vehicle_dynamics_with_slip}. Here, simply speaking, the assumption that tyre side-slip angles are neglected can be interpreted as the assumption that actual tyre moving directions are the same to tyre pointing directions.

\subsubsection*{Vehicle complete dynamics}

\begin{figure}[h!]
\begin{center}
\includegraphics[width=0.45\columnwidth]{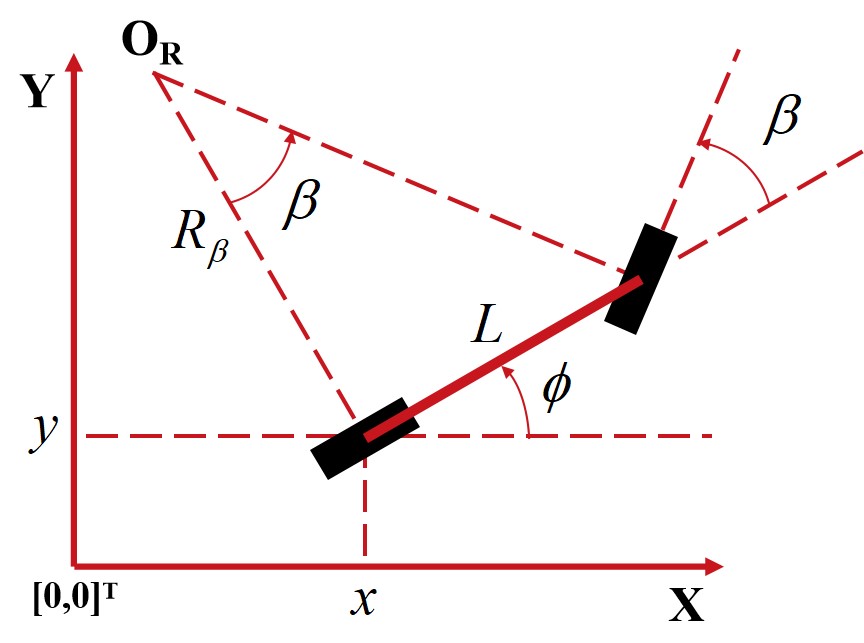}
\end{center}
\caption{Vehicle configuration without tyre side-slip angle}
\label{fig:vehicle_dynamics_no_slip}
\end{figure}

Consider vehicle configuration illustrated in Figure \ref{fig:vehicle_dynamics_no_slip}. The vehicle state
\begin{align*}
\mathbf{x} \equiv \begin{bmatrix} x & y & \phi & \beta \end{bmatrix}^\mathrm{T}
\end{align*}
consists of the vehicle longitudinal position $x$, the vehicle lateral position $y$, the vehicle orientation or heading angle $\phi$ (namely yaw angle), and the vehicle steering angle $\beta$. Besides, $L$ denotes the vehicle wheel-base.

For analysis of vehicle dynamics at constant speed $v$, draw two lines perpendicular to the two wheels respectively. The two lines intersect at the turning center $\mathbf{O}_\mathbf{R}$ (If the vehicle steering angle $\beta$ is zero or in other words if there is no turning action, then the turning center $\mathbf{O}_\mathbf{R}$ is imagined to be located at a virtual point infinitely far away). The turning radius associated with the steering angle $\beta$ is
\begin{align*}
R_{\beta} = \frac{L}{\tan \beta}
\end{align*}
and hence the yaw rate is
\begin{align}  \label{eq:vehicle_phi_dynamics}
\frac{\mathrm{d}}{\mathrm{d} t} \phi = \frac{v}{R_{\beta}} = \frac{v}{L} \tan \beta.
\end{align}
The velocity projections on the horizontal axis $\mathbf{X}$ and the vertical axis $\mathbf{Y}$ are respectively
\begin{align*}
v_x = v \cos \phi, \qquad v_y = v \sin \phi.
\end{align*}
So dynamics of the state elements $x$ and $y$ are described by
\begin{subequations}  \label{eq:vehicle_x+y_dynamics}
\begin{align}
\frac{\mathrm{d}}{\mathrm{d} t} x &= v_x = v \cos \phi, \\
\frac{\mathrm{d}}{\mathrm{d} t} y &= v_y = v \sin \phi.
\end{align}
\end{subequations}

Vehicle steering dynamics is normally described by a first-order differential equation
\begin{align}  \label{eq:vehicle_beta_dynamics}
\tau_{\beta} \frac{\mathrm{d}}{\mathrm{d} t} \beta + \beta = \beta_I \iff \frac{\mathrm{d}}{\mathrm{d} t} \beta = \frac{1}{\tau_{\beta}} (\beta_I - \beta),
\end{align}
where $\beta_I$ denotes the vehicle steering angle command that serves as control input and $\tau_{\beta}$ denotes the time-constant of the steer controller. Combine (\ref{eq:vehicle_phi_dynamics}), (\ref{eq:vehicle_x+y_dynamics}), and (\ref{eq:vehicle_beta_dynamics}) to obtain the state differential equation (\ref{eq:bicycle_kinematics_model})
\begin{align*}
\frac{\mathrm{d}}{\mathrm{d} t} \mathbf{x} \equiv \frac{\mathrm{d}}{\mathrm{d} t} \begin{bmatrix} x \\ y \\ \phi \\ \beta \end{bmatrix} = \begin{bmatrix} v \cos \phi \\ v \sin \phi \\ \frac{v}{L} \tan \beta \\ \frac{1}{\tau_{\beta}} (\beta_I - \beta) \end{bmatrix} \equiv f(\mathbf{x}, \beta_I).
\end{align*}
The model described in (\ref{eq:bicycle_kinematics_model}) is called the \textit{bicycle kinematics model}.

\subsubsection*{Vehicle lateral dynamics}

For analysis of vehicle lateral dynamics, the vehicle lateral position $y$ and the vehicle orientation angle $\phi$ actually refer to the lateral position and orientation angle of the vehicle with respect to certain local road reference. In other words, $y$ and $\phi$ here refer to the relative lateral position and orientation angle in certain local road reference, instead of absolute ones in the global world reference.

For sake of formalizing vehicle lateral dynamics, simply remove $x$ from the vehicle state
\begin{align*}
\begin{bmatrix} x & y & \phi & \beta \end{bmatrix}^\mathrm{T}
\end{align*}
and its corresponding equation from (\ref{eq:bicycle_kinematics_model}) to obtain the state differential equation (\ref{eq:vehicle_lateral_control})
\begin{align*}
\frac{\mathrm{d}}{\mathrm{d} t} \mathbf{x} \equiv \frac{\mathrm{d}}{\mathrm{d} t} \begin{bmatrix} y \\ \phi \\ \beta \end{bmatrix} = \begin{bmatrix} v \sin \phi \\ \frac{v}{L} \tan \beta \\ \frac{1}{\tau_{\beta}} (\beta_I - \beta) \end{bmatrix} \equiv f(\mathbf{x}, \beta_I).
\end{align*}
The vehicle lateral state
\begin{align*}
\mathbf{x} \equiv \begin{bmatrix} y & \phi & \beta \end{bmatrix}^\mathrm{T}
\end{align*}
specified in (\ref{eq:vehicle_lateral_control}) consists of the vehicle lateral position, the vehicle orientation or heading angle (namely yaw angle), and the vehicle steering angle only. The model described in (\ref{eq:vehicle_lateral_control}) is called the \textit{bicycle lateral kinematics model}.

\subsubsection*{Linearized vehicle lateral dynamics}

If the vehicle orientation angle $\phi$ (namely yaw angle) and the vehicle steering angle $\beta$ are close to zero, then following approximations
\begin{align*}
\sin \phi \approx \phi, & \qquad \tan \beta \approx \beta
\end{align*}
can be taken respectively and hence the state differential equation (\ref{eq:vehicle_lateral_control}) can be fairly linearized about the equilibrium state and simplified into the linear state differential equation (\ref{eq:vehicle_lateral_control_approximation})
\begin{align*}
\frac{\mathrm{d}}{\mathrm{d} t} \mathbf{x} = \begin{bmatrix} v \phi \\ \frac{v}{L} \beta \\ \frac{1}{\tau_{\beta}} (\beta_I - \beta) \end{bmatrix} = \begin{bmatrix} 0 & v & 0 \\ 0 & 0 & \frac{v}{L} \\ 0 & 0 & -\frac{1}{\tau_{\beta}} \end{bmatrix} \mathbf{x} + \begin{bmatrix} 0 \\ 0 \\ \frac{1}{\tau_{\beta}} \end{bmatrix} \beta_I \equiv \mathbf{A} \mathbf{x} + \mathbf{B} \beta_I.
\end{align*}
The vehicle lateral state specified in (\ref{eq:vehicle_lateral_control_approximation}) is the same to that specified in (\ref{eq:vehicle_lateral_control}). 

It is worth noting that we may also linearize vehicle complete dynamics described by (\ref{eq:bicycle_kinematics_model}). If so, the state transition matrix $\mathbf{A}$ and the control input matrix $\mathbf{B}$ would no longer be fixed, but will vary according to the concrete state about which the state differential equation (\ref{eq:bicycle_kinematics_model}) is linearized.

\subsubsection{Consideration of tyre side-slip angles}  \label{sec:vehicle_dynamics_with_slip}

\begin{figure}[h!]
\begin{center}
\includegraphics[width=0.45\columnwidth]{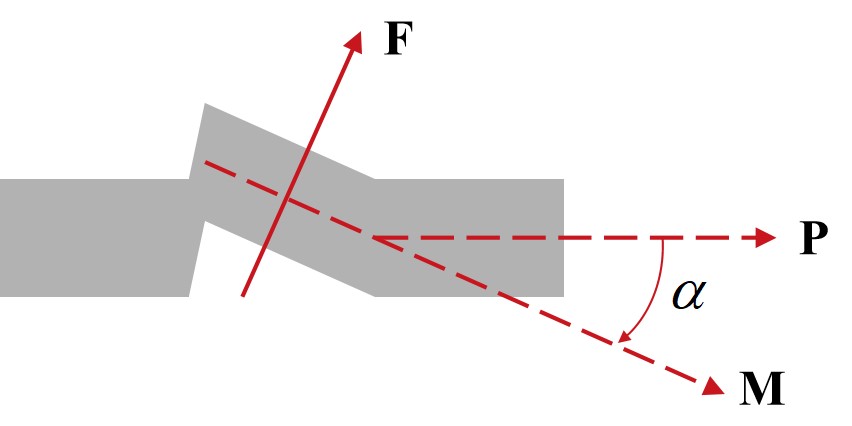}
\end{center}
\caption{Tyre side-slip angle}
\label{fig:tyre_side_slip_angle}
\end{figure}

In practical applications such as high-speed autonomous vehicle navigation, tyre side-slip angles had better be taken into account in vehicle dynamics modelling. For a tyre, the \textit{tyre side-slip angle} is the angle difference between the tyre pointing direction and the tyre moving direction. The tyre side-slip angle of a generic tyre is illustrated in Figure \ref{fig:tyre_side_slip_angle}, where $\mathbf{P}$ denotes the tyre pointing direction, $\mathbf{M}$ denotes the actual tyre moving direction, and $\alpha$ denotes the tyre side-slip angle.

Existence of the tyre side-slip angle is associated with tyre distortion, as illustrated in Figure \ref{fig:tyre_side_slip_angle}. It is right the tyre distortion that provides \textit{tyre force} to enable turning movements of the vehicle. The tyre force is represented by $\mathbf{F}$ in Figure \ref{fig:tyre_side_slip_angle} --- In fact, such function mechanism of the tyre dictates that the tyre side-slip angle always exists. So the assumption that actual tyre moving directions are the same to tyre pointing directions, which is adopted in Section \ref{sec:vehicle_dynamics_no_slip}, can never hold strictly. On the other hand, for the tyre side-slip angle, despite its universal existence, it can fairly be neglected in many practical applications. 

Existence of tyre side-slip angles mainly influences vehicle lateral dynamics. So in Section \ref{sec:vehicle_dynamics_with_slip}, we focus on vehicle lateral dynamics and analyse how tyre side-slip angles are taken into account in relevant modelling.

\subsubsection*{Vehicle lateral dynamics}

A general and comprehensive model that can compute tyre force especially the tyre lateral force part is the \textit{magic formula tyre model} \cite{Pacejka1992}
\begin{equation}  \label{eq:magic_formula_tyre_model}
y(x) = D \sin \{C \arctan [B (x + S_h) - E (B (x + S_h) - \arctan B (x + S_h))]\} + S_v,
\end{equation}
where $x$ denotes input including the side-slip angle, $y$ denotes output including the tyre lateral force, $B$ denotes the stiffness factor, $C$ denotes the shape factor, $D$ denotes the peak factor, $S_h$ denotes the horizontal shift, and $S_v$ denotes the vertical shift. It is worth noting that the magic formula tyre model described by (\ref{eq:magic_formula_tyre_model}) is an empirical model instead of a theoretically-derived model.

In many practical applications, we do not need to resort to the complicated model described by (\ref{eq:magic_formula_tyre_model}), though it enjoys generality and comprehensiveness. Instead, we can resort to a much more concise yet enough effective model \cite{Rajamani2012, Jazar2014}
\begin{equation}  \label{eq:slip_angle_to_force}
F = 2 C \alpha,
\end{equation}
where $\alpha$ denotes the tyre side-slip angle, $C$ denotes the tyre cornering stiffness
\footnote{For a car or moderate vehicle, the tyre cornering stiffness is usually in a range between $30000 \mathrm{N} / \mathrm{rad}$ and $90000 \mathrm{N} / \mathrm{rad}$. The tyre cornering stiffness is an important tyre parameter that determines vehicle manipulation stability: the higher it is, the better the stability is.}, 
and $F$ denotes the \textit{tyre lateral force} (we abuse the notation $F$ to denote tyre lateral force only, as we focus on vehicle lateral dynamics).

\begin{figure}[h!]
\begin{center}
\includegraphics[width=0.45\columnwidth]{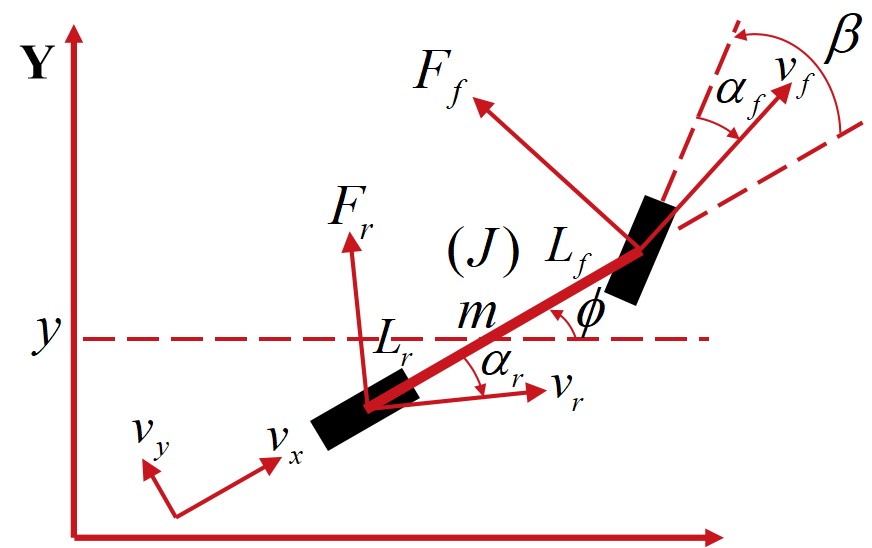}
\end{center}
\caption{Vehicle lateral configuration with tyre side-slip angles}
\label{fig:vehicle_dynamics_with_slip}
\end{figure}

Consider the vehicle lateral configuration with tyre side-slip angles, as illustrated in Figure \ref{fig:vehicle_dynamics_with_slip}. Here, $m$ denotes the vehicle mass, $J$ denotes the rotating inertia of the vehicle, $y$ denotes the vehicle lateral position with respect to certain local road reference, $\phi$ denotes the vehicle orientation or heading angle (namely yaw angle) with respect to the local road reference, $\beta$ denotes the vehicle steering angle, $L_f$ denotes the length between the vehicle mass center or gravity center and the front wheel, $L_r$ denotes the length between the vehicle gravity center and the rear wheel, the sum of $L_f$ and $L_r$ namely
\begin{align*}
L = L_f + L_r
\end{align*}
denotes the vehicle wheel-base, $\alpha_f$ denotes the front tyre side-slip angle, $F_f$ denotes the front tyre lateral force, $v_f$ denotes the actual velocity and moving direction of the front tyre, $\alpha_r$ denotes the rear tyre side-slip angle, $F_r$ denotes the rear tyre lateral force, $v_r$ denotes the actual velocity and moving direction of the rear tyre, and $v_x$ and $v_y$ denote respectively the longitudinal velocity and the lateral velocity of the vehicle gravity center with respect to the vehicle body itself. 

For practical applications that necessitate consideration of tyre side-slip angles, the yaw angle $\phi$ is usually small and following approximations
\begin{align*}
v_x \approx v, \qquad v_y \approx \dot y - v \phi, \qquad \dot v_y \approx \ddot y - v \dot \phi
\end{align*} 
tend to hold effectively. Establish two kinematics equations associated with the front tyre velocity $v_f$ and the rear tyre velocity $v_r$ as
\begin{subequations}  \label{eq:vehicle_slip_dynamics_vf+vr}
\begin{align}
\tan (\beta - \alpha_f) &= \frac{v_{fy}}{v_{fx}} = \frac{v_y + L_f \dot \phi}{v_x} = \frac{\dot y - v \phi + L_f \dot \phi}{v}  \nonumber \\ 
\iff \alpha_f &= \beta - \arctan \frac{\dot y - v \phi + L_f \dot \phi}{v},   \\
\tan (- \alpha_r) &= \frac{v_{ry}}{v_{rx}} = \frac{v_y - L_r \dot \phi}{v_x} = \frac{\dot y - v \phi - L_r \dot \phi}{v}  \nonumber \\ 
\iff \alpha_r &= - \arctan \frac{\dot y - v \phi - L_r \dot \phi}{v},
\end{align}
\end{subequations}
where $v$ denotes the vehicle velocity and 
\begin{align*}
\dot \phi \equiv \frac{\mathrm{d} \phi}{\mathrm{d} t}
\end{align*}
denotes the vehicle yaw rate.

According to the tyre lateral force model described by (\ref{eq:slip_angle_to_force}), the front tyre lateral force $F_f$ and the rear tyre lateral force $F_r$ are respectively
\begin{subequations}  \label{eq:tyre_lateral_force_Ff+Fr}
\begin{align}
F_f &= 2 C_f \alpha_f \implies F_{fy} = 2 C_f \alpha_f \cos (\beta - \alpha_f),   \\
F_r &= 2 C_r \alpha_r \implies F_{ry} = 2 C_r \alpha_r \cos \alpha_r,
\end{align}
\end{subequations}
where $C_f$ and $C_r$ denote the front tyre cornering stiffness and the rear tyre cornering stiffness respectively. The vehicle lateral acceleration consists of two parts: one is the translational acceleration $\dot v_y$ and the other is the centripetal acceleration $v_x \dot \phi$. Substitute (\ref{eq:vehicle_slip_dynamics_vf+vr}) and (\ref{eq:tyre_lateral_force_Ff+Fr}) into the force-acceleration relationship
\begin{align*}
m (\dot v_y + v_x \dot \phi) = F_{fy} + F_{ry}
\end{align*}
and obtain
\begin{align}  \label{eq:vehicle_slip_dynamics_ddy}
m (\ddot y - v \dot \phi + v \dot \phi) &= 2 C_f \alpha_f \cos (\beta - \alpha_f) + 2 C_r \alpha_r \cos \alpha_r \iff  \nonumber \\
\ddot y &= \frac{2 C_f}{m} (\beta - \arctan \frac{\dot y - v \phi + L_f \dot \phi}{v}) \cos (\arctan \frac{\dot y - v \phi + L_f \dot \phi}{v}) +  \nonumber \\
  &\qquad \frac{2 C_r}{m} (- \arctan \frac{\dot y - v \phi - L_r \dot \phi}{v}) \cos (\arctan \frac{\dot y - v \phi - L_r \dot \phi}{v}) \nonumber \\
  &= \frac{2 C_f}{m} \frac{v (\beta - \arctan \frac{\dot y - v \phi + L_f \dot \phi}{v})}{\sqrt{(\dot y - v \phi + L_f \dot \phi)^2 + v^2}} + \frac{2 C_r}{m} \frac{v (- \arctan \frac{\dot y - v \phi - L_r \dot \phi}{v})}{\sqrt{(\dot y - v \phi - L_r \dot \phi)^2 + v^2}}.
\end{align}

Substitute (\ref{eq:vehicle_slip_dynamics_vf+vr}) and (\ref{eq:tyre_lateral_force_Ff+Fr}) into the torque-angular acceleration relationship
\begin{align*}
J \ddot \phi = F_{fy} L_f - F_{ry} L_r
\end{align*}
and obtain
\begin{align}  \label{eq:vehicle_slip_dynamics_ddphi}
J \ddot \phi &= 2 C_f \alpha_f \cos (\beta - \alpha_f) L_f - 2 C_r \alpha_r \cos \alpha_r L_r \iff  \nonumber \\
\ddot \phi &= \frac{2 C_f L_f}{J} (\beta - \arctan \frac{\dot y - v \phi + L_f \dot \phi}{v}) \cos (\arctan \frac{\dot y - v \phi + L_f \dot \phi}{v}) -  \nonumber \\
  &\qquad \frac{2 C_r L_r}{J} (- \arctan \frac{\dot y - v \phi - L_r \dot \phi}{v}) \cos (\arctan \frac{\dot y - v \phi - L_r \dot \phi}{v}) \nonumber \\
  &= \frac{2 C_f L_f}{J} \frac{v (\beta - \arctan \frac{\dot y - v \phi + L_f \dot \phi}{v})}{\sqrt{(\dot y - v \phi + L_f \dot \phi)^2 + v^2}} -  \frac{2 C_r L_r}{J} \frac{v (- \arctan \frac{\dot y - v \phi - L_r \dot \phi}{v})}{\sqrt{(\dot y - v \phi - L_r \dot \phi)^2 + v^2}}.
\end{align}

Combine (\ref{eq:vehicle_beta_dynamics}), (\ref{eq:vehicle_slip_dynamics_ddy}), (\ref{eq:vehicle_slip_dynamics_ddphi}), and two trivial equations
\begin{align*}
\frac{\mathrm{d}}{\mathrm{d} t} y = \dot y \equiv \frac{\mathrm{d} y}{\mathrm{d} t}, \qquad \frac{\mathrm{d}}{\mathrm{d} t} \phi = \dot \phi \equiv \frac{\mathrm{d} \phi}{\mathrm{d} t}
\end{align*}
to obtain the state differential equation (\ref{eq:vehicle_lateral_dynamics_with_slip})
\begin{align*}
\frac{\mathrm{d}}{\mathrm{d} t} \mathbf{x} \equiv \frac{\mathrm{d}}{\mathrm{d} t} \begin{bmatrix} y \\ \frac{\mathrm{d} y}{\mathrm{d} t} \\ \phi \\ \frac{\mathrm{d} \phi}{\mathrm{d} t} \\ \beta \end{bmatrix} 
= \begin{bmatrix} \frac{\mathrm{d} y}{\mathrm{d} t} \\ 
\frac{2 C_f}{m} \frac{v (\beta - \arctan \frac{\dot y - v \phi + L_f \dot \phi}{v})}{\sqrt{(\dot y - v \phi + L_f \dot \phi)^2 + v^2}} + \frac{2 C_r}{m} \frac{v (- \arctan \frac{\dot y - v \phi - L_r \dot \phi}{v})}{\sqrt{(\dot y - v \phi - L_r \dot \phi)^2 + v^2}} \\ 
\frac{\mathrm{d} \phi}{\mathrm{d} t} \\ 
\frac{2 C_f L_f}{J} \frac{v (\beta - \arctan \frac{\dot y - v \phi + L_f \dot \phi}{v})}{\sqrt{(\dot y - v \phi + L_f \dot \phi)^2 + v^2}} -  \frac{2 C_r L_r}{J} \frac{v (- \arctan \frac{\dot y - v \phi - L_r \dot \phi}{v})}{\sqrt{(\dot y - v \phi - L_r \dot \phi)^2 + v^2}} \\
\frac{1}{\tau_{\beta}} (\beta_I - \beta) \end{bmatrix} \equiv f(\mathbf{x}, \beta_I),
\end{align*}
where the vehicle lateral state 
\begin{align*}
\mathbf{x} \equiv \begin{bmatrix} y & \frac{\mathrm{d} y}{\mathrm{d} t} & \phi & \frac{\mathrm{d} \phi}{\mathrm{d} t} & \beta \end{bmatrix}^\mathrm{T}
\end{align*}
namely
\begin{align*}
\mathbf{x} \equiv \begin{bmatrix} y & \dot y & \phi & \dot \phi & \beta \end{bmatrix}^\mathrm{T}
\end{align*}
consists of the vehicle lateral position with respect to certain local road reference, the vehicle lateral velocity in the local road reference, the vehicle orientation or heading angle (namely yaw angle) with respect to the local road reference, the vehicle yaw rate, and the vehicle steering angle. $\beta_I$ denotes the vehicle steering angle command which serves as control input.

\subsubsection*{Linearized vehicle lateral dynamics}

If relevant angles involved in (\ref{eq:vehicle_lateral_dynamics_with_slip}) are close to zero, then following approximations
\begin{align*}
\arctan \frac{\dot y - v \phi + L_f \dot \phi}{v} &= \beta - \alpha_f \approx \tan (\beta - \alpha_f) = \frac{\dot y - v \phi + L_f \dot \phi}{v},  \\
\arctan \frac{\dot y - v \phi - L_r \dot \phi}{v} &= - \alpha_r \approx \tan (- \alpha_r) = \frac{\dot y - v \phi - L_r \dot \phi}{v},  \\
\frac{v}{\sqrt{(\dot y - v \phi + L_f \dot \phi)^2 + v^2}} &= \cos (\beta - \alpha_f) \approx 1,  \\
\frac{v}{\sqrt{(\dot y - v \phi - L_r \dot \phi)^2 + v^2}} &= \cos \alpha_r \approx 1
\end{align*}
can be taken respectively and hence the state differential equation (\ref{eq:vehicle_lateral_dynamics_with_slip}) can be fairly linearized about the equilibrium state and simplified into the linear state differential equation (\ref{eq:vehicle_lateral_dynamics_with_slip_linear})
\begin{align*}
\frac{\mathrm{d}}{\mathrm{d} t} \mathbf{x} &= \begin{bmatrix} \frac{\mathrm{d} y}{\mathrm{d} t} \\ 
\frac{2 C_f}{m} (\beta - \frac{\dot y - v \phi + L_f \dot \phi}{v}) + \frac{2 C_r}{m} (- \frac{\dot y - v \phi - L_r \dot \phi}{v}) \\ 
\frac{\mathrm{d} \phi}{\mathrm{d} t} \\ 
\frac{2 C_f L_f}{J} (\beta - \frac{\dot y - v \phi + L_f \dot \phi}{v}) -  \frac{2 C_r L_r}{J} (- \frac{\dot y - v \phi - L_r \dot \phi}{v})  \\
\frac{1}{\tau_{\beta}} (\beta_I - \beta) \end{bmatrix}  \\
  &= \begin{bmatrix} 0 & 1 & 0 & 0 & 0 \\
0 & - \frac{2 C_f + 2 C_r}{m v} & \frac{2 C_f + 2 C_r}{m} & - \frac{2 C_f L_f - 2 C_r L_r}{m v} & \frac{2 C_f}{m} \\ 
0 & 0 & 0 & 1 & 0 \\
0 & - \frac{2 C_f L_f - 2 C_r L_r}{J v} & \frac{2 C_f L_f - 2 C_r L_r}{J} & - \frac{2 C_f L_f^2 + 2 C_r L_r^2}{J v} & \frac{2 C_f L_f}{J} \\
0 & 0 & 0 & 0 & - \frac{1}{\tau_{\beta}} \end{bmatrix} \mathbf{x} + \begin{bmatrix} 0 \\ 0 \\ 0 \\ 0 \\ \frac{1}{\tau_{\beta}} \end{bmatrix} \beta_I  \\ 
  &\equiv \mathbf{A} \mathbf{x} + \mathbf{B} \beta_I.
\end{align*}
The vehicle lateral state specified in (\ref{eq:vehicle_lateral_dynamics_with_slip_linear}) is the same to that specified in (\ref{eq:vehicle_lateral_dynamics_with_slip}).

\subsubsection{Constraint of vehicle steering dynamics}  \label{sec:vehicle_dynamics_constrain_steering}

When there is a considerable difference between the vehicle steering angle command and current vehicle steering angle, vehicle steering dynamics described by the first-order differential equation (\ref{eq:vehicle_beta_dynamics})
\begin{align*}
\tau_{\beta} \frac{\mathrm{d}}{\mathrm{d} t} \beta + \beta = \beta_I \iff \frac{\mathrm{d}}{\mathrm{d} t} \beta = \frac{1}{\tau_{\beta}} (\beta_I - \beta)
\end{align*}
is constrained by the maximum steering velocity and (\ref{eq:vehicle_beta_dynamics}) is augmented to
\begin{equation}  \label{eq:vehicle_beta_dynamics_constrained}
\frac{\mathrm{d}}{\mathrm{d} t} \beta = \max\{ \min\{ \frac{1}{\tau_{\beta}} (\beta_I - \beta), s_M \}, -s_M \},
\end{equation}
where $\beta$ denotes the vehicle steering angle, $\beta_I$ denotes the vehicle steering angle command that serves as control input, $\tau_{\beta}$ denotes the time-constant of the steer controller, and $s_M$ denotes the maximum steering velocity. If constraint of vehicle steering dynamics is considered, then vehicle dynamics models presented in Section \ref{sec:vehicle_dynamics_no_slip} and Section \ref{sec:vehicle_dynamics_with_slip} have their augmented versions respectively.

\subsubsection*{Negligence of tyre side-slip angles}

The augmented version of the vehicle complete dynamics model described by (\ref{eq:bicycle_kinematics_model}) is
\begin{align}  \label{eq:bicycle_kinematics_model_constrain_steering}
\frac{\mathrm{d}}{\mathrm{d} t} \mathbf{x} \equiv \frac{\mathrm{d}}{\mathrm{d} t} \begin{bmatrix} x \\ y \\ \phi \\ \beta \end{bmatrix} = \begin{bmatrix} v \cos \phi \\ v \sin \phi \\ \frac{v}{L} \tan \beta \\ \max\{ \min\{ \frac{1}{\tau_{\beta}} (\beta_I - \beta), s_M \}, -s_M \} \end{bmatrix} \equiv f(\mathbf{x}, \beta_I).
\end{align}
The vehicle state specified in (\ref{eq:bicycle_kinematics_model_constrain_steering}) is the same to that specified in (\ref{eq:bicycle_kinematics_model}).

The augmented version of the vehicle lateral dynamics model described by (\ref{eq:vehicle_lateral_control}) is
\begin{align}  \label{eq:vehicle_lateral_control_constrain_steering}
\frac{\mathrm{d}}{\mathrm{d} t} \mathbf{x} \equiv \frac{\mathrm{d}}{\mathrm{d} t} \begin{bmatrix} y \\ \phi \\ \beta \end{bmatrix} = \begin{bmatrix} v \sin \phi \\ \frac{v}{L} \tan \beta \\ \max\{ \min\{ \frac{1}{\tau_{\beta}} (\beta_I - \beta), s_M \}, -s_M \} \end{bmatrix} \equiv f(\mathbf{x}, \beta_I).
\end{align}
The vehicle lateral state specified in (\ref{eq:vehicle_lateral_control_constrain_steering}) is the same to that specified in (\ref{eq:vehicle_lateral_control}).

The constrained vehicle steering dynamics formalized as (\ref{eq:vehicle_beta_dynamics_constrained}) is by nature nonlinear, so the linearized vehicle lateral dynamics model described by (\ref{eq:vehicle_lateral_control_approximation}) has no augmented counterpart similar to those by (\ref{eq:bicycle_kinematics_model_constrain_steering}) and (\ref{eq:vehicle_lateral_control_constrain_steering}).

\subsubsection*{Consideration of tyre side-slip angles}

The augmented version of the vehicle lateral dynamics model described by (\ref{eq:vehicle_lateral_dynamics_with_slip}) is
\begin{align}  \label{eq:vehicle_lateral_dynamics_with_slip_constrain_steering}
\frac{\mathrm{d}}{\mathrm{d} t} \mathbf{x} \equiv \frac{\mathrm{d}}{\mathrm{d} t} \begin{bmatrix} y \\ \frac{\mathrm{d} y}{\mathrm{d} t} \\ \phi \\ \frac{\mathrm{d} \phi}{\mathrm{d} t} \\ \beta \end{bmatrix} 
= \begin{bmatrix} \frac{\mathrm{d} y}{\mathrm{d} t} \\ 
\frac{2 C_f}{m} \frac{v (\beta - \arctan \frac{\dot y - v \phi + L_f \dot \phi}{v})}{\sqrt{(\dot y - v \phi + L_f \dot \phi)^2 + v^2}} + \frac{2 C_r}{m} \frac{v (- \arctan \frac{\dot y - v \phi - L_r \dot \phi}{v})}{\sqrt{(\dot y - v \phi - L_r \dot \phi)^2 + v^2}} \\ 
\frac{\mathrm{d} \phi}{\mathrm{d} t} \\ 
\frac{2 C_f L_f}{J} \frac{v (\beta - \arctan \frac{\dot y - v \phi + L_f \dot \phi}{v})}{\sqrt{(\dot y - v \phi + L_f \dot \phi)^2 + v^2}} -  \frac{2 C_r L_r}{J} \frac{v (- \arctan \frac{\dot y - v \phi - L_r \dot \phi}{v})}{\sqrt{(\dot y - v \phi - L_r \dot \phi)^2 + v^2}} \\
\max\{ \min\{ \frac{1}{\tau_{\beta}} (\beta_I - \beta), s_M \}, -s_M \} \end{bmatrix} \equiv f(\mathbf{x}, \beta_I).
\end{align}
The vehicle lateral state specified in (\ref{eq:vehicle_lateral_dynamics_with_slip_constrain_steering}) is the same to that specified in (\ref{eq:vehicle_lateral_dynamics_with_slip}).

It is worth noting that consideration of tyre side-slip angles tends to be involved in high-speed vehicle lateral control, where the vehicle steering angle is normally small and changes smoothly for sake of driving safety. In such circumstance, we can fairly remove the constraints from the vehicle steering dynamics model and just adopt (\ref{eq:vehicle_beta_dynamics}).

\subsubsection{Negligence of vehicle steering dynamics}  \label{sec:vehicle_dynamics_neglect_steering}

Unlike modelling of motorcycle (or bicycle) dynamics which will be presented in Section \ref{sec:motorcycle_dynamics}, modelling of vehicle dynamics had better take vehicle steering dynamics into account. But if vehicle steering dynamics is neglected, then vehicle dynamics models presented in Section \ref{sec:vehicle_dynamics_no_slip} and Section \ref{sec:vehicle_dynamics_with_slip} have their reduced versions respectively. In all the reduced versions of vehicle dynamics models presented in Section \ref{sec:vehicle_dynamics_neglect_steering}, the steering angle $\beta$ is removed from the state and instead serves directly as control input, i.e. 
\begin{align*}
\beta \equiv \beta_I.
\end{align*}

\subsubsection*{Negligence of tyre side-slip angles}

The reduced version of the vehicle complete dynamics model described by (\ref{eq:bicycle_kinematics_model}) is
\begin{align}  \label{eq:bicycle_kinematics_model_neglect_steering}
\frac{\mathrm{d}}{\mathrm{d} t} \mathbf{x} \equiv \frac{\mathrm{d}}{\mathrm{d} t} \begin{bmatrix} x \\ y \\ \phi \end{bmatrix} = \begin{bmatrix} v \cos \phi \\ v \sin \phi \\ \frac{v}{L} \tan \beta \end{bmatrix} \equiv f(\mathbf{x}, \beta),
\end{align}
where the vehicle state
\begin{align*}
\mathbf{x} \equiv \begin{bmatrix} x & y & \phi \end{bmatrix}^\mathrm{T}
\end{align*}
consists of the vehicle longitudinal position, the vehicle lateral position, and the vehicle orientation or heading angle (namely yaw angle).

The reduced version of the vehicle lateral dynamics model described by (\ref{eq:vehicle_lateral_control}) is
\begin{align}  \label{eq:vehicle_lateral_control_neglect_steering}
\frac{\mathrm{d}}{\mathrm{d} t} \mathbf{x} \equiv \frac{\mathrm{d}}{\mathrm{d} t} \begin{bmatrix} y \\ \phi \end{bmatrix} = \begin{bmatrix} v \sin \phi \\ \frac{v}{L} \tan \beta \end{bmatrix} \equiv f(\mathbf{x}, \beta),
\end{align}
where the vehicle lateral state
\begin{align*}
\mathbf{x} \equiv \begin{bmatrix} y & \phi \end{bmatrix}^\mathrm{T}
\end{align*}
consists of the vehicle lateral position and the vehicle orientation or heading angle (namely yaw angle) only.

The reduced version of the linearized vehicle lateral dynamics model described by (\ref{eq:vehicle_lateral_control_approximation}) is
\begin{align}  \label{eq:vehicle_lateral_control_approximation_neglect_steering}
\frac{\mathrm{d}}{\mathrm{d} t} \mathbf{x} = \begin{bmatrix} 0 & v \\ 0 & 0 \end{bmatrix} \mathbf{x} + \begin{bmatrix} 0 \\ \frac{v}{L} \end{bmatrix} \beta \equiv \mathbf{A} \mathbf{x} + \mathbf{B} \beta,
\end{align}
where the vehicle lateral state is the same to that specified in (\ref{eq:vehicle_lateral_control_neglect_steering}).

\subsubsection*{Consideration of tyre side-slip angles}

The reduced version of the vehicle lateral dynamics model described by (\ref{eq:vehicle_lateral_dynamics_with_slip}) is
\begin{align}  \label{eq:vehicle_lateral_dynamics_with_slip_neglect_steering}
\frac{\mathrm{d}}{\mathrm{d} t} \mathbf{x} \equiv \frac{\mathrm{d}}{\mathrm{d} t} \begin{bmatrix} y \\ \frac{\mathrm{d} y}{\mathrm{d} t} \\ \phi \\ \frac{\mathrm{d} \phi}{\mathrm{d} t} \end{bmatrix} 
= \begin{bmatrix} \frac{\mathrm{d} y}{\mathrm{d} t} \\ 
\frac{2 C_f}{m} \frac{v (\beta - \arctan \frac{\dot y - v \phi + L_f \dot \phi}{v})}{\sqrt{(\dot y - v \phi + L_f \dot \phi)^2 + v^2}} + \frac{2 C_r}{m} \frac{v (- \arctan \frac{\dot y - v \phi - L_r \dot \phi}{v})}{\sqrt{(\dot y - v \phi - L_r \dot \phi)^2 + v^2}} \\ 
\frac{\mathrm{d} \phi}{\mathrm{d} t} \\ 
\frac{2 C_f L_f}{J} \frac{v (\beta - \arctan \frac{\dot y - v \phi + L_f \dot \phi}{v})}{\sqrt{(\dot y - v \phi + L_f \dot \phi)^2 + v^2}} -  \frac{2 C_r L_r}{J} \frac{v (- \arctan \frac{\dot y - v \phi - L_r \dot \phi}{v})}{\sqrt{(\dot y - v \phi - L_r \dot \phi)^2 + v^2}} \end{bmatrix} \equiv f(\mathbf{x}, \beta),
\end{align}
where the vehicle lateral state 
\begin{align*}
\mathbf{x} \equiv \begin{bmatrix} y & \frac{\mathrm{d} y}{\mathrm{d} t} & \phi & \frac{\mathrm{d} \phi}{\mathrm{d} t} \end{bmatrix}^\mathrm{T}
\end{align*}
namely
\begin{align*}
\mathbf{x} \equiv \begin{bmatrix} y & \dot y & \phi & \dot \phi \end{bmatrix}^\mathrm{T}
\end{align*}
consists of the vehicle lateral position with respect to certain local road reference, the vehicle lateral velocity in the local road reference, the vehicle orientation or heading angle (namely yaw angle) with respect to the local road reference, and the vehicle yaw rate.

The reduced version of the linearized vehicle lateral dynamics model described by (\ref{eq:vehicle_lateral_dynamics_with_slip_linear}) is
\begin{align}  \label{eq:vehicle_lateral_dynamics_with_slip_linear_neglect_steering}
\frac{\mathrm{d}}{\mathrm{d} t} \mathbf{x} &= \begin{bmatrix} 0 & 1 & 0 & 0 \\
0 & - \frac{2 C_f + 2 C_r}{m v} & \frac{2 C_f + 2 C_r}{m} & - \frac{2 C_f L_f - 2 C_r L_r}{m v} \\ 
0 & 0 & 0 & 1 \\
0 & - \frac{2 C_f L_f - 2 C_r L_r}{J v} & \frac{2 C_f L_f - 2 C_r L_r}{J} & - \frac{2 C_f L_f^2 + 2 C_r L_r^2}{J v} \end{bmatrix} \mathbf{x} + \begin{bmatrix} 0 \\ \frac{2 C_f}{m} \\ 0 \\ \frac{2 C_f L_f}{J} \end{bmatrix} \beta \\
  &\equiv \mathbf{A} \mathbf{x} + \mathbf{B} \beta.  \nonumber
\end{align}
The vehicle lateral state specified in (\ref{eq:vehicle_lateral_dynamics_with_slip_linear_neglect_steering}) is the same to that specified in (\ref{eq:vehicle_lateral_dynamics_with_slip_linear}).

\subsubsection{Vehicle longitudinal dynamics}  \label{sec:vehicle_longitudinal_dynamics}

Like vehicle steering dynamics is normally described by the first-order differential equation (\ref{eq:vehicle_beta_dynamics})
\begin{align*}
\tau_{\beta} \frac{\mathrm{d}}{\mathrm{d} t} \beta + \beta = \beta_I \iff \frac{\mathrm{d}}{\mathrm{d} t} \beta = \frac{1}{\tau_{\beta}} (\beta_I - \beta),
\end{align*}
vehicle longitudinal dynamics can also be described by a first-order differential equation
\begin{equation}  \label{eq:vehicle_longitudinal_dynamics}
\tau_v \frac{\mathrm{d}}{\mathrm{d} t} v + v = v_I \iff \frac{\mathrm{d}}{\mathrm{d} t} v = \frac{1}{\tau_v} (v_I - v),
\end{equation}
where $v$ denotes the vehicle velocity and $v_I$ denotes the vehicle velocity command.

If constraint of vehicle longitudinal dynamics namely the maximum vehicle acceleration (or deceleration) is taken into account, the vehicle longitudinal dynamics model (\ref{eq:vehicle_longitudinal_dynamics}) is augmented to the constrained version
\begin{equation}  \label{eq:vehicle_longitudinal_dynamics_constrain}
\frac{\mathrm{d}}{\mathrm{d} t} v = \max\{ \min\{ \frac{1}{\tau_v} (v_I - v), a_M \}, -a_M \}.
\end{equation}
Integrate (\ref{eq:vehicle_longitudinal_dynamics_constrain}) into (\ref{eq:bicycle_kinematics_model_constrain_steering}) namely the augmented version of (\ref{eq:bicycle_kinematics_model}) to obtain the even augmented version of the vehicle complete dynamics model
\begin{align}  \label{eq:IV_state_DE_complete_constrain}
\frac{\mathrm{d}}{\mathrm{d} t} \mathbf{x} \equiv \frac{\mathrm{d}}{\mathrm{d} t} \begin{bmatrix} x \\ y \\ \phi \\ \beta \\ v \end{bmatrix} = \begin{bmatrix} v \cos \phi \\ v \sin \phi \\ \frac{v}{L} \tan \beta \\ \max\{ \min\{ \frac{1}{\tau_{\beta}} (\beta_I - \beta), s_M \}, -s_M \} \\ \max\{ \min\{ \frac{1}{\tau_v} (v_I - v), a_M \}, -a_M \} \end{bmatrix} \equiv f(\mathbf{x}, \mathbf{u}),
\end{align}
where the state 
\begin{align*}
\mathbf{x} \equiv \begin{bmatrix} x & y & \phi & \beta & v \end{bmatrix}^\mathrm{T}
\end{align*}
consists of the vehicle longitudinal position, the vehicle lateral position, the vehicle orientation or heading angle (namely yaw angle), the vehicle steering angle, and the vehicle velocity. Besides, $L$ denotes the vehicle wheel-base, $\tau_{\beta}$ denotes the time-constant of the steer controller, $s_M$ denotes the maximum steering velocity, $\tau_v$ denotes the time-constant of the velocity controller, and $a_M$ denotes the maximum vehicle acceleration (or deceleration). The control input
\begin{align*}
\mathbf{u} \equiv \begin{bmatrix} \beta_I & v_I \end{bmatrix}^\mathrm{T}
\end{align*}
is the multiple-input of vehicle steering angle command $\beta_I$ and vehicle velocity command $v_I$.

The vehicle longitudinal dynamics models (\ref{eq:vehicle_longitudinal_dynamics}) and (\ref{eq:vehicle_longitudinal_dynamics_constrain}) can also be integrated into other vehicle dynamics models presented in Section \ref{sec:vehicle_dynamics_no_slip}, Section \ref{sec:vehicle_dynamics_with_slip}, Section \ref{sec:vehicle_dynamics_constrain_steering}, and Section \ref{sec:vehicle_dynamics_neglect_steering}, yet details are omitted here. It is worth noting that in practical applications, vehicle longitudinal dynamics is usually decoupled out from vehicle complete dynamics and accordingly the vehicle longitudinal control system is treated as an independent sub-component of the overall vehicle control system. This is why the vehicle velocity is by default treated as a pre-defined parameter in previously presented vehicle dynamics models. 

\subsection{Motorcycle (or bicycle) dynamics}  \label{sec:motorcycle_dynamics}

Motorcycle (or bicycle) dynamics belongs to the category of low-speed dynamics from the perspective of vehicle dynamics
\footnote{In daily life, the normal speed range for bicycles is between $15$ $km / h$ and $25$ $km / h$, whereas the normal speed range for motorcycles is between $30$ $km / h$ and $45$ $km / h$. Such speeds are rather low from the perspective of vehicles. Special motorcycles such as heavy motorcycles and racing motorcycles can achieve speeds much higher, yet they are beyond consideration here.}, 
so like modelling presented in Section \ref{sec:vehicle_dynamics_no_slip}, modelling of motorcycle (or bicycle) dynamics does not take tyre side-slip angles into account either. Besides, motorcycles (or bicycles) have two extra special points that distinguish themselves from vehicles: motorcycles (or bicycles) tend to have smooth steering operations
\footnote{This point will be further discussed in Section \ref{sec:motorcycle_dynamics_neglect_steering}.}, 
and motorcycles (or bicycles) are rather light in comparison with vehicles. The two extra special points further enhance fairness of neglecting tyre side-slip angles in modelling of motorcycle (or bicycle) dynamics.

\subsubsection{Consideration of motorcycle steering dynamics}  \label{sec:motorcycle_dynamics_consider_steering}

Consider dynamics of the motorcycle (or bicycle) illustrated in Figure \ref{fig:motorcycle_control}. Dynamics of the motorcycle is separated into two parts, namely the part of \textit{motorcycle horizontal dynamics} that is similar to vehicle dynamics and the part of \textit{motorcycle vertical dynamics} that is similar to single inverter pendulum dynamics.

\subsubsection*{Motorcycle complete dynamics}

The motorcycle state
\begin{align*}
\mathbf{x} \equiv \begin{bmatrix} x & y & \phi & \beta & \theta & \frac{\mathrm{d} \theta}{\mathrm{d} t} \end{bmatrix}^\mathrm{T}
\end{align*} 
consists of the motorcycle longitudinal position $x$, the motorcycle lateral position $y$, the motorcycle orientation or heading angle $\phi$ (namely yaw angle), the motorcycle steering angle $\beta$, the motorcycle vertical angle $\theta$ (namely roll angle), and the motorcycle vertical angular velocity 
\begin{align*}
\dot \theta \equiv \frac{\mathrm{d} \theta}{\mathrm{d} t}. 
\end{align*} 
The state elements $x$, $y$, and $\phi$ are involved only in motorcycle horizontal dynamics. The state elements $\theta$ and $\frac{\mathrm{d} \theta}{\mathrm{d} t}$ are involved only in motorcycle vertical dynamics. The state element $\beta$ is involved in both motorcycle horizontal dynamics and motorcycle vertical dynamics. The control input is the motorcycle steering angle command $\beta_I$. Besides, for motorcycle parameters, $L$ denotes the motorcycle wheel-base, $H$ denotes the height of the motorcycle gravity center, and $\tau_{\beta}$ denotes the time-constant of the steer controller.

\begin{figure}[h!]
\begin{center}
\includegraphics[width=0.15\columnwidth]{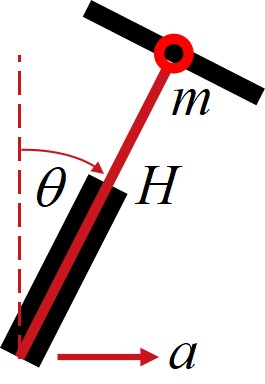}
\end{center}
\caption{Motorcycle lateral configuration}
\label{fig:motorcycle_lateral_configuration}
\end{figure}

Motorcycle horizontal dynamics at constant speed $v$ can be modelled by differential equations similar to those described in (\ref{eq:bicycle_kinematics_model}) as
\begin{subequations}  \label{eq:motorcycle_horizontal_dynamics}
\begin{align}
\frac{\mathrm{d}}{\mathrm{d} t} x &= v \cos \phi, \\
\frac{\mathrm{d}}{\mathrm{d} t} y &= v \sin \phi, \\
\frac{\mathrm{d}}{\mathrm{d} t} \phi &= \frac{v}{L} \tan \beta, \\
\frac{\mathrm{d}}{\mathrm{d} t} \beta &= \frac{1}{\tau_{\beta}} (\beta_I - \beta).
\end{align}
\end{subequations}

For analysis of motorcycle vertical dynamics at constant speed $v$, consider motorcycle lateral configuration illustrated in Figure \ref{fig:motorcycle_lateral_configuration}. The turning radius associated with the steering angle $\beta$ is
\begin{align*}
R_{\beta} = \frac{L}{\tan \beta}
\end{align*}
and hence the centripetal acceleration is
\begin{align*}
a = \frac{v^2}{R_{\beta}} = \frac{v^2}{L} \tan \beta.
\end{align*}
By comparing Figure \ref{fig:motorcycle_lateral_configuration} and Figure \ref{fig:inverted_pendulum_control}, we can observe that the motorcycle lateral configuration is essentially the same to that of a single inverted pendulum. So motorcycle vertical dynamics at constant speed $v$ can be modelled by differential equations similar to those described in (\ref{eq:SIP_state_DE}) as
\begin{subequations}  \label{eq:motorcycle_vertical_dynamics}
\begin{align}
\frac{\mathrm{d}}{\mathrm{d} t} \theta &= \dot \theta,  \\
\frac{\mathrm{d}}{\mathrm{d} t} \dot \theta &= \frac{\sin \theta}{H} g - \frac{\cos \theta}{H} a = \frac{\sin \theta}{H} g - \frac{\cos \theta}{H} \frac{v^2}{L} \tan \beta.
\end{align}
\end{subequations}

Combine (\ref{eq:motorcycle_horizontal_dynamics}) and (\ref{eq:motorcycle_vertical_dynamics}) to obtain the state differential equation (\ref{eq:motorcycle_state_DE})
\begin{align*}
\frac{\mathrm{d}}{\mathrm{d} t} \mathbf{x} \equiv \frac{\mathrm{d}}{\mathrm{d} t} \begin{bmatrix} x \\ y \\ \phi \\ \beta \\ \theta \\ \frac{\mathrm{d} \theta}{\mathrm{d} t} \end{bmatrix} 
= \begin{bmatrix} v \cos \phi \\ v \sin \phi \\ \frac{v}{L} \tan \beta \\ \frac{1}{\tau_{\beta}} (\beta_I - \beta) \\ \frac{\mathrm{d} \theta}{\mathrm{d} t} \\
\frac{\sin \theta}{H} g - \frac{\cos \theta}{H} \frac{v^2}{L} \tan \beta \end{bmatrix} \equiv f(\mathbf{x}, \beta_I).
\end{align*}

\subsubsection*{Motorcycle lateral dynamics}

For analysis of motorcycle lateral dynamics, the motorcycle lateral position $y$ and the motorcycle orientation angle $\phi$ actually refer to the lateral position and orientation angle of the motorcycle with respect to certain local road reference. In other words, $y$ and $\phi$ here refer to the relative lateral position and orientation angle in certain local road reference, instead of absolute ones in the global world reference. This is similar to how we handle vehicle lateral dynamics.

For sake of formalizing motorcycle lateral dynamics, simply remove $x$ from the motorcycle state 
\begin{align*}
\begin{bmatrix} x & y & \phi & \beta & \theta & \frac{\mathrm{d} \theta}{\mathrm{d} t} \end{bmatrix}^\mathrm{T}
\end{align*}
and its corresponding equation from (\ref{eq:motorcycle_state_DE}) to obtain the state differential equation (\ref{eq:motorcycle_lateral_control})
\begin{align*}
\frac{\mathrm{d}}{\mathrm{d} t} \mathbf{x} \equiv \frac{\mathrm{d}}{\mathrm{d} t} \begin{bmatrix} y \\ \phi \\ \beta \\ \theta \\ \frac{\mathrm{d} \theta}{\mathrm{d} t} \end{bmatrix}
= \begin{bmatrix} v \sin \phi \\ \frac{v}{L} \tan \beta \\ - \frac{1}{\tau_{\beta}} \beta \\ \frac{\mathrm{d} \theta}{\mathrm{d} t} \\
\frac{\sin \theta}{H} g - \frac{\cos \theta}{H} \frac{v^2}{L} \tan \beta \end{bmatrix} + \begin{bmatrix} 0 \\ 0 \\ \frac{1}{\tau_{\beta}} \\ 0 \\ 0 \end{bmatrix} \beta_I \equiv f(\begin{bmatrix} y \\ \phi \\ \beta \\ \theta \\ \frac{\mathrm{d} \theta}{\mathrm{d} t} \end{bmatrix}) + \begin{bmatrix} 0 \\ 0 \\ \frac{1}{\tau_{\beta}} \\ 0 \\ 0 \end{bmatrix} \beta_I.
\end{align*}
The motorcycle lateral state 
\begin{align*}
\mathbf{x} \equiv \begin{bmatrix} y & \phi & \beta & \theta & \frac{\mathrm{d} \theta}{\mathrm{d} t} \end{bmatrix}^\mathrm{T}
\end{align*}
specified in (\ref{eq:motorcycle_lateral_control}) consists of the motorcycle lateral position, the motorcycle orientation or heading angle (namely yaw angle), the motorcycle steering angle, the motorcycle vertical angle (namely roll angle), and the motorcycle vertical angular velocity only.

\subsubsection*{Linearized motorcycle lateral dynamics}

If the motorcycle orientation angle $\phi$ (namely yaw angle), the motorcycle vertical angle $\theta$ (namely roll angle), and the motorcycle steering angle $\beta$ are close to zero, then following approximations
\begin{align*}
\sin \phi \approx \phi, & \qquad \tan \beta \approx \beta,  \\
\sin \theta \approx \theta, & \qquad \cos \theta \approx 1
\end{align*}
can be taken respectively and hence the state differential equation (\ref{eq:motorcycle_lateral_control}) can be fairly linearized about the equilibrium state and simplified into the linear state differential equation (\ref{eq:motorcycle_lateral_control_approximation})
\begin{align*}
\frac{\mathrm{d}}{\mathrm{d} t} \mathbf{x} = \begin{bmatrix} 0 & v & 0 & 0 & 0 \\ 0 & 0 & \frac{v}{L} & 0 & 0 \\ 0 & 0 & -\frac{1}{\tau_{\beta}} & 0 & 0 \\ 0 & 0 & 0 & 0 & 1 \\ 0 & 0 & -\frac{v^2}{H L} & \frac{g}{H} & 0 \end{bmatrix} \mathbf{x} + \begin{bmatrix} 0 \\ 0 \\ \frac{1}{\tau_{\beta}} \\ 0 \\ 0 \end{bmatrix} \beta_I \equiv \mathbf{A} \mathbf{x} + \mathbf{B} \beta_I.
\end{align*}
The motorcycle lateral state specified in (\ref{eq:motorcycle_lateral_control_approximation}) is the same to that specified in (\ref{eq:motorcycle_lateral_control}).

It is worth noting that we may also linearize motorcycle complete dynamics described by (\ref{eq:motorcycle_state_DE}). If so, the state transition matrix $\mathbf{A}$ and the control input matrix $\mathbf{B}$ would no longer be fixed, but will vary according to the concrete state about which the state differential equation (\ref{eq:motorcycle_state_DE}) is linearized.

\subsubsection{Negligence of motorcycle steering dynamics}  \label{sec:motorcycle_dynamics_neglect_steering}

For the motorcycle state 
\begin{align*}
\mathbf{x} \equiv \begin{bmatrix} x & y & \phi & \beta & \theta & \frac{\mathrm{d} \theta}{\mathrm{d} t} \end{bmatrix}^\mathrm{T},
\end{align*}
evolution of the forth state element $\beta$ depends only on the fourth differential equation of (\ref{eq:motorcycle_state_DE})
\begin{align*}
\frac{\mathrm{d}}{\mathrm{d} t} \beta = \frac{1}{\tau_{\beta}} (\beta_I - \beta) \iff \tau_{\beta} \frac{\mathrm{d}}{\mathrm{d} t} \beta + \beta = \beta_I.
\end{align*}
In other words, motorcycle steering dynamics is independent of other state dynamics.

The motorcycle steering angle command $\beta_I$ is normally constant during a control period. Consider motorcycle steering dynamics in a generic control period. Let 
\begin{align*}  
\beta_0 \equiv \beta(0), \quad \beta_{\Delta T} \equiv \beta(\Delta T) 
\end{align*}
denote the steering angles at the beginning and end of the control period respectively. To solve the fourth differential equation, compute its characteristic equation namely
\begin{align*}
\tau_{\beta} \lambda + 1 = 0
\end{align*}
and obtain 
\begin{align*}
\lambda = -1/\tau_{\beta}. 
\end{align*}
Apply the following function transformation
\begin{align*}
\beta = \mathrm{e}^{\lambda t} \bar{\beta} = \mathrm{e}^{-\frac{1}{\tau_{\beta}} t} \bar{\beta}
\end{align*}
and obtain
\begin{align*}
\frac{\mathrm{d}}{\mathrm{d} t} \bar{\beta} = \frac{1}{\tau_{\beta}} \mathrm{e}^{\frac{1}{\tau_{\beta}} t} \beta_I,
\end{align*}
which implies that
\begin{align*}
\bar{\beta} = \mathrm{e}^{\frac{1}{\tau_{\beta}} t} \beta_I + \bar{\beta}_0 - \beta_I = \mathrm{e}^{\frac{1}{\tau_{\beta}} t} \beta_I + \beta_0 - \beta_I
\end{align*}
and
\begin{equation}  \label{eq:motorcycle_dynamics_beta}
\beta = \mathrm{e}^{-\frac{1}{\tau_{\beta}} t} \beta_0 +  (1 - \mathrm{e}^{-\frac{1}{\tau_{\beta}} t}) \beta_I.
\end{equation}
In fact, if the fourth differential equation is expressed in the state differential equation form as
\begin{align*}
\frac{\mathrm{d}}{\mathrm{d} t} \begin{bmatrix} \beta \end{bmatrix} = \begin{bmatrix} -\frac{1}{\tau_{\beta}} \end{bmatrix} \begin{bmatrix} \beta \end{bmatrix} + \begin{bmatrix} \frac{1}{\tau_{\beta}} \end{bmatrix} \beta_I,
\end{align*}
then (\ref{eq:motorcycle_dynamics_beta}) can also be derived by taking advantage of (\ref{eq:state_differential_equation_linear_solution}) as
\begin{align*}
\begin{bmatrix} \beta \end{bmatrix} = \mathrm{e}^{\begin{bmatrix} -\frac{1}{\tau_{\beta}} \end{bmatrix} t} \beta_0 + \int_0^t \mathrm{e}^{\begin{bmatrix} -\frac{1}{\tau_{\beta}} \end{bmatrix} (t - \tau)} \begin{bmatrix} \frac{1}{\tau_{\beta}} \end{bmatrix} \beta_I \mathrm{d} \tau = \mathrm{e}^{-\frac{1}{\tau_{\beta}} t} \beta_0 +  (1 - \mathrm{e}^{-\frac{1}{\tau_{\beta}} t}) \beta_I.
\end{align*}

The steering angle at the end of the control period, i.e. $\beta_{\Delta T}$, is computed via (\ref{eq:motorcycle_dynamics_beta}) as
\begin{align*}
\beta_{\Delta T} = \mathrm{e}^{-\frac{\Delta T}{\tau_{\beta}}} \beta_0 +  (1 - \mathrm{e}^{-\frac{\Delta T}{\tau_{\beta}}}) \beta_I.
\end{align*}
The mean steering angle during the control period is
\begin{align*}
\beta_M &= \frac{1}{\Delta T} \int_0^{\Delta T} \beta \mathrm{d} t = \frac{1}{\Delta T} \int_0^{\Delta T} [\mathrm{e}^{-\frac{1}{\tau_{\beta}} t} \beta_0 +  (1 - \mathrm{e}^{-\frac{1}{\tau_{\beta}} t}) \beta_I] \mathrm{d} t  \\
  &= \beta_I - \frac{\tau_{\beta}}{\Delta T} (\beta_I  - \beta_0) (1 - \mathrm{e}^{-\frac{\Delta T}{\tau_{\beta}}}).
\end{align*}
If motorcycle steering dynamics is negligible or in other words if
\begin{align*}
\tau_{\beta} \ll \Delta T,
\end{align*}
then
\begin{align*}
\mathrm{e}^{-\frac{\Delta T}{\tau_{\beta}}} \approx 0, \qquad 1 - \mathrm{e}^{-\frac{\Delta T}{\tau_{\beta}}} \approx 1,  \qquad \frac{\tau_{\beta}}{\Delta T} \approx 0
\end{align*}
and hence
\begin{align*}
\beta_{\Delta T} \approx \beta_I,  \qquad  \beta_M \approx \beta_I.
\end{align*}
The two equations above convey that both the final steering effect and the average steering effect of the control period can be approximated by the steering effect under constant $\beta_I$. So
\begin{equation}  \label{eq:motorcycle_dynamics_beta_tau=0}
\beta \approx \beta_I
\end{equation}
can be regarded to hold when motorcycle steering dynamics is negligible.

Substitute (\ref{eq:motorcycle_dynamics_beta_tau=0}) into (\ref{eq:motorcycle_state_DE}) and obtain
\begin{equation}  \label{eq:motorcycle_state_DE_beta_tau=0}
\frac{\mathrm{d}}{\mathrm{d} t} \mathbf{x} \equiv \frac{\mathrm{d}}{\mathrm{d} t} \begin{bmatrix} x \\ y \\ \phi \\ \theta \\ \frac{\mathrm{d} \theta}{\mathrm{d} t} \end{bmatrix} 
= \begin{bmatrix} v \cos \phi \\ v \sin \phi \\ \frac{v}{L} \tan \beta \\ \frac{\mathrm{d} \theta}{\mathrm{d} t} \\
\frac{\sin \theta}{H} g - \frac{\cos \theta}{H} \frac{v^2}{L} \tan \beta \end{bmatrix} \equiv f(\mathbf{x}, \beta),
\end{equation}
where the steering angle 
\begin{align*}
\beta \approx \beta_I
\end{align*}
is removed from the motorcycle state
\begin{align*}
\begin{bmatrix} x & y & \phi & \beta & \theta & \frac{\mathrm{d} \theta}{\mathrm{d} t} \end{bmatrix}^\mathrm{T}
\end{align*}
and serves directly as control input to the motorcycle. The reduced motorcycle state 
\begin{align*}
\mathbf{x} \equiv \begin{bmatrix} x & y & \phi & \theta & \frac{\mathrm{d} \theta}{\mathrm{d} t} \end{bmatrix}^\mathrm{T}
\end{align*}
consists of the motorcycle longitudinal position, the motorcycle lateral position, the motorcycle orientation or heading angle (namely yaw angle), the motorcycle vertical angle (namely roll angle), and the motorcycle vertical angular velocity only.

Similarly, substitute (\ref{eq:motorcycle_dynamics_beta_tau=0}) into (\ref{eq:motorcycle_lateral_control}) and obtain
\begin{equation}  \label{eq:motorcycle_lateral_control_beta_tau=0}
\frac{\mathrm{d}}{\mathrm{d} t} \mathbf{x} \equiv \frac{\mathrm{d}}{\mathrm{d} t} \begin{bmatrix} y \\ \phi \\ \theta \\ \frac{\mathrm{d} \theta}{\mathrm{d} t} \end{bmatrix} 
= \begin{bmatrix} v \sin \phi \\ \frac{v}{L} \tan \beta \\ \frac{\mathrm{d} \theta}{\mathrm{d} t} \\
\frac{\sin \theta}{H} g - \frac{\cos \theta}{H} \frac{v^2}{L} \tan \beta \end{bmatrix} \equiv f(\mathbf{x}, \beta),
\end{equation}
where the steering angle $\beta$ is removed from the motorcycle lateral state
\begin{align*}
\begin{bmatrix} y & \phi & \beta & \theta & \frac{\mathrm{d} \theta}{\mathrm{d} t} \end{bmatrix}^\mathrm{T}
\end{align*}
and serves directly as control input to motorcycle lateral control. The reduced motorcycle lateral state 
\begin{align*}
\mathbf{x} \equiv \begin{bmatrix} y & \phi & \theta & \frac{\mathrm{d} \theta}{\mathrm{d} t} \end{bmatrix}^\mathrm{T}
\end{align*}
consists of the motorcycle lateral position, the motorcycle orientation or heading angle (namely yaw angle), the motorcycle vertical angle (namely roll angle), and the motorcycle vertical angular velocity only. Like (\ref{eq:motorcycle_lateral_control_approximation}) is the linearized counterpart of (\ref{eq:motorcycle_lateral_control}), the following state differential equation
\begin{align}  \label{eq:motorcycle_lateral_control_approximation_beta_tau=0}
\frac{\mathrm{d}}{\mathrm{d} t} \mathbf{x} = \begin{bmatrix} 0 & v & 0 & 0 \\ 0 & 0 & 0 & 0 \\ 0 & 0 & 0 & 1 \\ 0 & 0 & \frac{g}{H} & 0 \end{bmatrix} \mathbf{x} + \begin{bmatrix} 0 \\ \frac{v}{L} \\ 0 \\ -\frac{v^2}{H L} \end{bmatrix} \beta \equiv \mathbf{A} \mathbf{x} + \mathbf{B} \beta
\end{align}
is the linearized counterpart of (\ref{eq:motorcycle_lateral_control_beta_tau=0}). The state specified in (\ref{eq:motorcycle_lateral_control_approximation_beta_tau=0}) is the same to that specified in (\ref{eq:motorcycle_lateral_control_beta_tau=0}).

\subsubsection*{Negligible motorcycle steering dynamics thanks to smooth steering operations}

The condition that the time-constant of the steer controller (i.e. $\tau_{\beta}$) is small enough with respect to the control period (i.e. $\beta_{\Delta T}$) is not the only condition under which motorcycle steering dynamics can be neglected. In fact, as demonstrated by the application example of simplified motorcycle lateral control presented in Section 2.2.3 in Chapter 2, 
\footnote{Namely Chapter 2 of the author's works \cite{Li2026ACTPA_SJTU_2, Li2026ACTPA_SJTU_1}. Note that this article is Chapter 1 of the works.}
the simplified version of motorcycle lateral control method still works even when the configured time-constant $\tau_{\beta}$ is by no means small enough to be directly negligible.

Some intuitive explanations hover over the phenomena. We had better realize that motorcycle lateral control is inclined towards smooth steering operations. Here, smooth steering operations include two aspects of smoothness: first, change of the steering angle is smooth, and second, the steering angle is moderate. It is like when we ride a motorcycle or bicycle, we tend to take smooth steering operations instead of drastic steering operations. If we did take drastic steering operations such as an abrupt change of the steering angle or a turning with some large steering angle, then we would very likely get unbalanced and fall down --- From this we can see that motorcycle lateral control and vehicle lateral control are essentially different, though they share a common or similar part of dynamics. For vehicle lateral control, after all, we do not need to worry about the issue of keeping balance and can take drastic steering operations (though not recommended so in normal circumstances). In contrast, for motorcycle lateral control, we need to worry about the issue of keeping balance.

\begin{figure}[h!]
\begin{center}
\includegraphics[width=0.6\columnwidth]{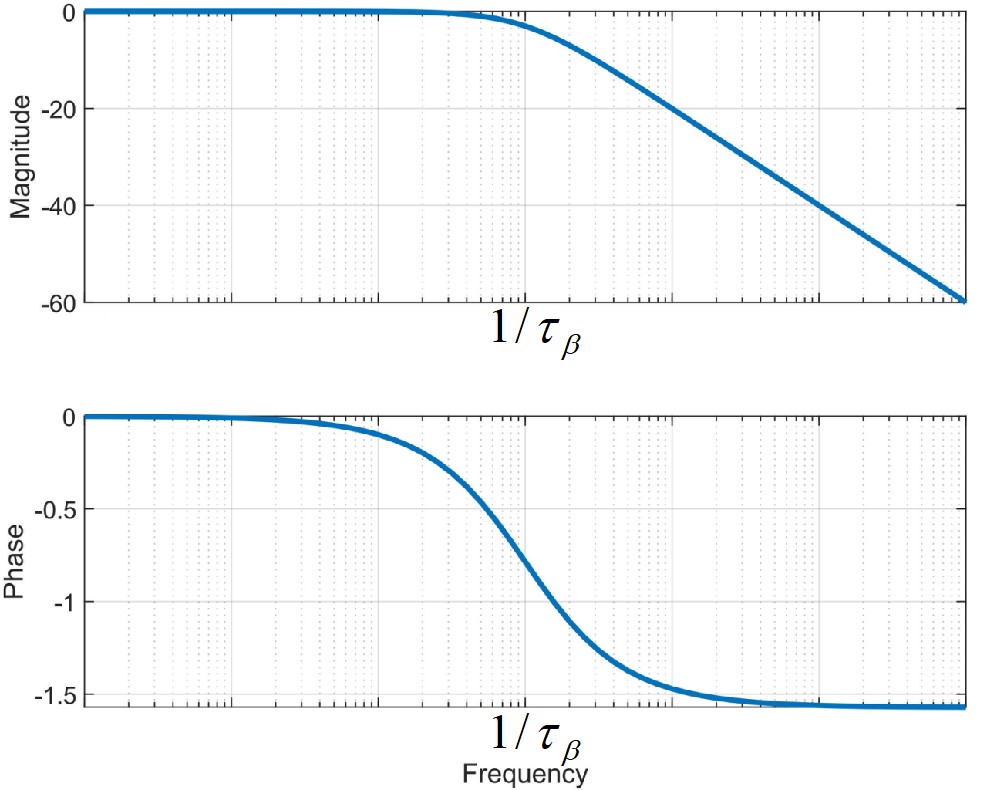}
\end{center}
\caption{Bode plots of the frequency transfer function}
\label{fig:LPF_bode_plot}
\end{figure}

Then on one hand, smooth motorcycle steering angle commands appear somehow like low-frequency signals. On the other hand, the fourth differential equation of (\ref{eq:motorcycle_state_DE})
\begin{align*}
\frac{\mathrm{d}}{\mathrm{d} t} \beta = \frac{1}{\tau_{\beta}} (\beta_I - \beta) \iff \tau_{\beta} \frac{\mathrm{d}}{\mathrm{d} t} \beta + \beta = \beta_I,
\end{align*}
which describes motorcycle steering dynamics, functions somehow like a low-pass filter \cite{Oppenheim1997} between the filter input namely the motorcycle steering angle command $\beta_I$ and the filter output namely the motorcycle steering angle $\beta$. To understand this last point, perform the Laplace transform on both sides of the fourth differential equation of (\ref{eq:motorcycle_state_DE}) and obtain
\begin{align*}
(\tau_{\beta} s + 1) \beta (s) = \beta_I (s) \iff \frac{\beta (s)}{\beta_I (s)} = \frac{1}{1 + \tau_{\beta} s}
\end{align*}
the frequency transfer function of which is
\begin{align*}
T(\omega \mathrm{j}) = \frac{1}{1 + \tau_{\beta} \omega \mathrm{j}}.
\end{align*}
When 
\begin{align*}
\omega \ll \frac{1}{\tau_{\beta}}
\end{align*}
or loosely 
\begin{align*}
\omega < \frac{1}{\tau_{\beta}}, 
\end{align*}
the magnitude of $T(\omega \mathrm{j})$ is 
\begin{align*}
| T(\omega \mathrm{j}) | \approx 1
\end{align*}
and the phase or phase angle of $T(\omega \mathrm{j})$ is 
\begin{align*}
\angle T(\omega \mathrm{j}) \approx 0, 
\end{align*}
which implies that low-frequency signals can almost be passed without losing fidelity. In contrast, when 
\begin{align*}
\omega \gg \frac{1}{\tau_{\beta}}
\end{align*}
or loosely 
\begin{align*}
\omega > \frac{1}{\tau_{\beta}}, 
\end{align*}
the magnitude of $T(\omega \mathrm{j})$ is 
\begin{align*}
| T(\omega \mathrm{j}) | \approx 0,
\end{align*}
which implies that high-frequency signals are suppressed. The low-pass filtering characteristic of the frequency transfer function can also be intuitively reflected by its Bode plots \cite{Bode1940} illustrated in Figure \ref{fig:LPF_bode_plot}.

So smooth motorcycle steering angle commands are passed through such low-pass filter, as if the approximated relationship (\ref{eq:motorcycle_dynamics_beta_tau=0})
\begin{align*}
\beta \approx \beta_I
\end{align*}
holds or in other words as if motorcycle steering dynamics is neglected.


\newpage
\addcontentsline{toc}{chapter}{Bibliography}

\fancyhf{} 

\bibliographystyle{unsrt}
\bibliography{LI_Hao_Refs_ACTPA}

@article{Anagnost1991,
author={J. Anagnost and C. Desoer},
title={An elementary proof of the Routh-Hurwitz stability criterion},
journal={Circuits Systems Signal Process},
volume={10},
number={1},
pages={101-114},
year={1991}
}

@book{Arnold1989,
author={V. Arnold},
title={Mathematical methods of classical mechanics},
publisher={Springer Science \& Business Media},
year={1989}
}

@book{Baker2005,
author={G. Baker and J. Blackburn},
title={The pendulum: a case study in physics},
publisher={Oxford University Press},
year={2005}
}

@article{Bode1940,
author={H. Bode},
title={Relations between attenuation and phase in feedback amplifier design},
journal={The Bell System Technical Journal},
volume={19},
number={3},
pages={421-454},
year={1940}
}

@book{Boyd1994,
author={S. Boyd and L. Ghaoui and E. Feron and V. Balakrishnan},
title={Linear matrix inequalities in system and control theory},
publisher={Society for Industrial and Applied Mathematics},
year={1994}
}

@book{Dorf2008,
author={R. Dorf and R. Bishop},
title={Modern control systems},
publisher={Pearson Prentice Hall},
year={2008}
}

@book{Feynman2004,
author={R. Feynman},
title={The Feynman lectures on physics (commemorative issue)},
publisher={Pearson Education},
year={2004}
}

@article{Foucault1851,
author={L. Foucault},
title={Démonstration physique du mouvement de rotation de la Terre au moyen du pendule},
journal={Comptes Rendus Hebdomadaires des Séances de l’Académie des Sciences},
volume={32},
number={},
pages={135-138},
year={1851}
}

@book{Golub1996,
author={G. Golub and C. Van Loan},
title={Matrix computations},
publisher={Johns Hopkins University Press},
year={1996}
}

@book{Hermite,
author={C. Hermite},
title={Oeuvres de Charles Hermite},
publisher={Cambridge university press},
year={2009}
}

@article{Hilbert1904,
author={D. Hilbert},
title={Grundzüge einer allgemeinen Theorie der linearen Integralgleichungen},
journal={Nachrichten von der Gesellschaft der Wissenschaften zu Göttingen, Mathematisch-Physikalische Klasse},
volume={},
number={},
pages={213-259},
year={1904}
}

@book{Horn1991,
author={R. Horn and C. Johnson},
title={Topics in Matrix Analysis},
publisher={Cambridge University Press},
year={1991}
}

@book{Horn2012,
author={R. Horn and C. Johnson},
title={Matrix analysis},
publisher={Cambridge University Press},
year={2012}
}

@book{Jazar2014,
author={R. Jazar},
title={Vehicle dynamics: theory and application},
publisher={Springer},
year={2014}
}

@article{Kleinman1968,
author={D. Kleinman},
title={On an iterative technique for {R}iccati equation computations},
journal={IEEE Transactions on Automatic Control},
volume={13},
number={1},
pages={114 - 115},
year={1968}
}

@book{LaValle2006,
author={S. LaValle},
title={Planning algorithms},
publisher={Cambridge university press},
year={2006}
}

@book{Li2024CTPA_SJTU_1,
author={\begin{CJK}{UTF8}{gbsn}李颢\end{CJK}},
title={\begin{CJK}{UTF8}{gbsn}面向实际应用的控制理论（英文版）\end{CJK}},
publisher={\begin{CJK}{UTF8}{gbsn}上海交通大学出版社\end{CJK}},
year={2024}
}

@book{Li2024CTPA_Springer,
author={H. Li},
title={Control theory for practical applications: with {MATLAB} demonstration programs},
publisher={Springer},
year={2024}
}

@book{Li2026ACTPA_SJTU_1,
author={\begin{CJK}{UTF8}{gbsn}李颢\end{CJK}},
title={\begin{CJK}{UTF8}{gbsn}面向实际应用的高级控制理论（英文版）\end{CJK}},
publisher={\begin{CJK}{UTF8}{gbsn}上海交通大学出版社\end{CJK}},
year={2026}
}

@book{Li2026ACTPA_SJTU_2,
author={H. Li},
title={Advanced control theory for practical applications},
publisher={Shanghai Jiao Tong University Press},
year={2026}
}

@article{Liapounoff1900,
author={A. Liapounoff},
title={Sur une proposition de la théorie des probabilités},
journal={Bulletin de l'Académie Impériale des Sciences de Saint-Pétersbourg},
volume={13},
number={4},
pages={359-386},
year={1900}
}

@book{Liapounoff1907,
author={A. Liapounoff},
title={Problème général de la stabilité du mouvement (traduit du russe)},
publisher={Princeton University Press},
year={1907}
}

@article{Mason1956,
author={S. Mason},
title={Feedback theory-further properties of signal flow graphs},
journal={Proceedings of the IRE},
volume={44},
number={7},
pages={920-926},
year={1956}
}

@book{Mitrinovic1970,
author={D. Mitrinovic and P. Vasic},
title={Analytic inequalities},
publisher={Springer-Verlag Berlin Heidelberg},
year={1970}
}

@book{Oppenheim1997,
author={A. Oppenheim and A. Willsky and S. Nawab},
title={Signals and systems},
publisher={Pearson Education},
year={1997}
}

@article{Pacejka1992,
author={H. Pacejka and E. Bakker},
title={The magic formula tyre model},
journal={Vehicle System Dynamics},
volume={21},
number={S1},
pages={},
year={1992}
}

@book{Rajamani2012,
author={R. Rajamani},
title={Vehicle dynamics and control},
publisher={Springer Science \& Business Media},
year={2012}
}

@article{Samad2017,
author={T. Samad},
title={A Survey on Industry Impact and Challenges Thereof [Technical Activities]},
journal={IEEE Control Systems Magazine},
volume={37},
number={1},
pages={17-18},
year={2017}
}

@book{Tao2010,
author={T. Tao},
title={An epsilon of room, I: real analysis},
publisher={American Mathematical Society},
year={2010}
}

@book{Tao2011,
author={T. Tao},
title={An introduction to measure theory},
publisher={American Mathematical Society},
year={2011}
}

\fancyhead[LE,RO]{\thepage}
\fancyhead[RE]{\textit{ \nouppercase{\leftmark}} }
\fancyhead[LO]{\textit{ \nouppercase{\rightmark}} }

\end{document}